\documentclass[a4paper,fleqn,usenatbib]{mnras}

\usepackage{multicol}
\usepackage[T1]{fontenc}
\usepackage{ae,aecompl}

\usepackage{graphicx}	% Including figure files
\usepackage{amsmath}	% Advanced maths commands
\usepackage{amssymb}	% Extra maths symbols

\title[TESS light curves of cataclysmic variables -- III]
{TESS light curves of cataclysmic variables -- III -- More superhump
systems among old novae and novalike variables}

\author[A. Bruch]{Albert Bruch
\\
Laborat\'orio Nacional de Astrof\'{\i}sica, Rua Estados Unidos, 154, 
CEP 37500-364, Itajub\'a, MG, Brazil
}

\date{Accepted XXX. Received YYY; in original form ZZZ}

\pubyear{2019}

\begin{document}
\label{firstpage}
\pagerange{\pageref{firstpage}--\pageref{lastpage}}
\maketitle

% Abstract of the paper
\begin{abstract}
Continuing previous work on the identification and 
characterization of periodic and non-periodic variations in long and almost 
uninterrupted high cadence light curves of cataclysmic variables observed 
by the TESS mission, the
results on 23 novalike variables and old novae out of sample of 127 such
systems taken from the Ritter \& Kolb catalogue are presented. All of
them exhibit at least at some epochs either positive or negative (or both)
superhumps, and in 19 of them superhumps were detected for the first time.
The basic properties of the superhumps such as their periods, their
appearance and disappearance, and their waveforms are explored.
Together with recent reports in the literature, this elevates the number
of known novalike variables and old novae with superhumps by more than 50\%.
The previous census of superhumps and the Stolz-Schoembs
relation for these stars are updated. Attention is drawn to superhump 
properties in some stars which behave differently from the 
average, as well as to positive superhumps in high mass ratio systems
which defy theory. As a byproduct, the orbital periods of 13 stars are either
improved or newly measured, correcting previously reported erroneous values.
\end{abstract}

% Select between one and six entries from the list of approved keywords.
% Don't make up new ones.

%%%%%%%%%%%%%%%%%%%%%%%%%%%%%%%%%%%%%%%%%%%%%%%%%%

%%%%%%%%%%%%%%%%% BODY OF PAPER %%%%%%%%%%%%%%%%%%

\section{Introduction}
\label{Introduction}

The Kepler space mission \citep{Borucki10} has opened up new horizons for 
time resolved astronomical photometry. For the first time long (often
extending over years) almost uninterrupted light curves with a high temporal
resolution of up to 58~sec became available for many stars. Although mainly
designed for the detection of exoplanets the Kepler data also proved to be of 
extraordinary value for the study and characterization of variable stars
in unprecedented detail. Kepler was retired in late 2018.

In many respects the Transiting Exoplanet Survey Telescope
\citep[TESS,][]{Ricker14}, 
also mainly focussed on exoplanet detection, is a continuation of
the Kepler mission. It has the advantage that, in contrast to Kepler who
observed only a limited part of the sky, over the course of its lifetime
TESS covered almost the entire sky. Thus, many more targets could be observed
with TESS as compared to Kepler. On the other hand, due to the observing
strategy\footnote{https://tess.mit.edu/observations} the time base of
the light curves of a single star is limited to about a month or small
multiples thereof, but the same object may have been observed at different
epochs, months or years apart. 
Moreover, additional limitations are imposed by the smaller
size of the telescopes onboard TESS and the coarse pixel resolution of 
about 20'' of its detectors. Even so, the TESS database is a treasure trove 
for the study of variability in many different classes of astronomical
objects.  

Here, I focus on TESS (and to a small degree on Kepler) observations of
a subgroup of cataclysmic variables (CVs), i.e., old novae and novalike
variables. As is well known, CVs are close interacting binaries, where
a Roche lobe overflowing late type star -- mostly on or close to the main 
sequence -- transfers matter to a white dwarf (WD). In the absence of strong
magnetic fields of the latter an accretion disk is formed around the compact
object before the transferred matter is finally deposited onto its surface.
See \citet{Warner95} for a comprehensive review of all aspects of CVs.

Disk accreting CVs can roughly be divided into two large groups. The dwarf
novae exhibit more or less frequently outbursts in their light curves which
can be as small as 1~mag in some, or can reach up to 6-7~mag in other systems.
In dwarf novae the average mass transfer rate from the late type secondary
star remains below a given threshold, but for much of the time is higher 
than the accretion rate onto the WD, meaning that matter accumulates in the 
disk over time. In the context of the disk-instability model 
\citep[see, e.g.,][]{Hameury20} the outbursts are then explained by an 
instability in the accretion disk which sets in after enough matter has 
accumulated. In consequence much more matter is dumped onto the WD than 
before. The sudden release of gravitational energy explains
the outburst. After the disk has been depleted of much of its matter the
outburst subsides and the cycle begins anew. 

In contrast, the members of the second large group have an average mass 
transfer rate above the threshold which leads to dwarf novae outbursts.
They are called novalike variables and include also CVs which have 
suffered from a classical (or recurrent) nova outburst in recent times, 
i.e., old novae (for simplicity, I will hereafter lump together systems
traditionally named novalike variables and old novae under the common
abbreviation NL). They remain in a more or less stable high brightness 
state which can be viewed as a permanent dwarf nova outburst (although members 
of a particular subgroup, the VY~Scl stars, sometimes drop into a fainter
low state for a limited time interval). Only NLs are topic of this study.

Among the numerous variable phenomena observed in CVs are superhumps (SHs),
i.e., variations with a period slightly longer (positive superhumps, pSHs) 
or shorter (negative superhumps, nSHs) than the orbital period. pSHs are
routinely observed in the longer and brighter than usual outbursts
(i.e., superoutbursts) that occur in the SU~UMa type subclass of dwarf
novae, but are much rarer in NLs. They are explained as variations of the
disk brightness in an apsidally precessing excentric accretion disk. 
nSHs have been observed in different classes of CVs but in individual 
systems they remain an exception. They are thought to be caused by an 
inclination or tilt of the accretion disk with respect to the orbital plane
together with a retrograde nodal disk precession. For a brief exposition 
of the respective mechanisms and for further references, see the introduction
of \citet{Bruch23}.

TESS light curves of NLs have been very useful to identify and characterize
superhumps in NLs. In the first paper of this series 
\citep[][hereafter referred to as Paper~I]{Bruch22} I used TESS data to
identified SHs in several systems which were hitherto not known to be 
superhumpers. This motivated a second study \citep[][Paper~II]{Bruch23}, 
this time focussed on a detailed characterization of SHs in TESS data of 
known superhump systems, leading to the most complete census of SHs in
NLs performed so far. In continuation of this work, I have analysed the
available TESS light curves of all objects classified as NLs in the final 
version (December 2016) of the Ritter \& Kolb catalogue 
\citep[][RK hereafter]{Ritter03} which have not yet been dealt with in 
previous publications. However, being mainly interested in phenomena
occurring in the accretion disk, I excluded the highly magnetic AM~Her stars 
which are also classified as NLs by RK but do not possess disks. This 
yielded an ensemble of 127 targets. Among these, I found many systems 
exhibiting SHs. The characterization of their basic properties and
temporal behaviour is the topic of this exploratory study. The investigation 
of other interesting aspects and additional features found in the 127 
stars from the RK catalogue will be published separately.

\section{Data and data handling}
\label{Data and data handling}

The details of the data used in this study and their handling are largely
the same as in Papers~I and II to which the reader is referred for more 
details. TESS data with a time resolution of 2~min were downloaded from the 
Barbara A.\ Mikulski Archive for Space Telescopes 
(MAST)\footnote{https://archive.stsci.edu}. According to the motivation
outlined in Paper~I in most cases SAP data are used. Only when the SAP 
light curves contained features which apparently are not real such as strong
gradients or discontinuities across data gaps, PDS-SAP
light curves were preferred. The difference between SAP and PDS-SAP data
affects mainly variations on time scales of days but not of hours which
are or interest here. In view of a possible contamination of the light
curves caused by neighbouring stars or an inadequate background subtracton, 
given the coarse spatial resolution of the TESS telescopes, in no case the 
absolute flux values or the absolute amplitude of variations are used to 
infer scientific conclusions. For one object (NS~Cnc)
data from the Kepler mission, retrieved from the same source quoted above, 
are also used. 

TESS observes different sectors of the sky continuously for about 27 days 
during two spacecraft orbits. These observations are, however, interrupted for
a few days after each orbit in order to download the data to Earth. Thus,
each 27~day light curve contains a gap in the middle. Further gaps
may be introduced due to the exclusion of intervals with bad data. 

Depending on their location on the sky different TESS sectors may overlap 
each other. This means that a given object may be included in more than one
sector. If these sectors are observed in immediate succession, the light
curves of these objects can be combined and then extend over a longer time 
interval. Moreover,
some of the sectors have been observed more than once, generating light
curves at different epochs. Here, I designate light curves of a given star, 
derived from observations in one or more sectors adjacent in time
as LC\#1, LC\#2, etc. A list of target stars is provided in 
% Table~\ref{Table: Target stars} 
Table~1 where the second column contains the CV subtype as defined in the
AAVSO International Variable Star Index, using standard notation. 
It may differ from the type given
by RK. The third column give the range of variability of the system, taken
from the same source. The other column contain, for each TESS light curve,
the average $g$ band magnitude as derived from ASAS-SN light curves. The
error estimated from the standard deviation of the data points is of the
order of 0.1 -- 0.2~mag. In some
cases the TESS observations fall into a gap of the ASAS-SN data. Then, some 
ASAS-SN magnitudes just before and after the gap were used to estimate
the brightness of the target during the TESS observations.
A list of all light curves is given in 
% Table~\ref{Table: Light curves} 
Table~2 where for each object and light curve
the respective TESS sectors and the time interval in Julian Dates are
listed. When comparing TESS data to observations taken 
with other instruments it should be kept in mind that its passband encompasses 
a wide range between 6\,000 and 10\,000~\AA, centred on the Cousins $I$-band.
Kepler has a similarly broad passband, but offset by roughly 1\,000~\AA\ to
the blue. 

%--------------------------------------------------------------
\begin{table*}
\label{Table: Target list}	
\centering
	\caption{List of target stars (column~1) together with their respective
CV types (column~2), total range of variability (column~3) and average
$g$ band magnitudes during each TESS light curve (columns 4--8).}

\begin{tabular}{llllllll}
\hline
Name &
Type &
magnitude range &
\multicolumn{5}{l}{agerage $g$ magnitude} \\
     &
     &
     &
LC\#1 &
LC\#2 &
LC\#3 &
LC\#4 &
LC\#5 \\
\hline
OR And &
NL/VY  &
14.5 -- 19.0 V &
14.7 &
14.6 &
-- &
-- &
-- \\
LS Cam &
NL     &
16.7 -- 19.5 V &
16.2 &
16.8 &
16.3 &
16.2 &
16.3 \\
NS Cnc &
NL+E  &
15.2 -- 17.7 CV &
16.2 &
15.9 &
15.7 &
-- &
-- \\
V425 Cas &
NL/VY  &
14.4 -- 18.0 V &
14.4 &
14.7 &
-- &
-- &
-- \\
V1024 Cep &
NL/VY+E  &
14.7 -- 20.7 CV &
15.3 &
15.3 &
15.5 &
15.1 &
15.1 \\ [1ex]
DN Gem &
Na    &
3.6 -- 16.0 B &
14.1 &
14.1 &
14.1 &
-- &
-- \\
V1084 Her &
NL  &
12.48 -- 12.75 V &
12.7 &
12.7 &
-- &
-- &
-- \\
CP Lac &
NA/VY  &
2.1 V -- 20.4 CV &
16.6 &
16.4 &
-- &
-- &
-- \\
DK Lac &
NA+NL/VY &
5.0 p - 19.4 V &
16.7 &
17.3 &
-- &
-- &
-- \\
KQ Mon &
NL &
12.1 -- 13.0 p &
13.3 &
-- &
-- &
-- &
-- \\ [1ex]
LZ Mus &
NA  &
8.5 -- $<$18 V &
?  &
-- &
-- &
-- &
-- \\
FY Per &
NL/VY  &
11.9 -- 14.5 V &
12.6 &
12.8 &
-- &
-- &
-- \\
LX Ser &
NL/VY+E  &
13.3 - 17.4 B &
15.0 &
15.1 &
-- &
-- &
-- \\
EI UMa &
UG/DQ  &
12.5 CV -- 15.8 V &
14.4 &
14.3 &
-- &
-- &
-- \\
LN UMa &
UGZ/IW+VY  &
14.6 -- 18 V &
15.3 &
15.1 &
15.2 &
15.1 &
-- \\ [1ex]
CN Vel &
NB  &
9.8 -- $<$16.5 p &
?  &
-- &
-- &
-- &
-- \\
HS 0229+8016 &
NL|UGZ:  &
13.4 -- 15.1 V &
14.2 &
14.1 &
14.1 &
13.9 &
-- \\
HS 0506+7725 &
NL/VY  &
14.6 V -- 18.5 B &
14.8 &
14.9 &
14.8 &
15.0 &
-- \\
HS 0642+5049 &
NL  &
15.2 - 16.0 V &
14.8 &
15.0 &
-- &
-- &
-- \\
IGR J08390-4833 &
DQ  &
16.1 -- ? R &
16.6 &
16.7 &
-- &
-- &
-- \\ [1ex]
H$\alpha$ 1039-4701 &
CV  &
16.4 -- ? R &
16.3 &
-- &
-- &
-- &
-- \\
H$\alpha$ 1129-5355 &
CV  &
15.5 -- ? R &
15.7 &
-- &
-- &
-- &
-- \\
ASASS-14ix &
UG+E  &
15.4 -- 19.6 CV &
16.7 &
-- &
-- &
-- &
-- \\
\hline
\end{tabular}
\end{table*}
%------------------------------------------------------------------

%--------------------------------------------------------------
\begin{table*}
\label{Table: Light curves}	\
\centering
	\caption{List of TESS light curves.}

\begin{tabular}{lllllllllll}
\hline
Name &
LC\#1 &
      &
LC\#2 &
      &
LC\#3 &
      &
LC\#4 &
      &
LC\#5 &
      \\
      &
TESS &
Time (JD) &
TESS &
Time (JD) &
TESS &
Time (JD) &
TESS &
Time (JD) &
TESS &
Time (JD) \\
      &
Sector &
2450000+ &
Sector &
2450000+ &
Sector &
2450000+ &
Sector &
2450000+ &
Sector &
2450000+ \\
\hline
OR And &
16--17 &
8738--8789 &
57    &
9853--9883 &
-- &
-- &
-- &
-- &
-- &
-- \\
LS Cam &
19--20 &
8816--8869 &
26    &
9010--9036 &
40    &
9390--9419 &
53    &
9743--9769 &
59--60 &
9910--9963 \\
NS Cnc &
5$^*$      &
7139--7215 &
18$^*$     &
8251--8303 &
44--46     &
9500--9579 &
-- &
-- &
-- &
-- \\
V425 Cas &
17         &
8764--8788 &
57         &
9853--9883 &
-- &
-- &
-- &
-- &
-- &
-- \\
V1024 Cep &
19--20     &
8816--8869 &
25--26     &
8983--9036 &
40         &
9390--9419 &
52-53      &
9718--9769 &
59-60 &
9910--9963 \\ [1ex]
DN Gem &
20         &
8842--8869 &
45         &
9525--9551 &
47 &
9579--9607 &
-- &
-- &
-- &
-- \\ [1ex]
V1084 Her &
51         &
9692--9718 &
52         &
9718--9742 &
-- &
-- &
-- &
-- &
-- &
-- \\
CP Lac &
16--17     &
8738--8789 &
56--57     &
9825--9883 &
-- &
-- &
-- &
-- &
-- &
-- \\
DK Lac &
16--17     &
8738--8789 &
34         &
9853--9883 &
-- &
-- &
-- &
-- &
-- &
-- \\
KQ Mon &
34 &
9928--9955 &
-- &
-- &
-- &
-- &
-- &
-- &
-- &
-- \\ [1ex]
LZ Mus &
37--38     &
9307--9355 &
-- &
-- &
-- &
-- &
-- &
-- &
-- &
-- \\ [1ex]
FY Per &
19         &
8816--8842 &
59         &
9910--9937 &
-- &
-- &
-- &
-- &
-- &
-- \\
LX Ser &
24         &
8955--8983 &
51         &   
9692--9718 &
-- &
-- &
-- &
-- &
-- &
-- \\
EI UMa &
20         &
8842--8869 &
47         &
9579--9607 &
-- &
-- &
-- &
-- &
-- &
-- \\
LN UMa &
14         &
8683--8711 &
20--21     &
8842--8898 &
40--41     &
9390--9447 &
47         &
9579--9607 &
-- &
-- \\ [1ex]
CN Vel &
36--37 &
9582--9333 &
-- &
-- &
-- &
-- &
-- &
-- &
-- &
-- \\ [1ex]
HS 0229+8016 &
19--22     &
8790--8842 &
25--26     &
8983--9036 &
52-53      &
9718--9769 &
59         &
9910--9937 &
-- &
-- \\
HS 0506+7725 &
19--20     &
8816--8869 &
25--26     &
8983--9036 &
52--53     &
9718--9769 &
59--60     &
9910--9963 &
-- &
-- \\
HS 0642+5049 &
20         &
8842--8869 &
60         &
9938--9963 &
-- &
-- &
-- &
-- &
-- &
-- \\
IGR J08390--4833 &
8--9       &
8517--8569 &
35--36     &
9254--9306 &
-- &
-- &
-- &
-- &
-- &
-- \\ [1ex]
H$\alpha$ 1039--4701 &
36--37     &
9280--9333 &
-- &
-- &
-- &
-- &
-- &
-- &
-- &
-- \\
H$\alpha$ 1129--5355 &
37         &
9308--9333 &
-- &
-- &
-- &
-- &
-- &
-- &
-- &
-- \\
ASASSN--14ix &
28        &
9061-9088 &
-- &
-- &
-- &
-- &
-- &
-- &
-- &
-- \\
\hline
\end{tabular}
$^*$Kepler K2 campain
\end{table*}
%------------------------------------------------------------------

The main purpose of this study is the search of periodic variations in
the target stars. Therefore, I make ample use of Fourier techniques to
calculate periodograms (hereafter also termed power spectra) applying
the Lomb-Scargle algorithm \citep{Lomb76, Scargle82} or following
\citet{Deeming75}. Both yield equivalent results. In systems with deep 
eclipses, these are masked before calculating power spectra in order to
avoid that the signals due to the eclipses and their overtones dominate
the spectra.

No attempt is made to formally quantify the false alarm probability or the
significance of peaks in the power spectra. Methods to do so in one way or 
another assume that the ``continuum'' is caused by random (white) noise. 
Due to flickering or random variations on longer time scales this 
assumption is at most approximately justified in CVs at very high frequencies. 
Therefore, I rely
on a visual assessment of the strength of a power spectrum peak. In most
of the examples shown in the figures of this paper there is no doubt about
their significance. In other cases previous knowledge is available. This 
refers sometimes to the orbital frequency if the orbital signal is weak in
the power spectra. Similarly, there is previous knowledge about overtone
frequencies or linear combinations of frequencies of different periodicities.
This helps to identify fainter signals. In doubtful cases I will use cautious 
terms upon identifying possible signals.
Absolute power levels in power spectra are not considered. They can be very
different for different light curves. Therefore, the vertical scale of the
various power spectrum plots is in general not labelled. Unless otherwise said
the lower limit is always 0, and power is plotted on a linear scale.

Waveforms of periodical variations contain much information about their
causes. However, the construction of waveforms (i.e., folding the light
curve on the respective period) of often small amplitude
is complicated by the presence of other modulations in the light curves.
If these occur on time scales much longer than the considered periodic
variations, they can be eliminated by subtracting a \citet{Savitzky64} 
filtered version of the light curve from the original one. Here, this
filter is used with a cut-off time scale of 2~d and a 4$^{\rm th}$ order
smoothing polynomial. But this does not solve problems for the construction
of waveforms due to the superposition of more than one periodicity with
periods of similar order of magnitude, e.g., orbital and superhump periods.
In principal, variations caused by one phenomenon should be subtracted from 
the light curve before constructing the waveform of the other. Fortunately,
thanks to the long TESS light curves which cover many cycles this is not
necessary. Folding the data on one period, variations of the other(s) are
evenly distributed in phase. Binning the phase folded light curves in
intervals (phase intervals of 0.01 are used throughout this paper) then
cancels out other periodicities to a very high degree. The minimum of the
waveforms is arbitrarily chosen as the zero point of phase.

In the figures of this paper I use the notation $\Omega$ for the orbital
frequency, and $\Sigma^-$ and $\Sigma^+$, respectively, for negative and
positive SH frequencies (and in one case $\omega$ for the WD spin frequency). 
In the text, the corresponding periods are termed $P_{\rm orb}$, $P_{\rm nSH}$, 
and $P_{\rm pSH}$. Most of the periods measured here are derived from
the frequency of peaks in power spectra. Their errors are estimated according
to the recipe in Sect.~4.4 of \citet{Schwarzenberg-Czerny91}. Or, when
several independent measurements are available (e.g., the orbital 
frequency measured in more than one light curve), the standard deviation
of the mean is taken. 
To simplify notation, errors are given in brackets in units of the
last decimal digits of the nominal value of the respective quantity.

In some systems periodic signals may not persist over the entire duration
of the light curve or their strength in power spectra may change 
significantly with time. Even variations of their frequencies occur. Such
properties can best be traced using time resolved (or dynamical) power
spectra. In these cases power spectra of the light curve in sliding windows 
of width $W$ are calculated and plotted as a function of the mid-points of
the windows. The step width between subsequent windows is $\Delta W$. 
This means that structures in the dynamical power spectra separated in 
time by less than $W$ are not independent. Unless otherwise specified, I use
$W = 4$~d and $\Delta W = 0.4$~d. 

\section{Results}
\label{Results}

% Table~\ref{Table: Masterlist} 
Table~3 contains an overview of the results of the
present study. It lists all stars with newly detected superhumps and is 
organized similarly to Table~4 of Paper~II. The orbital period is given
in italics whenever a more precise values than hitherto known could be
measured in the TESS data. Otherwise, literature values are quoted and
the reference is given in square brackets and specified at the end of the
table. The superhump period excess is defined as 
$\epsilon = \left( P_{\rm SH} - P_{\rm orb} \right) / P_{\rm orb}$. The table also
provides information about the SH waveform which is here classified as either S
for sinusoidal or NS (non-sinusoidal) if clear deviations from a sinusoidal 
shape exist. But
note that deviations from a pure sine wave may be hidden in noise when
superhumps are weak or the star is faint. The detection or non-detection 
of variations on the disk precession period is also indicated. In contrast
to Paper~II, I do not try to classify the superhumps as permanent of transient
because the limited observational coverage does not permit to be reasonably
certain that any of the stars exhibits SHs permanently.

%--------------------------------------------------------------
\begin{table*}
\label{Table: Masterlist}	\
\centering
	\caption{Summary of superhump properties of the target stars.}

\begin{tabular}{llllllllll}
\hline
Name &
orb. period &
\multicolumn{4}{l}{negative superhump} &
\multicolumn{4}{l}{positive superhump} \\
  &
(d) &
period (d) &
$\epsilon^a_{\rm nSH}$ &
WF$^b$ &
Prec$^c$ &
period (d) &
$\epsilon^a_{\rm pSH}$ &
WF$^b$ &
Prec$^c$ \\
\hline
OR And                     &     % Name of star
{\it 0.13569(7)}           &     % Orbital period (d)
0.125240(6)                &     % Negative superhump period
$-0.077$                   &     % period excess (nSH)
S                          &     % Waveform
Y                          &     % Precession (Y/N)
--                         &     % Positive superhump period
--                         &    % period excess (pSH)
--                         &     % Waveform
--                         \\    % Precession (Y/N)
% % 
% V405 Aur (?)               &     % Name of star
% 0.1726196 [1]              &     % Orbital period (d)
% 0.1501(1)                  &     % Negative superhump period
% $-0.131$                   &     % period excess (nSH)
% --                         &     % Waveform
% --                         &     % Precession (Y/N)
% --                         &     % Positive superhump period
% --                         &    % period excess (pSH)
% --                         &     % Waveform
% --                         \\    % Precession (Y/N)
% 
LS Cam                     &     % Name of star
0.1423853 [1]              &     % Orbital period (d)
0.13747(7)                 &     % Negative superhump period
$-0.035$                   &     % period excess (nSH)
NS                         &     % Waveform
Y                          &     % Precession (Y/N)
0.1547(4)                  &     % Positive superhump period
0.086                      &    % period excess (pSH)
NS                          &     % Waveform
N                          \\    % Precession (Y/N)
NS Cnc                     &     % Name of star
0.1600525 [7]              &     % Orbital period (d)
0.152042(6)                &     % Negative superhump period
$-0.050$                   &     % period excess (nSH)
NS                         &     % Waveform
Y                          &     % Precession (Y/N)
0.17463(3)                 &     % Positive superhump period
0.091                      &     % period excess (pSH)
NS                          &     % Waveform
Y                          \\    % Precession (Y/N)
V425 Cas                   &     % Name of star
{\it 0.14896(7)}           &     % Orbital period (d)
0.14509(3)                 &     % Negative superhump period
$-0.026$                   &     % period excess (nSH)
S                          &     % Waveform
N                          &     % Precession (Y/N)
0.1654(2)                  &     % Positive superhump period
0.110                      &    % period excess (pSH)
S                          &     % Waveform
N                          \\    % Precession (Y/N)
V1024 Cep                  &     % Name of star
0.14872403 [2]             &     % Orbital period (d)
0.142990(9)                &     % Negative superhump period
$-0.039$                   &     % period excess (nSH)
NS                         &     % Waveform
Y                          &     % Precession (Y/N)
--                         &     % Positive superhump period
--                         &    % period excess (pSH)
--                         &     % Waveform
--                         \\ [1ex]   % Precession (Y/N)
DN Gem                     &     % Name of star
{\it 0.12783(8)}           &     % Orbital period (d)
0.1224(2)                  &     % Negative superhump period
$-0.044$                   &     % period excess (nSH)
NS                         &     % Waveform
N                          &     % Precession (Y/N)
--                         &     % Positive superhump period
--                         &     % period excess (pSH)
--                         &     % Waveform
--                         \\ [1ex]   % Precession (Y/N)
V1084 Her                  &     % Name of star
0.120560 [3]               &     % Orbital period (d)
0.11692(2)                 &     % Negative superhump period
$-0.030$                   &     % period excess (nSH)
NS                         &     % Waveform
Y                          &     % Precession (Y/N)
--                         &     % Positive superhump period
--                         &     % period excess (pSH)
--                         &     % Waveform
--                         \\    % Precession (Y/N)
CP Lac                     &     % Name of star
0.145143 [4]               &     % Orbital period (d)
0.13886(1)                 &     % Negative superhump period
$-0.043$                   &     % period excess (nSH)
S                          &     % Waveform
Y                          &     % Precession (Y/N)
--                         &     % Positive superhump period
--                         &     % period excess (pSH)
--                         &     % Waveform
--                         \\    % Precession (Y/N)
DK Lac                     &     % Name of star
{\it 0.116644(4)}          &     % Orbital period (d)
--                         &     % Negative superhump period
--                         &     % period excess (nSH)
--                         &     % Waveform
--                         &     % Precession (Y/N)
0.12963(5)                 &     % Positive superhump period
0.113                      &     % period excess (pSH)
S                          &     % Waveform
N                          \\    % Precession (Y/N)
KQ Mon                     &     % Name of star
{\it 0.13450(3)}           &     % Orbital period (d)
0.12895(1)                 &     % Negative superhump period
$-0.041$                   &     % period excess (nSH)
S                          &     % Waveform
Y                          &     % Precession (Y/N)
--                         &     % Positive superhump period
--                         &     % period excess (pSH)
--                         &     % Waveform
--                         \\ [1ex]   % Precession (Y/N)
LZ Mus                     &     % Name of star
{\it 0.16348(5)}           &     % Orbital period (d)
0.15098(7)                 &     % Negative superhump period
$-0.077$                   &     % period excess (nSH)
S                          &     % Waveform
N                          &     % Precession (Y/N)
--                         &     % Positive superhump period
--                         &     % period excess (pSH)
--                         &     % Waveform
--                         \\ [1ex]   % Precession (Y/N)
FY Per                     &     % Name of star
0.2585 [3]                 &     % Orbital period (d)
0.24400(3)                 &     % Negative superhump period
$-0.056$                   &     % period excess (nSH)
NS                         &     % Waveform
N                          &     % Precession (Y/N)
--                         &     % Positive superhump period
--                         &     % period excess (pSH)
--                         &     % Waveform
--                         \\    % Precession (Y/N)
LX Ser                     &     % Name of star
0.158432491 [5]            &     % Orbital period (d)
0.15177(8)                 &     % Negative superhump period
$-0.042$                   &     % period excess (nSH)
NS                         &     % Waveform
Y                          &     % Precession (Y/N)
--                         &     % Positive superhump period
--                         &     % period excess (pSH)
--                         &     % Waveform
--                         \\    % Precession (Y/N)
EI UMa                     &     % Name of star
{\it 0.2684(1)}            &     % Orbital period (d)
--                         &     % Negative superhump period
--                         &     % period excess (nSH)
--                         &     % Waveform
--                         &     % Precession (Y/N)
0.317 -- 0.354             &     % Positive superhump period
0.181 -- 0.319             &     % period excess (pSH)
NS                         &     % Waveform
N                          \\    % Precession (Y/N)
LN UMa                     &     % Name of star
{\it 0.14471(4)}           &     % Orbital period (d)
0.138339(5)                &     % Negative superhump period
$-0.044$                   &     % period excess (nSH)
NS                         &     % Waveform
N                          &     % Precession (Y/N)
--                         &     % Positive superhump period
--                         &     % period excess (pSH)
--                         &     % Waveform
--                         \\ [1ex]   % Precession (Y/N)
CN Vel                     &     % Name of star
{\it 0.22338(6)}           &     % Orbital period (d)
--                         &     % Negative superhump period
--                         &     % period excess (nSH)
--                         &     % Waveform
--                         &     % Precession (Y/N)
0.24513(7)                 &     % Positive superhump period
0.097                      &     % period excess (pSH)
S                          &     % Waveform
N                          \\ [1ex]   % Precession (Y/N)
HS 0229+8016               &     % Name of star
0.161439(9)                &     % Orbital period (d)
0.15369(3)                 &     % Negative superhump period
$-0.048$                   &     % period excess (nSH)
S                          &     % Waveform
N                          &     % Precession (Y/N)
--                         &     % Positive superhump period
--                         &     % period excess (pSH)
--                         &     % Waveform
--                         \\    % Precession (Y/N)
HS 0506+7725               &     % Name of star
0.1477 [6]                 &     % Orbital period (d)
0.1417(1)                  &     % Negative superhump period
$-0.041$                   &     % period excess (nSH)
S                          &     % Waveform
N                          &     % Precession (Y/N)
--                         &     % Positive superhump period
--                         &     % period excess (pSH)
--                         &     % Waveform
--                         \\    % Precession (Y/N)
HS 0642+5049               &     % Name of star
{\it 0.15791(2)}           &     % Orbital period (d)
--                         &     % Negative superhump period
--                         &     % period excess (nSH)
--                         &     % Waveform
--                         &     % Precession (Y/N)
0.18471(6)                 &     % Positive superhump period
0.170                      &     % period excess (pSH)
S                          &     % Waveform
Y                          \\    % Precession (Y/N)
IGR J08390-4833            &     % Name of star
{\it 0.25408(2)}           &     % Orbital period (d)
0.24013(4)                 &     % Negative superhump period
$-0.055$                   &     % period excess (nSH)
NS                         &     % Waveform
Y                         &     % Precession (Y/N)
--                         &     % Positive superhump period
--                         &     % period excess (pSH)
--                         &     % Waveform
--                         \\ [1ex]   % Precession (Y/N)
H$\alpha$ 1039-4701        &     % Name of star
{\it 0.15769(2)}           &     % Orbital period (d)
0.15286(2)                 &     % Negative superhump period
$-0.031$                   &     % period excess (nSH)
NS                         &     % Waveform
Y                          &     % Precession (Y/N)
--                         &     % Positive superhump period
--                         &    % period excess (pSH)
--                         &     % Waveform
--                         \\    % Precession (Y/N)
H$\alpha$ 1129-5355        &     % Name of star
0.153546 [8]               &     % Orbital period (d)
0.14796(8)                 &     % Negative superhump period
$-0.036$                   &     % period excess (nSH)
S                          &     % Waveform
Y                          &     % Precession (Y/N)
0.1571(1)                  &     % Positive superhump period
0.023                      &     % period excess (pSH)
--                         &     % Waveform
--                         \\    % Precession (Y/N)
ASASSN-14ix                &     % Name of star
0.1444610954 [9]           &     % Orbital period (d)
0.1342(2)                  &     % Negative superhump period
$-0.044$                   &     % period excess (nSH)
S                          &     % Waveform
N                          &     % Precession (Y/N)
--                         &     % Positive superhump period
--                         &     % period excess (pSH)
--                         &     % Waveform
--                         \\    % Precession (Y/N)
\hline
\multicolumn{10}{l}{$^a$ Period excess defined as 
            $\epsilon = \left( P_{\rm SH}-P_{\rm orb} \right)/P_{\rm orb}$} \\
\multicolumn{10}{l}{$^b$ Waveform: S = sinusoidal; 
                                  NS = not (strictly) sinusoidal} \\
\multicolumn{10}{l}{$^c$ Precession period detected (Yes/No)} \\
\multicolumn{10}{l}{
References: 
%[1] \citet{Still98};
[1] \citet{Thorstensen17};
[2] \citet{Rodriguez-Gil07a};
% [3] \citet{Lujan07};
[3] \citet{Patterson02};}\\
\multicolumn{10}{l}{
[4] \citet{Peters06};
%[4] \citet{Gaensicke09};
[5] \citet{Li17};
[6] \citet{Aungwerojwit05};
[7] \citet{Warner02};}\\
\multicolumn{10}{l}{
[8] \citet{Pretorius08};
[9] \citet{Hambsch14b}}.
\end{tabular}
\end{table*}
%------------------------------------------------------------------

Subsequently, the individual systems are discussed in more detail.

\subsection{OR Andromedae}
\label{OR And}

OR~And is a poorly observed CV. An orbital period of 0.1359~d is quoted by
RK on the basis of an internet communication by J.\ Patterson which is no 
longer available. Without giving details \citet{Barlow22}, using a part of 
the same TESS data investigated here, found a dominant period of 0.125246~d. 
They do not specify its nature.

\begin{figure}
	\includegraphics[width=\columnwidth]{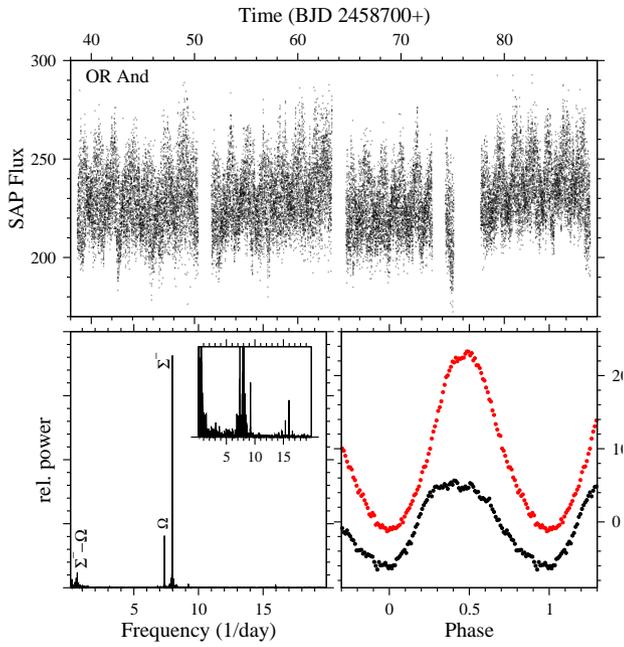}
%      \fbox{\rule[14cm]{14cm}{0cm}}
\caption{Light curve (top) and power spectrum (lower left) of LC\#1 of OR~And,
and waveforms (lower right) of the orbital (black) and superhump (red) 
variations averaged over LC\#1 and LC\#2.}
\label{orand}
\end{figure}

TESS observed OR~And in two subsequent sectors in 2019 and then again in 2022
in one sector. The combined 2019 light curve
(upper frame of Fig.~\ref{orand}) is characterized by strong periodic 
variations on the time scale of 1--2~d. It generates a (suprisingly
faint) signal in the power spectrum (lower left frame of the figure) 
at a frequency equal to the frequency difference between two stronger signals,
one being compatible with the orbital period quoted by RK and the other (the 
dominant signal) at a slightly higher frequency. This is the signal already 
mentioned by \citet{Barlow22}. Fainter peaks 
(inset in the figure) can all be identified as overtones or linear 
combinations of the main signals. Thus, the 0.125~d modulation can be 
interpreted as a negative superhump. The 2022 light curve yields almost
identical results. The only noteworthy difference is the replacement of the
orbital -- superhump beat signal by a signal at exactly twice (within the
formal error limits) the beat frequency.
Both, the orbital (black in the lower right frame of the figure) and the 
superhump waveform (red), averaged over both light curves, are 
nearly sinusoidal. The orbital power spectrum signal permits a slight 
revision of the period as listed in Table~3. 
% Table~\ref{Table: Masterlist}.
% Orbital period (LC#1): 0.1357279 pm 0.000013919    (Gewicht 2)
% SH      period (LC#1): 0.12524359 pm 0.000006515   (Gewicht 2)
% Orbital period (LC#2): 0.1355981827 pm 0.000034474 (Gewicht 1)
% SH      period (LC#2): 0.125232622 pm 0.0000111689 (Gewicht 1)

%--------------------------------------------------------------

\subsection{LS Camelopardalis}
\label{LS Cam}

The only detailed study of LS~Cam was published by \citet{Dobrzycka98}. They
found radial velocity, flux and equivalent width variations in the
He~II~$\lambda$4686 emission line with a period of $\approx$50~min which they 
attribute to the Alv\'en radius of the WD. This apparently caused RK to classify
the system as an intermediate polar (IP) candidate. The TESS light curves do not
contain indications for an IP nature of LS~Cam. The tentative orbital period
cited by \citet{Dobrzycka98} was later revised by \citet{Thorstensen17} to 
be 0.1423853(5)~d.

TESS observed LS~Cam in seven sectors. The data of the first and the last 
two sectors can be combined. Thus, five light 
curves are available. The subsequent frequency analysis leads to conclusions
similar to those obtained by \citet{Rawat22} based on the same TESS data
(except LC\#5), but addresses also some additional aspects. 

\begin{figure}
	\includegraphics[width=\columnwidth]{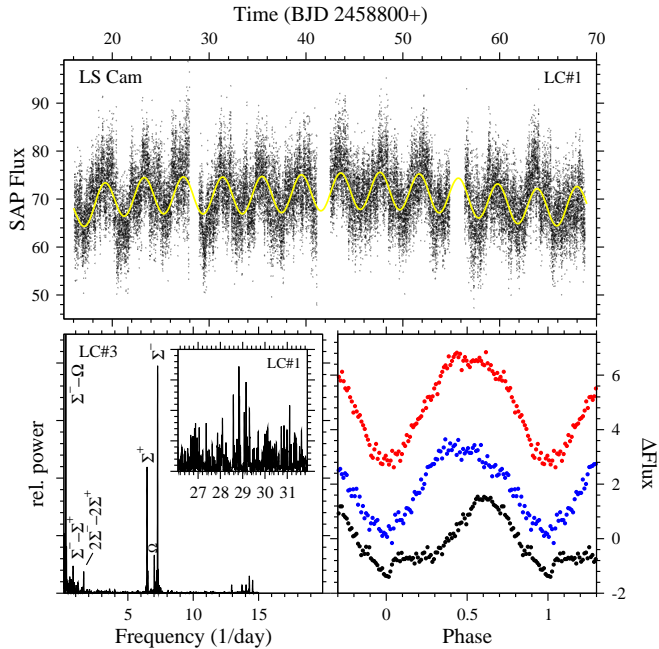}
%      \fbox{\rule[14cm]{14cm}{0cm}}
\caption{{\it Top:} Light curve LC\#1 of LS~Cam. The yellow curve is the 
sum of a low order polynominal (to follow more gradual variations) and a 
sine curve fitted to the data. {\it Bottom left:} Power 
spectrum of LC\#3. The inset shows a part of the power spectrum of LC\#1 
with possible QPOs consistent with $\approx$50~min variations reported 
by \citet{Dobrzycka98}. {\it Bottom right:} The waveforms
(average of LC\#1 -- LC\#5 for orbit and nSH, and of LC\#2 -- LC\#4
for pSH), referring to (from bottom to top) the orbital, pSH and 
nSH variations.}
\label{lscam}
\end{figure}

\begin{figure}
	\includegraphics[width=\columnwidth]{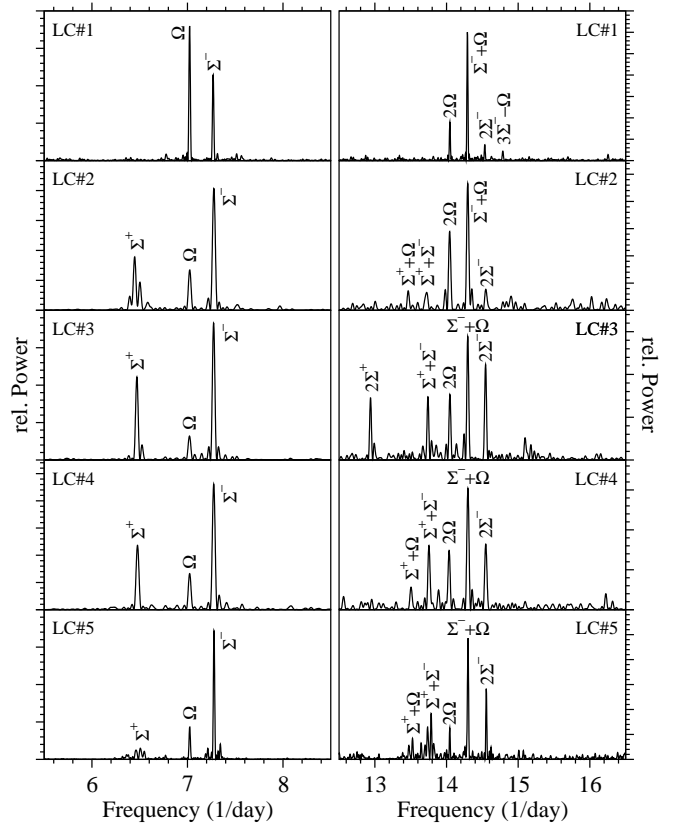}
%      \fbox{\rule[14cm]{14cm}{0cm}}
\caption{Power spectra of the five light curves of LS~Cam in the
range around the orbital frequency (left) and its first overtone (right).} 
\label{lscam-ps}
\end{figure}

All light curves exhibit a well expressed periodic modulation close to 
4 days. The upper frame of Fig.~\ref{lscam} shows light curve LC\#1.
The yellow curve is the sum of a low order polynominal (to follow more
gradual variations) and a sine curve fitted to the data. The obvious 
periodicity suggests to be the beat between the orbital and a superhump 
modulation. This is confirmed by the power spectra. The lower left frame of
Fig.~\ref{lscam}
contains the spectrum of LC\#3. Fig.~\ref{lscam-ps} provides a  more 
detailed view. The power spectra of all light curves in a small range 
around the orbital frequency (left) and its first overtone (right) are 
displayed. Apart from the orbital signal LC\#1 clearly exhibits a negative
superhump together with signals at the frequencies of simple arithmetic 
combinations of the orbital and superhump frequencies.
While in LC\#1 the superhump signal is still weaker than the orbital one,
it outshines the latter in all other light curves. In addition to the negative
superhump, in the later light curves a positive superhump is present which
leads to some complexity of the power spectra around the overtone of the
orbital frequency. In LC\#2 the pSH signal is weak and appears to split up
into three components. In fact, a dynamical power spectrum reveals a signal
at slightly different frequencies in the first three quarters of the light
curve (being weaker in the second quarter) which then vanishes totally in
the last quarter. The exact period of the superhumps varies slightly from
one epoch to the next. The values cited in 
%Table~\ref{Table: Masterlist}.
Table~3 are averages.
Thus, LS~Cam belongs to the growing group of CVs which exhibit positive and
negative superhumps simultaneously. As expected considering the strong 4~day 
variability all power spectra contain a well expressed peak at the (negative) 
superhump -- orbital beat frequency (extreme left hand edge in the
lower left frame of Fig.~\ref{lscam}), corresponding to a period of 4.004~d. 
On the other hand, a beat signal with the orbital period caused by the 
positive superhump is not seen. However in LC\#3 and LC\#4 (and marginally
also in LC\#2) the beat between the two superhumps and its first overtone 
appear.

% LC#1 orb: 0.1423990279 +- 0.0000131838
% LC#2 orb: 0.1423769295 +- 0.0000399160
% LC#3 orb: 0.1424280256 +- 0.0000439976
% LC#4 orb: 0.1423919201 +- 0.0000495296
% LC#5 orb: 0.1423795670 +- 0.0000155533

% LC#1 nSH: 0.1375775039 +- 0.0000151987
% LC#2 nSH: 0.1374283284 +- 0.0000224110
% LC#3 nSH: 0.1374800950 +- 0.0000159117
% LC#4 nSH: 0.1374578029 +- 0.0000253938
% LC#5 nSH: 0.1374011487 +- 0.0000072018

% LC#2 pSH: 0.1551032066 +- 0.0000392203
% LC#3 pSH: 0.1545513719 +- 0.0000262990
% LC#4 pSH: 0.1543936580 +- 0.0000461994

%--------------------------------------------------------------
%______________________________________________________________

%--------------------------------------------------------------
%______________________________________________________________

The waveforms of the orbital (black), positive (blue) and negative (red) 
superhump variations, averaged over all light curves 
(but restricted to LC\#2 -- LC\#4 for the pSH), are shown in 
the lower right frame of Fig.~\ref{lscam}. The orbital waveform 
consists of a hump encompassing 60--70\% of the orbit. The slight depression
after the hump may indicated shallow eclipses in LS~Cam. 

\begin{figure}
	\includegraphics[width=\columnwidth]{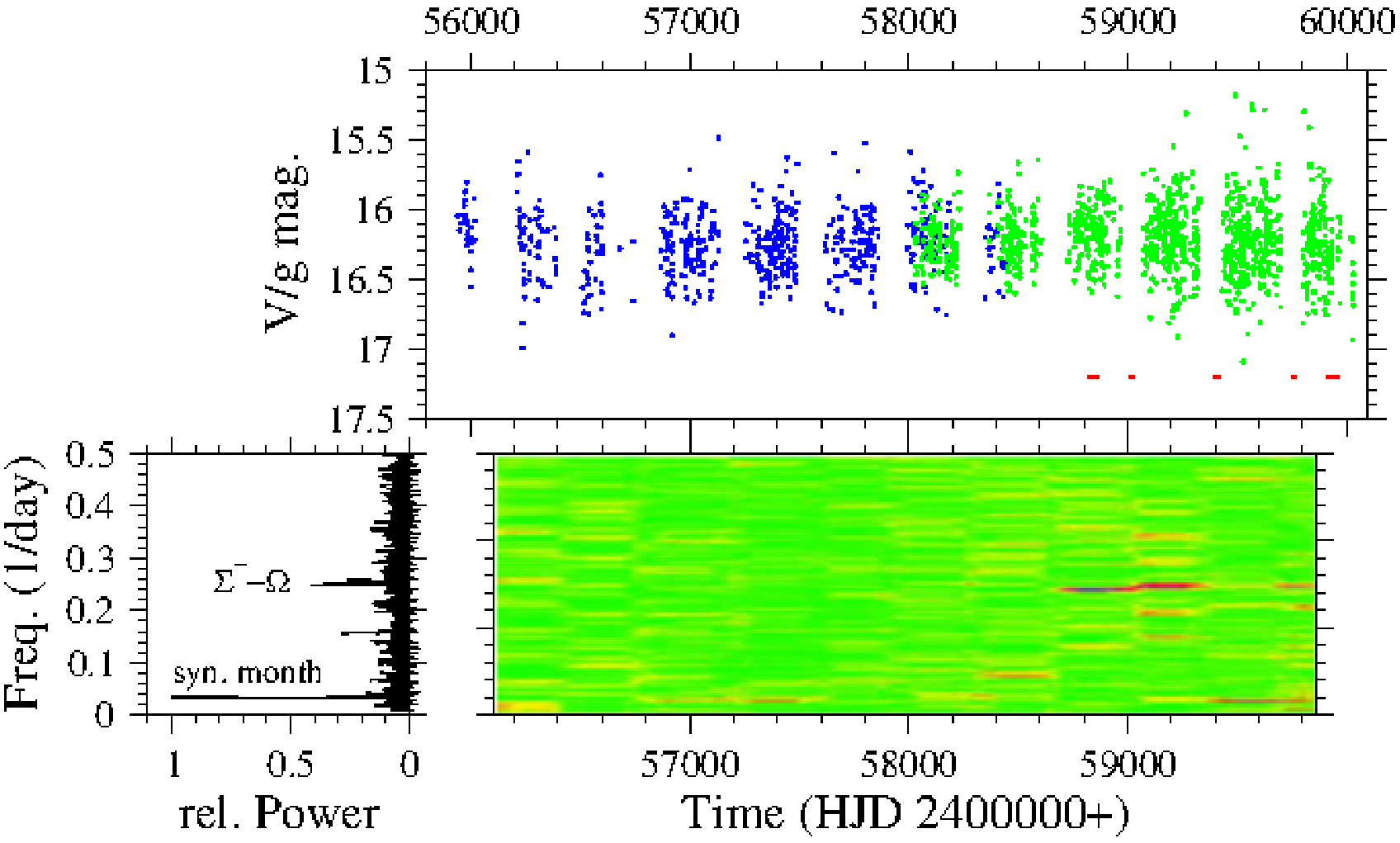}
%      \fbox{\rule[14cm]{14cm}{0cm}}
\caption{{\it Top:} ASAS-SN long term light curve of LS~Cam. The blue and
green dots refer to the $V$ and $g$ bands, respectively. The red bars
beneath the light curve indicate the intervals when TESS observed LS~Cam.
{\it Bottom:} Power spectrum of the ASAS-SN light curve in the conventional
(left) and time resolved (right) forms.}
\label{lscam-stacked}
\end{figure}

Although all five TESS light curves clearly exhibit the negative superhump
this is not a permanent feature of LS~Cam but only developed in recent
years. The upper frame of Fig.~\ref{lscam-stacked} contains the ASAS-SN
long term light curve of the system (blue dots: $V$ band; green dots: $g$ 
band). The small red bars indicate the intervals covered by the TESS data. 
The lower left frame shows the power spectrum of the light curve. The peak 
at low frequencies corresponds to a synodic month and can thus be attributed 
to data sampling. More importantly, two closely spaced signals at 0.2503
and 0.2568~d$^{-1}$ have frequencies which are almost identical to the
difference between the orbital and nSH frequencies (ranging between
0.245 and 0.255~d$^{-1}$ in LC\#1 -- LC\#5). Thus, in spite of the much lower
cadence the ASAS-SN data doubtlessly exhibit the beat between orbit and
superhump which is so prominent in the TESS data. However, as the dynamical
power spectrum in the lower right frame of Fig.~\ref{lscam-stacked} 
(calculated with a sliding window of 1~yr width) shows, it only started
shortly before the first TESS observation.

%--------------------------------------------------------------
%______________________________________________________________

Finally, the light curves do not contain evidence for a consistent 50~min
variation as reported by \citep{Dobrzycka98} in their spectroscopic
observations. However, the power spectrum of LC\#1 does have some faint
peaks in the frequency range between 28.5 -- 29.5~d$^{-1}$ (inset in
the lower right frame of Fig.~\ref{lscam}) which may indicated 
quasi-periodic oscillations (QPOs) with a period close to 50~min.
But there are no similar signals in the other light curves.

\subsection{NS Cancri}
\label{NS Cnc}

Identified spectroscopically as a cataclysmic variable by \citet{Szkody06} 
among the stars observed in the Sloan Digital Sky Survey, NS~Cnc 
(= SDSS~J081256.85+191157.8) was found to be deeply 
eclipsing by \citet{Gulsecen14}. The most accurate value of 0.1600525(30)~d 
for the orbital 
period was determined by \citet{Thorstensen15}. \citet{Gulsecen14} attribute
another period of 0.148159~d to a negative superhump. The ASAS-SN long term
light curve (not shown) reveals that from about mid-2016 on NS~Cnc 
steadily increased its brightness at a rate of 0.084~mag/yr. Starting from a
previously constant level of $V \approx 16.1$ it has reached $g \approx 15.6$ 
by early 2023.

\begin{figure*}
	\includegraphics[width=\textwidth]{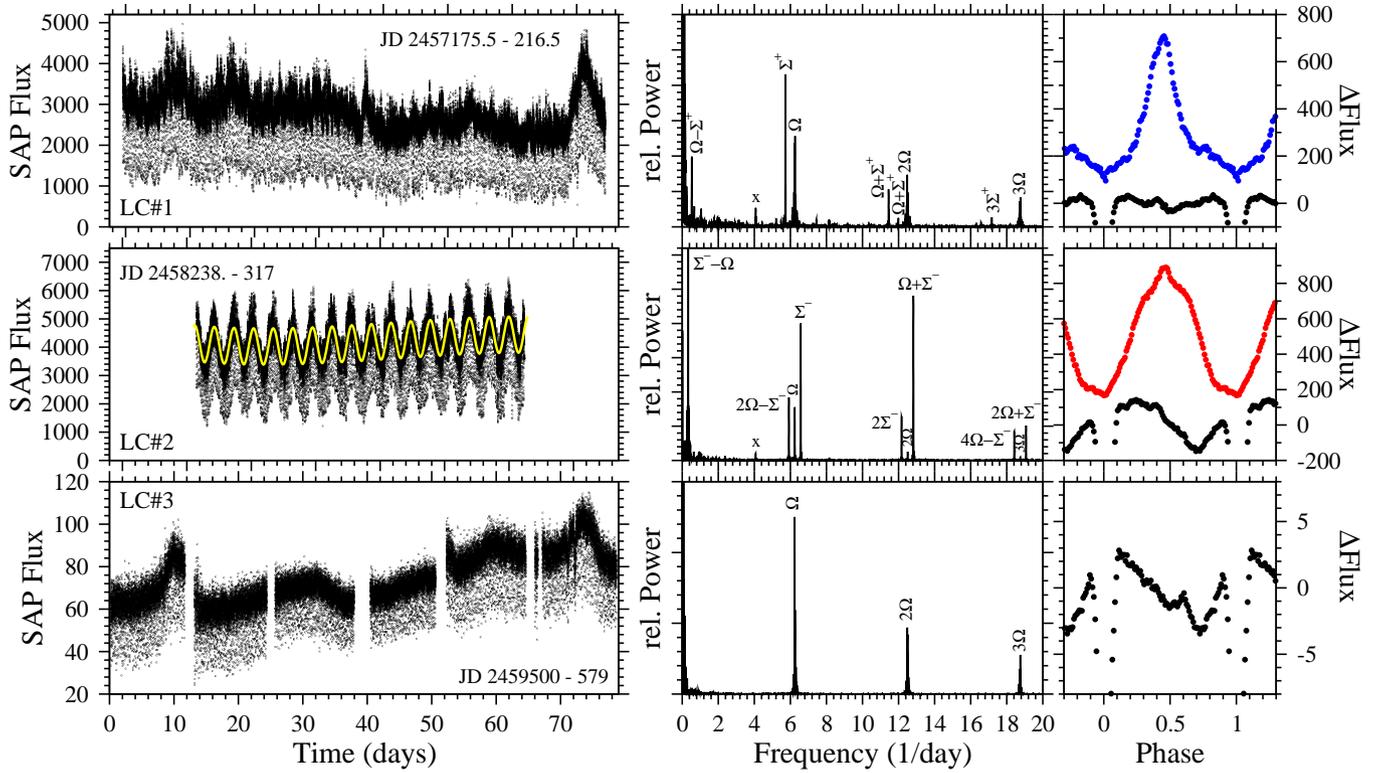}
%      \fbox{\rule[14cm]{14cm}{0cm}}
\caption{{\it Left:} Light curves of NS~Cnc. LC\#1 and LC\#2 were
observed by Kepler, LC\#3 by TESS. {\it Middle:} Power spectra of
the light curves. 
{\it Right:} Waveforms of the orbital variations (black), and of the positive 
(blue) and negative (red) superhumps.}
\label{j0812}
\end{figure*}

NS~Cnc was observed by the Kepler satellite as part of the K2 mission 
during campaigns 5 and 18 with an interval of about 3~yrs between them. 
Additionally, TESS observed the star in three consecutive sectors
after another 3~yrs. All light curves are reproduced in the left column
of Fig.~\ref{j0812}. Considering that the TESS data are noisier and therefore
cannot reveal as many details as the Kepler data, the general structure
of LC\#1 and LC\#3 is similar with the out-of-eclipse variations largely 
being dominated by modulations on time scales of several days and shorter
variations superposed (better seen in LC\#1). In contrast, LC\#2 is 
characterize by a very clear, strict periodicity with a period
of about 3~d which immediately suggests to be the beat between two other
periods. The yellow curve is the best fit to the out-of-eclipse light
curve of the sum of a low order polynomial (to fit slight long term 
variations) and a sine function with a period of 3.046~d.

%--------------------------------------------------------------
%______________________________________________________________

%%--------------------------------------------------------------
%\begin{figure}
%      \includegraphics[width=\columnwidth]{j0812-lc.eps}
%      \fbox{\rule[14cm]{14cm}{0cm}}
%\caption{Kepler K2 light curves of SDSS~J0812, both plotted on the same
%time and flux scale.}
%\label{j0812-lc}
%\end{figure}
%%______________________________________________________________

The power spectra of all light curves (eclipses masked) is reproduced in the 
central column of Fig.{\ref{j0812}\footnote{The peak marked with an ``x'' in
the two upper frames appears in the power spectra of the K2 light curves 
of all object of the ensemble of stars from the RK catalogue mentioned
in Sect.~\ref{Introduction} and therefore must be instrumental.}.
Apart from the orbital signal the power spectrum of LC\#1  contains a 
strong peak at a frequency corresponding to a period of
% 0.1746307015 +- 0.0000274472
0.17463(3)~d. It can be interpreted as being due to a positive superhump.
Overtones and combinations of the orbital and superhump signal, in 
particular the beat frequency $\Omega - \Sigma^+$, are identified in the 
figure. The superhump does not persist over the entire duration of the light 
curve. Instead, it suddenly vanishes right in the middle. This is seen in 
Fig.~\ref{j0812-stacked} where the power spectrum of the original data
(i.e., including the eclipses) in a small frequency range encompassing the
superhump and the orbital signal is shown in the conventional form at the
left, and time resolved in the right frame. A sliding window of 10~d width
was used in this case to calculate the dynamical power spectrum. The
superhump waveform (right column of Fig.~\ref{j0812}) is far from sinusoidal 
and consists basically of a narrow hump.  

%%--------------------------------------------------------------
%\begin{figure}
%	\includegraphics[width=\columnwidth]{j0812-ps.eps}
%      \fbox{\rule[14cm]{14cm}{0cm}}
%\caption{Power spectra of Kepler K2 light curves LC\#1 (top) and LC\#2 (bottom)
%of SDSS~J0812 (eclipses masked). Significant peak are identified. The signal
%marked as ``x'' is caused by an instrumental effect.} 
%\label{j0812-ps}
%\end{figure}
%%______________________________________________________________

%--------------------------------------------------------------
\begin{figure}
	\includegraphics[width=\columnwidth]{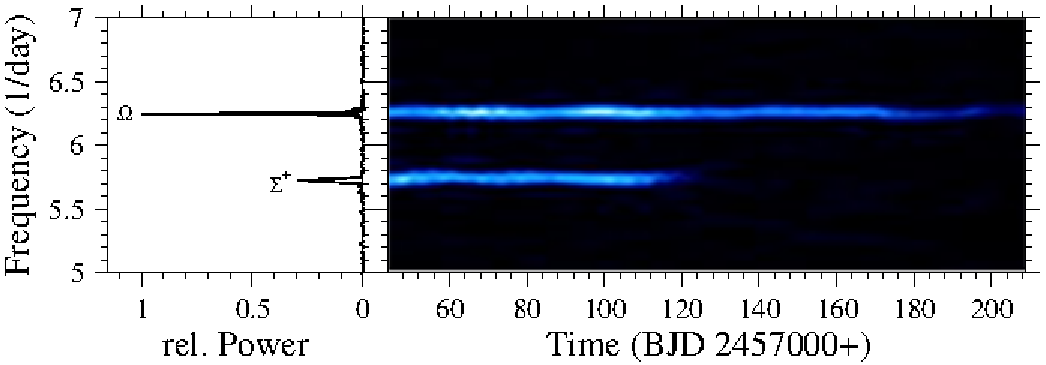}
%      \fbox{\rule[14cm]{14cm}{0cm}}
\caption{Power spectrum of LC\#1 of NS~Cnc in a narrow frequency range 
around the positive superhump and orbital frequency in conventional (left) and
time resolved form (right).}
\label{j0812-stacked}
\end{figure}
%______________________________________________________________

Turning the attention now to LC\#2, the power spectrum shows that the 
positive superhump has gone and made room for a negative one. The strongest 
peak in the spectrum (truncated in the Fig.~\ref{j0812}) is caused by
the beat between the superhump and the orbit. Other combinations and overtones
are identified in the figure, and many more can be detected at higher 
frequencies up to the Nyquist frequency. The superhump period is
% 0.1520417631 +- 0.0000058311
0.152042(6)~d. In contrast to
the positive superhump in LC\#1 the nSH remains active over
the entire duration of the light curve. The waveform is almost sinusoidal 
with a curious extra hump upon the maximum. It is noteworthy that its period 
is much longer than the 0.148159~d period seen by \citet{Gulsecen14}. The 
latter implies a much higher period excess as will be discussed in 
Sect.~\ref{Discussion}.

Finally, the power spectrum of LC\#3 shows that the superhumps have vanished
altogether. Only the orbital signals and its overtones remain. 

The light curves also reveal an evolution of the orbital waveform of NS~Cnc
(black curves in the right column of Fig.~\ref{j0812}). In all cases, out of
eclipse it consists of a hump with some superposed structure. However, the
phase of its maximum changes from epoch to epoch. A secondary eclipse is
also seen in all light curves, although in LC\#2 it causes only a slight
depression on the downward slope of the orbital hump.

\subsection{V425 Cassiopeiae}
\label{V425 Cas}

Our knowledge of the novalike variable V425~Cas is quite limited. Long-term 
observations of \citet{Wenzel87} revealed low states, permitting a 
classification as a VY~Scl star. A spectroscopic period of 0.14964(36)~d was 
reported in a short notice of \citet{Shafter82}.  Large amplitude 
variations (up to 1.5~mag) with a
period of 2.65~d were seen by \citet{Kato01} who interpret them -- not without
difficulties -- as being due to a combination of disk instabilities and
irradiation of the secondary star.

Such variations are not present in the two single sector light curves observed 
by TESS with an interval of 3~yr between them. LC\#1 is reproduced in the 
upper frame
of Fig.~\ref{v425cas-stacked}. The power spectra of the light curves contain 
considerable noise due to irregular variability on time scales longer than
0.5~d. At higher frequencies they are quite different from
each other (left frame of Fig.~\ref{v425cas}). In LC\#1 (black graph) two
significant signals are present. One of them corresponds
to a period just two standard deviations shorter than the orbital period
measured by \citet{Shafter82}. The revised value is listed in Table~3.
% Table~\ref{Table: Masterlist}. 
The lower frequency peak ($P_{\rm pSH}=0.1654$~d) 
can then be interpreted being due to a positive superhump. In contrast, 
in LC\#2 a single peak slightly above the orbital frequency 
($P_{\rm nSH} = 0.14509(3)$~d) appears to be due to a negative superhump. The
orbital signal itself is not seen, and neither is the pSH. Thus, in the interim
between the light curves V425~Cas has changed from a positive to a negative
superhump regime. The orbital (black; LC\#1), pSH (blue; LC\#1) and nSH 
(red; LC\#2) waveforms are shown in the right frame of the figure. They are
all very nearly sinusoidal.

%--------------------------------------------------------------
\begin{figure}
	\includegraphics[width=\columnwidth]{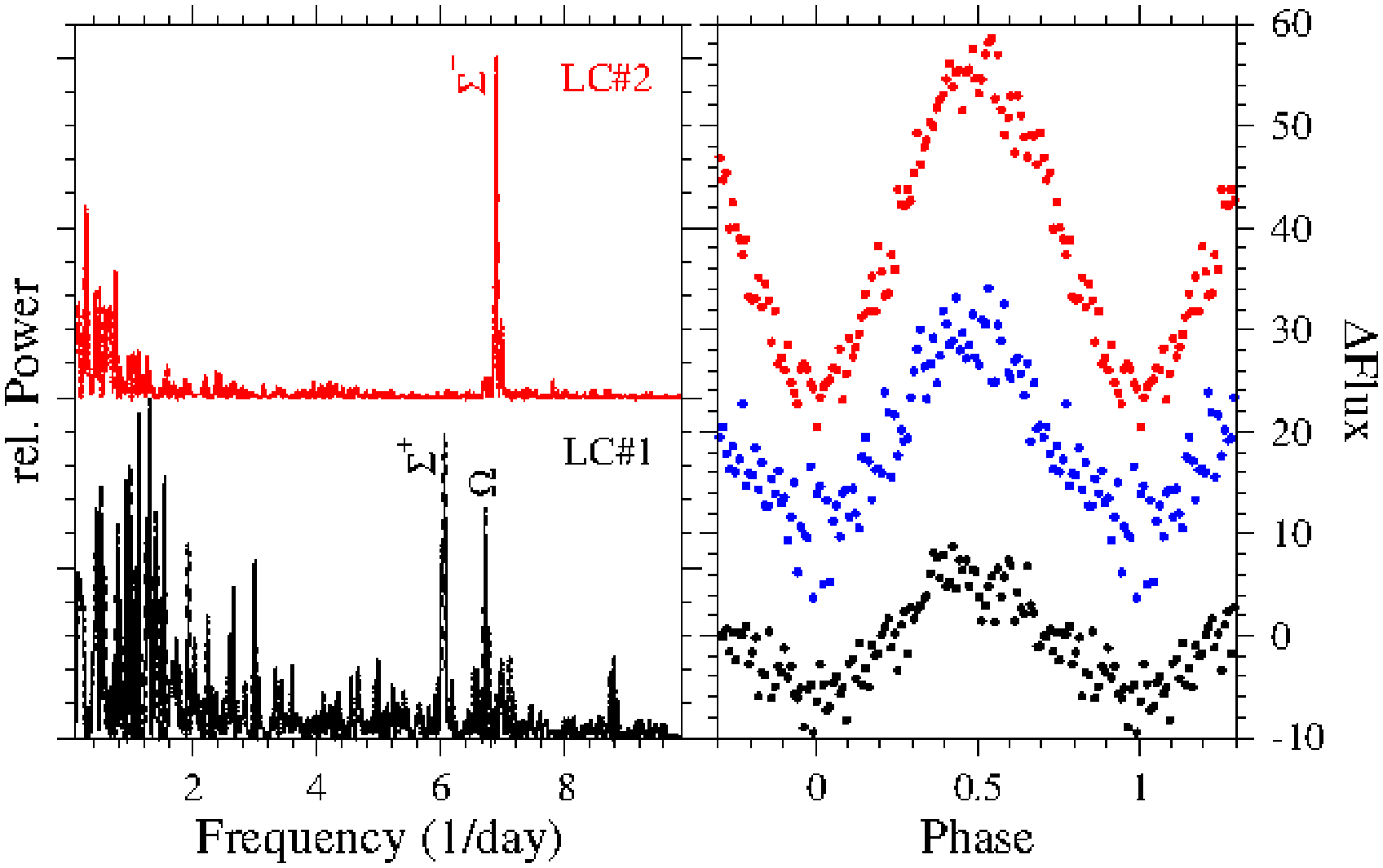}
%      \fbox{\rule[14cm]{14cm}{0cm}}
\caption{{\it Left:} Power spectra of LC\#1 and LC\#2 of V425~Cas.
{\it Right:} Waveforms of the orbital (black, LC\#1), pSH
(blue; LC\#1) and nSH (red; LC\#2) variations.}
\label{v425cas}
\end{figure}
%______________________________________________________________

While dynamical power spectra reveal that the negative superhump is persistent
over the entire length of LC\#2, this is not the case for the pSH in LC\#1, 
as is shown in the lower frames of Fig.~\ref{v425cas-stacked} which contains 
a small part of the power spectrum in the range close to the orbital and
superhump frequencies in the conventional form on the left, and in time 
resolved form on the right. It reveals that the 
superhump is completely absent in the first half of the light curve and 
appears only after the gap in the middle of the TESS observations. 

% LC#1: pSH 
% LC#1: orb
% LC#2: nSH 0.1450905055 +- 0.0000303148

%--------------------------------------------------------------
\begin{figure}
	\includegraphics[width=\columnwidth]{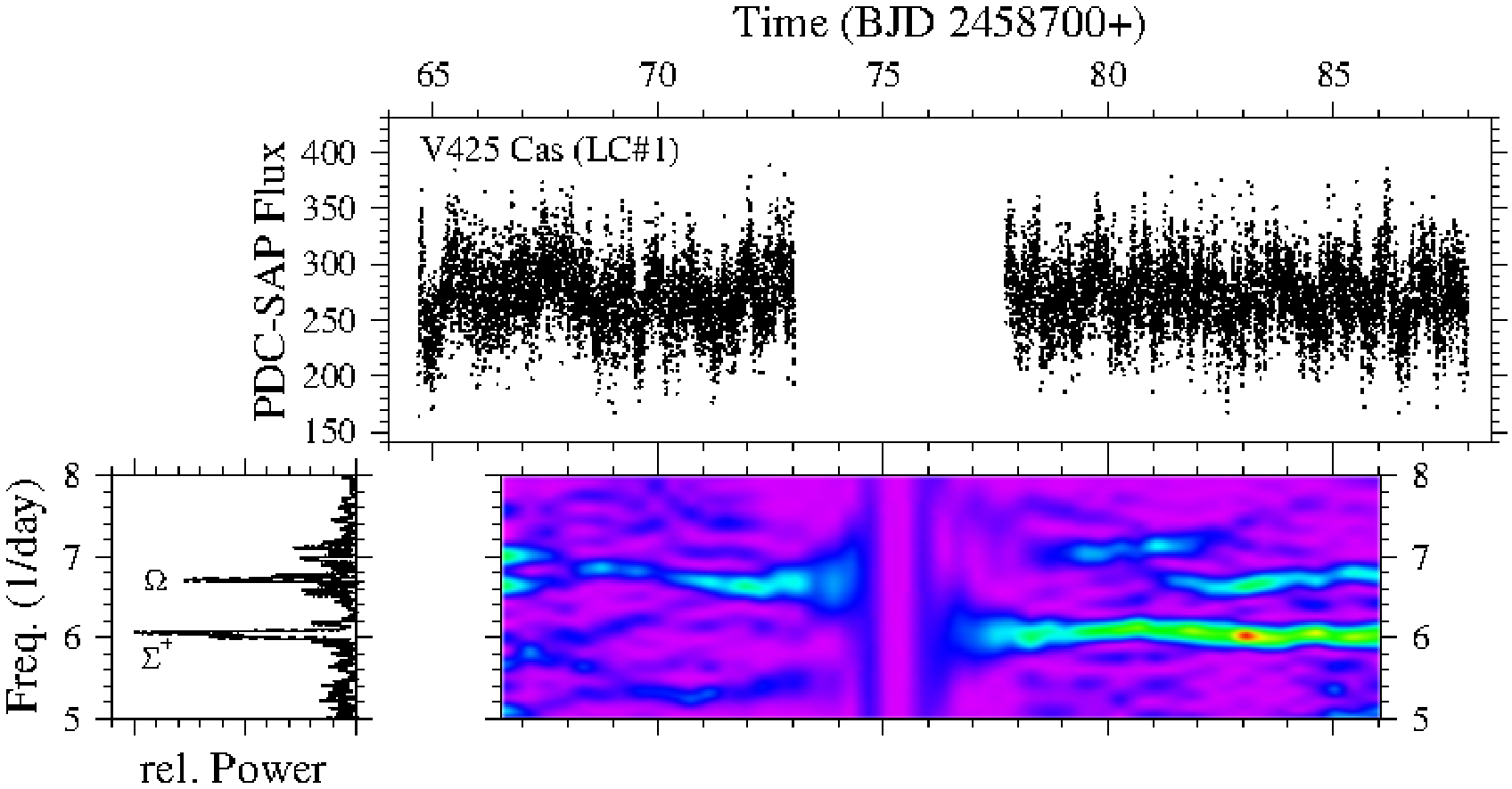}
%      \fbox{\rule[14cm]{14cm}{0cm}}
\caption{{\it Top:} Light curve LC\#1 of V425~Cas. {\it Bottom:} Power 
spectrum of the light curve in a small range around the orbital and 
pSH frequencies in the conventional (left) and dynamical 
form (right).}
\label{v425cas-stacked}
\end{figure}
%______________________________________________________________

In order to avoid possible misinterpretations of a marginally significant
signal at 126.6~d$^{-1}$ in LC\#1 (not shown), which may at first glance 
point at an
intermediate polar nature of V425~Cas, I note that a close inspection reveals
it to be not persistent but caused by an isolated event in the 
light curve.

\subsection{V1024 Cep}
\label{V1024 Cep}

V1024 Cep (originally named HS~0455+08315) was discovered as an
eclipsing cataclysmic variables by \citet{Gaensicke02}. They
suspected it to be a SW~Sex star. This classification was confirmed
by \citet{Rodriguez-Gil07a} who also measured a precise orbital period.
\citet{Shears16} detected deep low states in the long-term light curve
of V1024 Cep. It is thus also a VY~Scl star. 

TESS observed V1024~Cep multiple times. The data can be combined into 
five light curves, four of which consist of two sector observations and
one comprises only a single sector.
Light curves LC\#1, LC\#4 and LC\#5 are reproduced in
the left column of Fig.~\ref{v1024cep}. The yellow curve layed over
the LC\#5 light curve is the sum of a low order polynominal (to follow more 
gradual variations) and a sine curve fitted to the out-of-eclipse data.
The light curves are obviously of quite different
character. Their power spectra are shown in the central column of the
figure. Just as the spectrum of LC\#1, those of LC\#2 and LC\#3 (not shown), 
all taken
before 2022, contain only signals at the orbital frequency and overtones.
This changed in 2022 (LC\#4 and LC\#5), when a negative superhump developed
and grew stronger over time. In the spectrum of LC\#5 many combinations of
the orbital and superhump frequencies can be identified (also at higher
frequencies beyond the limits of the figure). While variations on the
beat period between orbit and superhump start to develop in the latter part
of LC\#4 (but are not clearly identified in the power spectrum), they are
prominent in LC\#5 and appear to increase in amplitude over time. 
The superhump period is
% 0.1429897100 +- 0.0000093643
0.142990(9)~d in LC\#5 (and slightly longer in LC\#4). 

%--------------------------------------------------------------
\begin{figure*}
	\includegraphics[width=\textwidth]{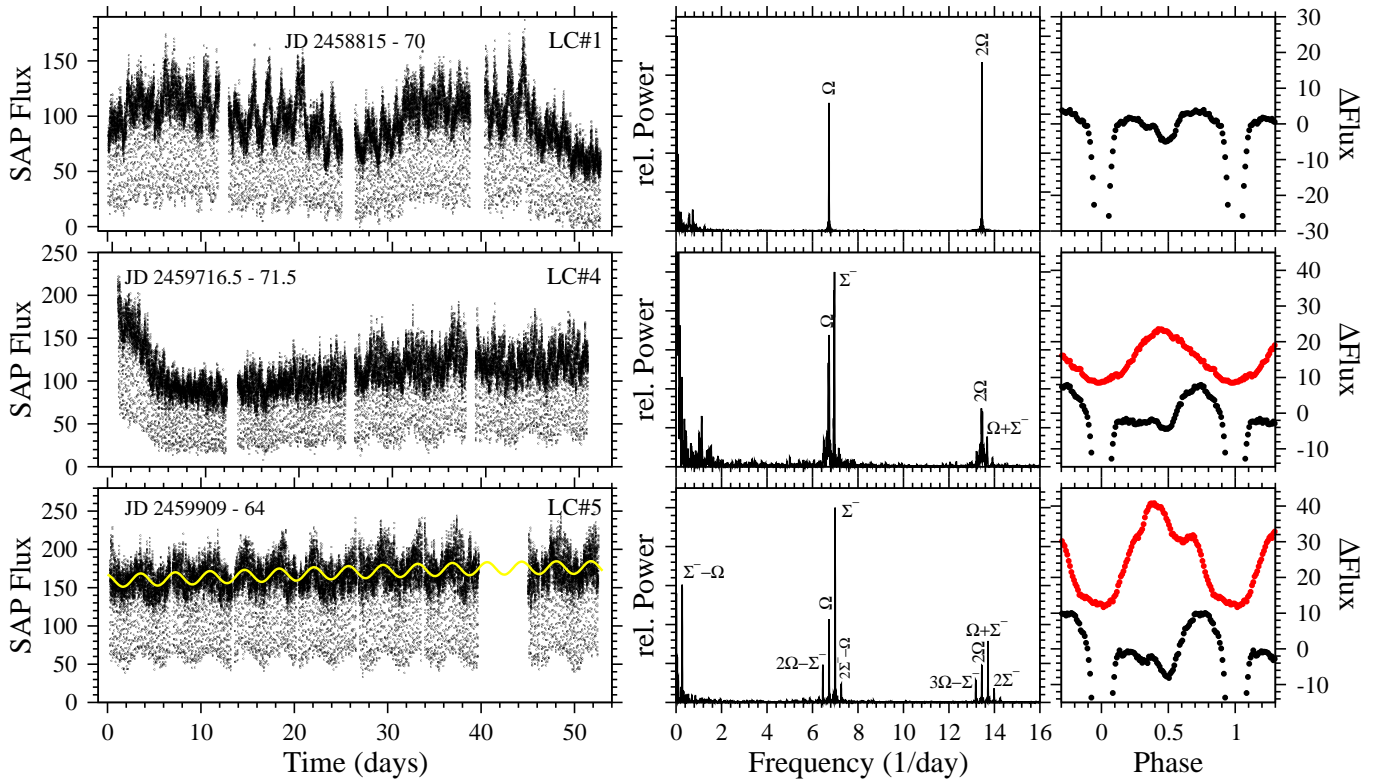}
%      \fbox{\rule[14cm]{14cm}{0cm}}
\caption{{\it Left:} Three of the five available TESS list curves of 
V1024~Cep. The yellow curve in the lower frame is the 
sum of a low order polynominal (to follow more gradual variations) and a 
sine curve fitted to the out-of-eclipse data.{\it Middle:} Power spectra of
the light curves. 
{\it Right:} Waveforms of the orbital (black), and superhump (red)
variations.}
\label{v1024cep}
\end{figure*}
%______________________________________________________________

Orbital and superhump waveforms are shown in black and red, respectively,
in the right column of Fig.~\ref{v1024cep}. The orbital waveform contains
a clear secondary eclipse. Before the superhump developped, only a small
hump before the primary eclipse is seen. The waveform is quite stable
in the three light curves LC\#1 -- LC\#3. As a curious feature I note the 
small dip at phase 0.34 which is also seen in LC\#5. During the presence 
of the superhump the orbital
waveform changes significantly in the sense that the hump grew much stronger.
The superhump waveform also underwent an evolution. While it is still almost
sinusoidal in LC\#4, it is much more structured and has a secondary maximum
in LC\#5.

\subsection{DN Geminorum}
\label{DN Geminorum}

The only available time resolved photometric study of the 1912 nova
DN~Gem revealed a period of 0.126744(5)~d \citep{Retter99} which the
authors consider to be orbital, although they also discuss (but discart) 
alternatives, i.e, the WD spin period in an intermediate polar scenario 
or a superhump period. The orbital nature of the variations was later 
confirmed spectroscopically by \citet{Peters06}

Three TESS light curves of DN~Gem are available. The first is separated from 
the second by almost 2~yrs, but only a single sector separates the second 
from the third light curve. As an example LC\#2 is shown in the upper frame
of Fig.~\ref{dngem}. Just as the others it contains only smooth variations on
time scales of many days. The power spectra in a narrow range around the
orbital frequency are shown in three of the lower frames of the figure. 
LC\#1 and LC\#2 (and rather
weakly also LC\#3) show a clear signal corresponding to a mean period of 
0.12783(8)~d, only slightly -- but significantly, considering the error 
margins -- different from the orbital period of \citet{Retter99}, but within 
the limits of the spectroscopic period of \citet{Peters06}. I take this
as the true value. In all light curves, the orbital waveform is very nearly
sinusoidal (black dots in the lower right frame of the figure).

%--------------------------------------------------------------
\begin{figure}
	\includegraphics[width=\columnwidth]{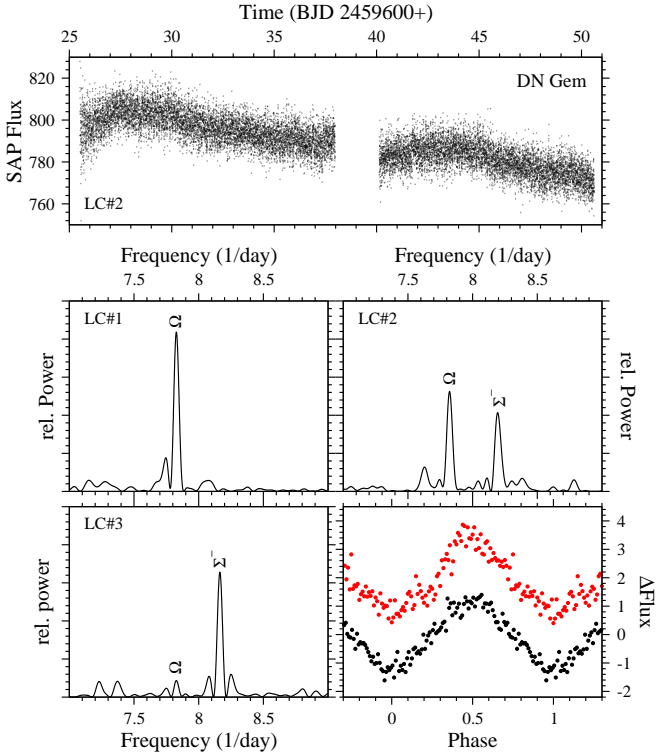}
%      \fbox{\rule[14cm]{14cm}{0cm}}
\caption{Light curve LC\#2 of DN~Gem (top) together with the power spectra 
of the three light curves of DN~Gem in a narrow
range around the orbital and the nSH frequencies. The lower right frame
contains the orbital (black) and superhump (red) waveforms.}
\label{dngem}
\end{figure}
%______________________________________________________________

No other significant signal is present in the power spectrum of LC\#1.
But both other light curves, closely spaced in time, contain a signal at a
somewhat higher frequency. While its strength rivals with that of the
orbital signal in LC\#2 it is much stronger in LC\#3.  
Obviously, between the first and the later observing epochs a negative
superhump has developed in DN~Gem. Its period and the period excess are
listed in 
% Table~\ref{Table: Masterlist},
Table~3, and it average waveform is shown
in lower right frame of Fig.~\ref{dngem} as red dots. Note that details 
such as the small intermediate
maximum at phase 0.15 are present in the waveforms of both, LC\#2 and LC\#3,
and may therefore be real and persistent structures.

\subsection{V1084 Herculis}
\label{V1084 Her}

V1084~Her is known as a negative superhump system. Since it was not
included in the study of TESS light curves of superhumpers in Paper~II
its properties are characterized here. First detected by
\citet{Mickaelian02} the nSHs were further studied by 
\citet{Patterson02} who also measured the orbital period of 0.120560(14)~d.
On shorter time scales \citet{Mickaelian02} reported variation of the 
order of 15~min. Similarly, \citet{Patterson02} mention QPOs near 1000~s.
Citing a more specitic number, \citet{Rodriguez-Gil09} claim the presence
of a polarimetric period of 19.38(39)~min and emission-line flaring on
twice this period which they relate to the rotation of the WD. This
intermediate polar hypothesis was, however, refuted by \citet{Worpel20}
who did not see corresponding variations in X-rays and concluded that
V1084~Her is a non-magnetic CV.

The two available single sector
TESS light curves confirm the presence of superhumps. The
wiggles seen in LC\#1 (upper frame of Fig.~\ref{v1084her}) give rise to a 
low frequency signal corresponding to a period of 3.86(4)~d in the power 
spectrum (left frame in the central row of the figure) which can readily 
be interpreted as the beat between the faint orbital signal which is barely 
resolved from the strong superhump peak in the figure (but see the inset). 
In LC\#1 the superhump waveform is approximately sinusoidal (right middle
frame; red dots) while the lower amplitude and noisier orbital waveform 
(black) seems to have some more structure.
Some differences are seen in LC\#2. The orbital signal cannot clearly
be detected in the power spectrum which is reproduced on an expanded
vertical scale in the lower left frame of Fig.~\ref{v1084her}. But its
first overtone is clearly present, as is the overtone of the SH signal.
Correspondingly, the orbital waveform (lower right frame, in black) is
now double humped. The SH waveform (red) also changed considerably, having
a flat minimum encompassing about 40\% of the cycle. 

%--------------------------------------------------------------
\begin{figure}
	\includegraphics[width=\columnwidth]{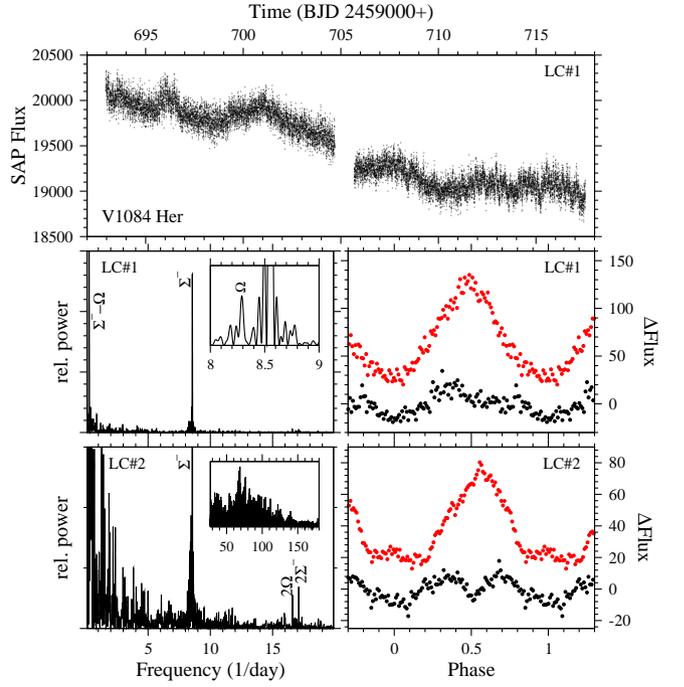}
%      \fbox{\rule[14cm]{14cm}{0cm}}
\caption{{\it Top:} Light curve LC\#1 of V1084~Her. {\it Middle left:}
Power spectrum of LC\#1, dominated by a nSH signal and the beat signal
between orbit and nSH. The expanded scale used in the inset permits a
better identification of the orbital signal. {\it Middle right:} Waveforms
of the orbital (black) and superhump (red) variations in LC\#1.
{\it Lower left:} Power spectrum of LC\#2 on an expanded vertical scale
permitting the clear identification of the overtones of the orbital and
superhump signals. In inset shows a broad range of QPO signals. 
{\it Lower right:} Waveforms of orbital (black) and 
superhump (red) variations in LC\#2.}  
\label{v1084her}
\end{figure}
%______________________________________________________________

At higher frequencies, the power spectra show somewhat enhanced power in a 
broad range between 50 and 130~d$^{-1}$ (more so in LC\#2 than in LC\#1; inset
in the lower left frame of Fig.~\ref{v1084her}), but 
they do not indicate a strong preference for variations to occur on the times 
scales mentioned by \citet{Mickaelian02} and \citet{Patterson02}.

% \subsection{V524 Hydrae}
% \label{V524 Hya}
% 
% (Appears to me now to be an artefact of data reduction!. Delete star from 
% paper!)
% 
% \citet{Szkody03} first recognized the little studied variable V524~Hya as an 
% eclipsing CV. The orbital period of 0.147878(8)~d was measured by 
% \citet{Gaensicke09}. Out of eclipse \cite{ZenginCamurdan10} found little
% variability. In particular, they note the absence of an orbital hump.
% In contrast, folding the the single sector TESS light curve on the orbital 
% period reveals a noticable hump just prior to eclipse ingress 
% (Fig.~\ref{wf-orb-2}(a)). More interesting, however, it the power spectrum.
% After masking the eclipses the orbital signal stands out only moderately
% strong (Fig.~\ref{ps-2}(n). The amplified part of the spectrum in the
% inset of the figure shows that the orbital signal is accompanied at
% a slightly higher frequency by an equally strong peak corresponding to
% a period of 0.14627(6)~d. Its (noisy) waveform is approximately sinusoidal
% (Fig~\ref{wf-nsh-1}(g)). This signal may tentatively be identified as being
% due to a negative superhump. The period excess would then be 
% $\epsilon = -0.011$ which would be uncomfortably small for the orbital
% period of V524~Hya (see Paper~II). On the other hand, the disk precession
% period would then be $\approx$13~d, which is compatible with the distance
% of the two maxima in the light curve (Fig.~\ref{lc-1}(w)).

\subsection{CP Lacertae}
\label{CP Lac}

The old nova (1935) CP~Lac did not attract much attention in the past. In
a limited amount of high time resolution photometry \citet{Rodriguez-Gil05}
suspect the presence of a 0.127~d periodicity together with a slight dip
in the phase folded light curve and wonder whether it may be due to an 
eclipse or not. This photometric period is, however, not orbital since
radial velocity measurements of \citet{Peters06} yielded a period of
0.145143(1)~d. 

TESS observations in four sectors can be combined into two light curves 
separated in time by about 2~yrs. No strong variations on time scales of 
days are present.
The power spectrum of LC\#1 (left frame of Fig.~\ref{cplac}, in black) 
contains no signal but the orbital one. The 0.127~d period seen by 
\citet{Rodriguez-Gil05} can be explained as a 1/day alias
of the true orbital period, introduced by the temporal distribution of their
observations. However, another period of 62.7(3)~min seen by them (beyond the
limits of the figure) is definitely not present in the TESS data.

%--------------------------------------------------------------
\begin{figure}
	\includegraphics[width=\columnwidth]{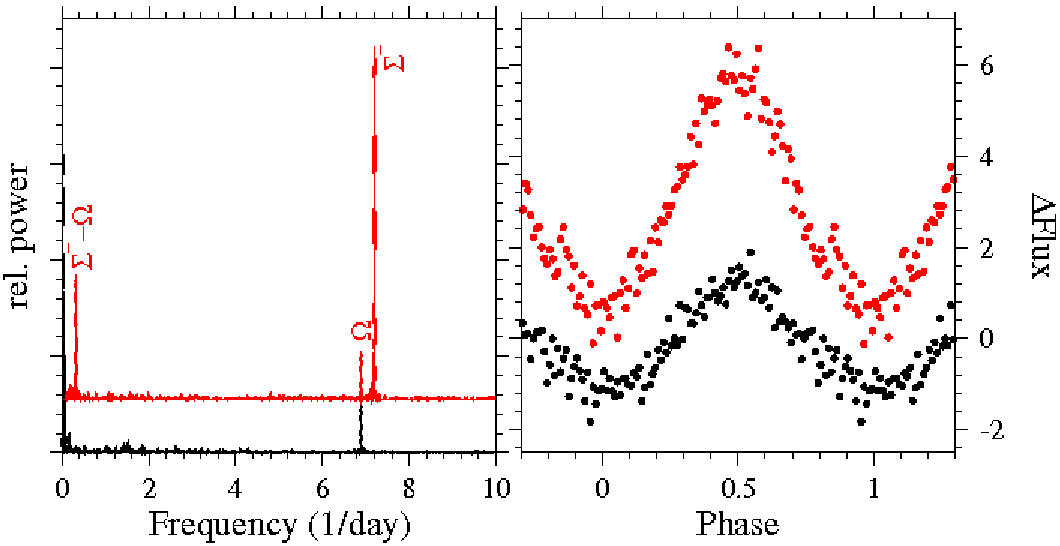}
%      \fbox{\rule[14cm]{14cm}{0cm}}
\caption{{\it Left:} Power spectra of LC\#1 (black) and LC\#2 (red) of
CP~Lac. {\it Right:} Orbital (black; average of LC\#1 and LC\#2) and
superhump (red; LC\#2) waveforms of CP~Lac.} 
\label{cplac}
\end{figure}
%______________________________________________________________

This is also true for LC\#2, the power spectrum of which otherwise contains 
a strong peak at a frequency somewhat higher than the orbital frequency (red 
graph in the left frame of Fig.~\ref{cplac}). This is undoubtedly due to a
negative superhump. A clear signal is also present at the beat frequency
between the orbit and the superhump. The SH waveform (red dots in the 
right frame of Fig.~\ref{cplac}) is very nearly sinusoidal. The same is true 
for the orbital waveform (black dots) which does not contain any indication
of the eclipse suspected by \citet{Rodriguez-Gil05}.

\subsection{DK Lacertae}
\label{DK Lac}

Only little time resolved photometry of DK~Lac (Nova Lac 1950) has been 
published. \citet{Katysheva08} found a possible period of 0.1296~d in 2003 
which is, however, rejected to be orbital by \citet{Schaefer22}. Also,
variations on time scale of 0.11 -- 0.13~d appear to be present in 2004 
\citep{Katysheva08}. On longer time scales, the system exhibited at least 
one episode of a VY~Scl-like low state.

Data from two TESS sectors can be combined into a single light curve. Three
years later TESS observed the system again in one sector. LC\#1 is displayed
in the upper frame of Fig.~\ref{dklac}. It does
not contain strong variations above the noise level. However, the power
spectrum (lower left frame of the figure) contains, apart from some low 
frequency noise, a pair of signals which suggest to be due to orbital and 
superhump variations. The corresponding periods are
% P_1 = 0.12963400972 +- 0.0000481075
$P_1 = 0.12963(5)$~d and
%P_2 = 0.1164427559 +- 0.0000366782
$P_2 = 0.11644(4)$~d.
$P_1$ is very close to the period seen by \citet{Katysheva08}. If this is the
orbital period, $P_2$ must be considered a negative superhump, leading to an
uncomfortable high period excess of $\epsilon = -0.103$. The alternative,
$P_2$ being orbital and $P_1$ the period of a positive superhump, gives
$\epsilon = 0.113$, still quite high for the period 
(see Fig.~\ref{sh-epsilon-rel}), but not 
to such an extend. Therefore, I tentatively identify DK~Lac as a positive
superhump system. The quite noisy superhump waveform (blue dots in the lower
left frame of the figure) cannot be distinguished from a simple sine curve. 
The same is true for the orbital waveform (black dots). 
The power spectrum of LC\#2 does not exhibit the pair of superhump and orbital
signals. However, an otherwise inconspicuous (in view of the surrounding
noise) peak is present at the orbital frequency.  

%--------------------------------------------------------------
\begin{figure}
	\includegraphics[width=\columnwidth]{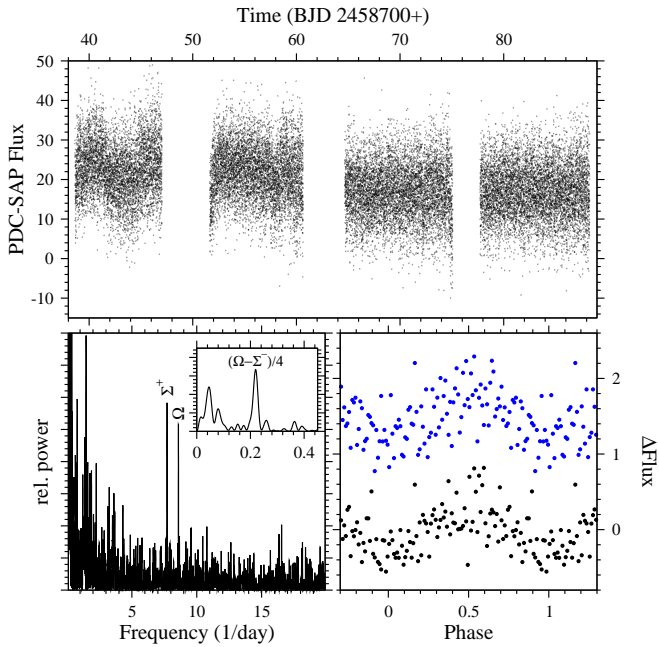}
%      \fbox{\rule[14cm]{14cm}{0cm}}
\caption{Light curve LC\#1 (top), power spectrum (lower left) and waveforms
(lower right) of the orbital (back) and superhump (blue) variations of
DK Lac.}
\label{dklac}
\end{figure}
%______________________________________________________________

Finally, it is worth mentioning that the power spectrum of LC\#1 contains 
a strong low frequency peak corresponding to 
% P_3 = 4.5595474243 +- 0.0336836353
4.56(3)~d (inset in the lower left frame of Fig.~\ref{dklac}).
Within the formal $1\sigma$ error limits this is equal to 4 times
the beat between the orbital and superhump periods. Thus, DK~Lac appears to
join the select group of superhumping CVs with a modulation on an
integral multiple or fraction of
the accretion disk precession period (see Sect.~\ref{Discussion}).

\subsection{KQ Monocerotis}
\label{KQ Mon}

KQ~Mon was originally classified as a CV by \citet{Bond79}. While the
ultraviolet spectral properties of the star have been reasonably well
studied \citep{Sion82, Wolfe13}, time resolved photometry of the system
is completely missing in the literature. Based on optical spectroscopy in 
two nights \citet{Schmidtobreick05} determined an orbital period of
3.08(4)~h (= 0.128(2)~d), but cannot exclude several alias periods.

When I had already analyzed the single sector TESS light curve of KQ~Mon
\citet{Stefanov23} published their study of the same data. I therefore 
restrict myself here to summarize the results and to put forward an additional
argument. The power spectrum contains two signals corresponding to periods of
%0.1345042437 +- 0.0000264795 and 0.1289480627 +- 0.0000145094
$P_1 =  0.13450(3)$~d and $P_2 = 0.12895(1)$~d together with a signal at
the beat frequency. Thus, it is 
obvious that the light curve contains the orbital and a superhump signal, 
together with the disk precession period. But which of $P_1$ and $P_2$ is
orbital and which is due to the SH? $P_2$ lies within the $1\sigma$
error margin of the spectroscopic orbital period of \citet{Schmidtobreick05},
but the same is true for $P_1$, if an alias period at 0.134(2)~d is taken.
\citet{Stefanov23} argue in favour of $P_1$ being orbital and $P_2$ being due
to a negative superhump. I agree with this interpretation and add as an
additional argument a comparison between the observed period excess
$\epsilon_{\rm obs}$ and its expected value $\epsilon_{\rm SS}$ from the
Stolz-Schoembs relation, i.e., the relation between the superhump period 
and the period excess (Sect.~\ref{Discussion}). We then have 
$\epsilon_{\rm obs} = -0.041$ versus $\epsilon_{\rm SS} = -0.36$. The 
alternative case of $P_2$ being the orbital and $P_1$ the (positive) 
superhump period yields $\epsilon_{\rm obs} = 0.043$ versus 
$\epsilon_{\rm SS} = 0.065$. Thus, the
nSH hypothesis is in better agreement with the expectations.

\subsection{LZ Muscae}
\label{LZ Mus}

Practically nothing is known about the quiescent state of Nova Muscae 1998 = 
LZ~Mus. Even the (not well documented) orbital period quoted by
\citet{Retter98} in a short IAU Circular as $P_{\rm RLG} = 0.16930(15)$~d 
(or an alias at 0.20390(25)~d) may not be trustworthy because it refers
to variations observed soon after outburst during the nova transition 
phase when the system was far from its quiescent state. RK quote the first 
of these values as the orbital period of LZ~Mus but mark it as uncertain.

TESS observed the star in two consecutive sectors. Thus, the data can be
combined into a single light curve which, however, is quite noisy due to the
faintness of LZ~Mus (upper frame of Fig.~\ref{lzmus}). Correspondingly, 
the power spectrum (lower left frame of the figure) is also noisy, but it 
contains a clear signal at 
% 0.1634812355 +- 0.0000514093
$P_1 = 0.16348(5)$~d, somewhat shorter than $P_{\rm RLG}$ It is accompanied by 
a less prominent signal at a slightly shorter period of
%0.1509716511 +- 0.000665737
$P_2 = 0.15097(7)$~d. How to interpret the different periods? The prominence
of the $P_1$ signal suggests that this is the orbital period. This assumption
is strengthened by what appears to be a shallow eclipse suggested by the 
phase folded light curve shown in the lower right frame of the figure as 
black dots. However, in view of the noise in
the waveform, this requires confirmation. Assuming $P_1 = P_{\rm orb}$ one
might speculate that $P_{\rm RLG}$ is due to a positive superhump in the
newly formed accretion disk just after the nova outburst. Although at
$\epsilon = 0.036$ the period excess would then be much smaller than 0.082 as 
predicted by the Stolz-Schoembs relation (Sect.~\ref{Discussion}), the scatter
around that relation (Fig.~\ref{sh-epsilon-rel}) would still encompass the low
period excess. 

The significance of the $P_2$ signal may be doubted in view of its low 
strength. Considering it to be real and interpreting it as being
caused by a negative superhump, the situation is reversed: The period excess
of $-0.077$ is much larger than predicted by the Stolz-Schoembs relation 
($-0.040$). However, other systems with higher nSH credentials have a 
similarly excessive period excess; an issue which will be discussed in
Sect.~\ref{Discussion}. Therefore, I tentatively assume LZ~Mus to exhibit
nSHs. Their waveform (red dots in the
lower right frame of Fig.~\ref{lzmus}) is very noisy and can only be 
distinguished from noise after binning the data in larger phase bins. 
I emphasize, however, that the
identification of $P_{RLG}$ and $P_2$ as positive and negative superhump
periods, respectively, is only tentative.

%--------------------------------------------------------------
\begin{figure}
	\includegraphics[width=\columnwidth]{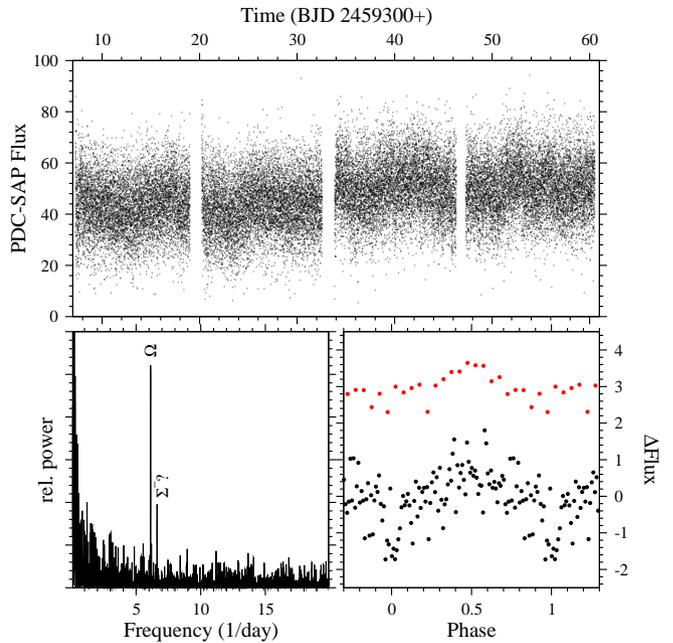}
%      \fbox{\rule[14cm]{14cm}{0cm}}
\caption{Light curve (top), power spectrum (lower left) and waveforms
(lower right) of the orbital (back) and superhump variations (red) of
LZ~Mus.}
\label{lzmus}
\end{figure}
%______________________________________________________________

\subsection{FY Persei}
\label{FY Per}

No detailed studies of FY~Per resolving short time scales are available in 
the literature. The two single sector TESS light curves, separated in time
by 3~yrs (displayed in the top frames of Fig.~\ref{fyper} on the same flux 
scale) contain well expressed outburst-like events.
This is consistent with the long-term behaviour of the system investigated
in some detail by \citet{Honeycutt01} \citep[see also][]{Honeycutt04} who
observed the ``common occurrence of 0.6~mag oscillations with a characteristic
interval of 20--25~days''. This aspect of FY~Per will be analyzed in a 
separate paper. \citet{Honeycutt01} also identified the system as
a VY~Scl star.

%--------------------------------------------------------------
\begin{figure*}
	\includegraphics[width=\textwidth]{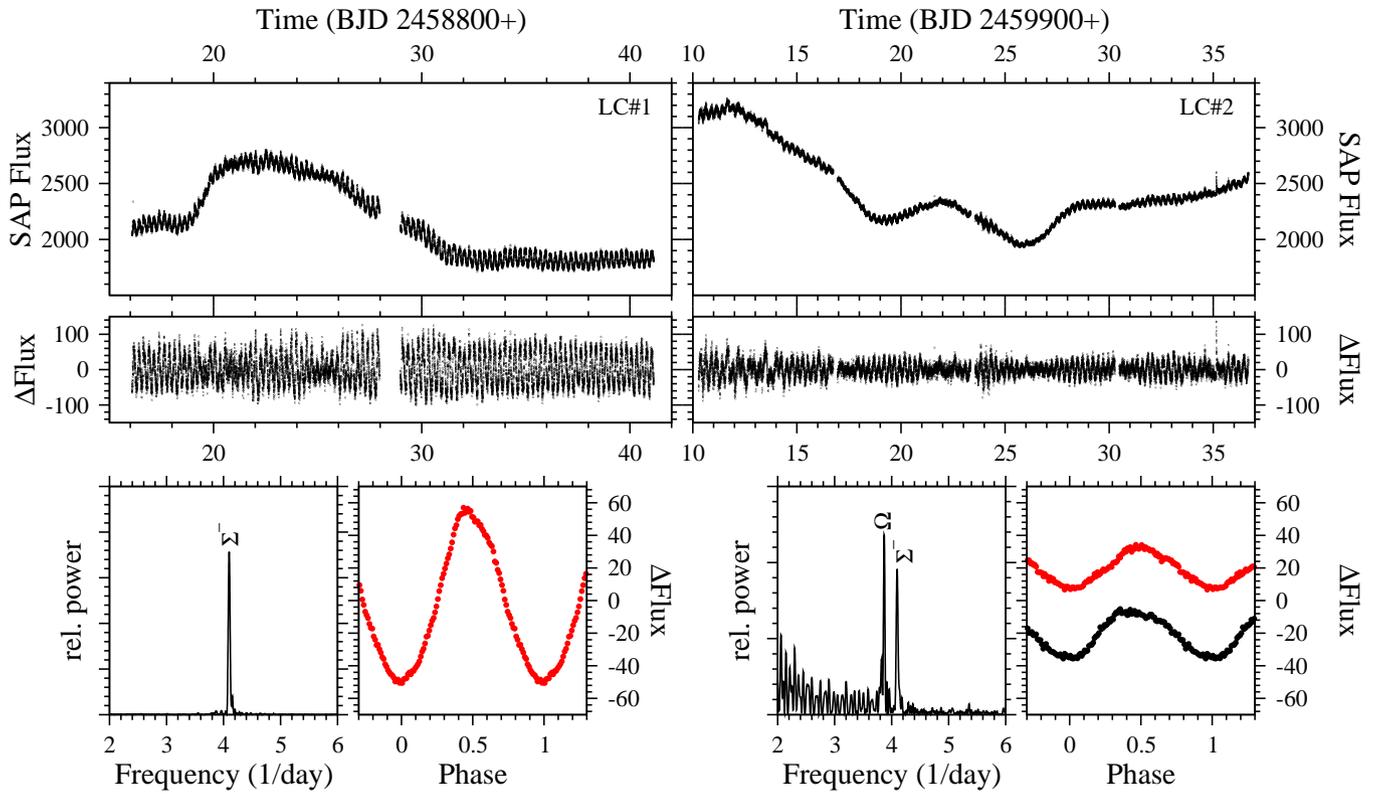}
%      \fbox{\rule[14cm]{14cm}{0cm}}
\caption{{\it Top:} Light curves of FY~Per drawn on the same flux scale.
{\it Middle:} Light curves after subtraction of variations on 
time scales longer than 2~d, dominated by negative superhumps in LC\#1 and
the interplay between superhump and orbital modulations in LC\#2.
{\it Bottom:} Power spectra (left), and superhump (red)
and orbital (black) waveforms (right) for both light curves.}
\label{fyper}
\end{figure*}
%______________________________________________________________

The most impressive features of the TESS light curves are, however, strong
periodic modulations on the time scale of hours, easily visible to the unaided 
eye. They are persistent over the entire duration of the light curves and have 
a much higher amplitude in LC\#1 than in LC\#2,
as is obvious from the second row of frames in
Fig.~\ref{fyper} where the residuals after subtracting variations on time
scales longer than two days are plotted on the same flux scale. Not
surprisingly, these variations cause strong signals in the power spectra
displayed in the lower frames of the figure. 

In the power spectrum of LC\#1 only one peak appears (together with a very 
weak first overtone). It corresponds to a period of
% 0.2439958900 +- 0.0000316526
0.24400(3)~d. Trusting that the spectroscopic period of 0.2584(3) reported by 
\citet{Thorstensen17} is indeed orbital, I conclude that FY~Per exhibits
negative superhumps. The SH waveform (see lower frames 
of Fig.~\ref{fyper}) is nearly sinusoidal with a slightly skewed maximum.
The orbital period is not detected in the power spectrum. However, the 
strongest low frequency peak (after removing the outburst-like structure) has 
a frequency different by just 1.3 times the formal $1\sigma$ error margin 
of the beat period calculated from the orbital and SH frequencies. It 
thus most probably reflects the precession period of the accretion disk. 

In contrast to LC\#1, the power spectrum of LC\#2 contains two clear peaks.
One is only slightly offset from the superhump frequency observed in LC\#1,
corresponding to a period of 
% 0.2437983900 +- 0.0000840768
0.24380(8)~d and can thus considered to be due to the same phenomenon. The 
lower frequency peak corresponds to a period of 
% 0.2583718896 +- 0.0000668069
0.25837(7)~d. This is within the error margin of the orbital period
measured by \citet{Thorstensen17}. The orbital and superhump waveforms are
of almost equal amplitude which is, however, much smaller than the superhump
waveform in LC\#1. Even so, a close inspection reveals that the shape of the
superhump remained remarkably stable: Just after maximum the same deviations 
from a sinusoid can be detected in both waveforms. 

As can be seen in the left central frame of Fig.~\ref{fyper} the amplitude
of the superhumps remains roughly constant. Only during
two small time intervals close to BJD 2458819.5 and 2458821 they vanish and
make room for incoherent variability. A few
days later (BJD 2458825) their amplitude decreases significantly for
just over one day. It is remarkable, however, that the outburst-like event
has apparently no bearing on the behaviour of the superhumps. This suggests
that the light source responsible for the brightening is independent of the
superhump light source. The same is true for LC\#2 (right central frame of
the figure), but in this case the amplitude of the variations varies
significantly over time. This can be explained by the interplay between
the orbital and superhump modulations. Indeed, an analysis of the rms-scatter
of the residual light curve in bins of 0.5~d shows a clear periodicity of
4.35~d, compatible with the beat period between orbit and superhump.

\subsection{LX Serpentis}
\label{LX Ser}

LX~Ser, also known as Stepanyan's star, is a deeply eclipsing novalike 
variable. The orbital period of 0.158432491(2)~d has been steadily refined 
by many authors over the years and was last updated by \citet{Li17}. 
\citet{Liller80} reported low states in the long term light curves 
suggesting a VY~Scl nature for LX~Ser. No other periodic brightness
variations of the star have been reported.

LX Ser is a negative superhumper, but the superhumps are not permanent. 
The first of two single sector TESS light curve
is shown in the upper frame of Fig.~\ref{lxser}. It contains irregular and 
quite rapid ($<$1~d) out-of-eclipse variation superposed upon a clear 
periodic modulation. A least squares sine fit (yellow curve) returns a 
period of 3.54~d. The power spectrum has a low frequency peak corresponding
to the same period within the formal error limits. Otherwise, the low
frequency range contains many peaks which I attribute to irregular 
variations in the light curve. However, at not so low frequencies
(second frame of Fig.~\ref{lxser}; the eclipses were masked before 
calculating the power spectrum) some peaks stand
out. I interpret a signal corresponding to a period of  
% 0.1517662555 +- 0.0000812027
$P_1 = 0.15177(8)$~d as being due to a negative superhump. Its beat with 
the orbital period is compatible with the 3.5~d supraorbital period. 
This interpretation is reinforced by other signals
which can be identified as simple arithmetic combinations of the SH
and orbital frequencies. Others are weakly present beyond the limits of
the figure. 
The somewhat structured superhump waveform is reproduced as red dots in
the lower left frame of Fig.~\ref{lxser} together with the orbital
waveform (in black). Away from the primary eclipse the latter consists of an
asymmetrical hump, interrupted by a secondary eclipse. This shape explains
why the orbital period itself only appears weakly in the (eclipses masked) 
power spectrum.

%--------------------------------------------------------------
\begin{figure}
	\includegraphics[width=\columnwidth]{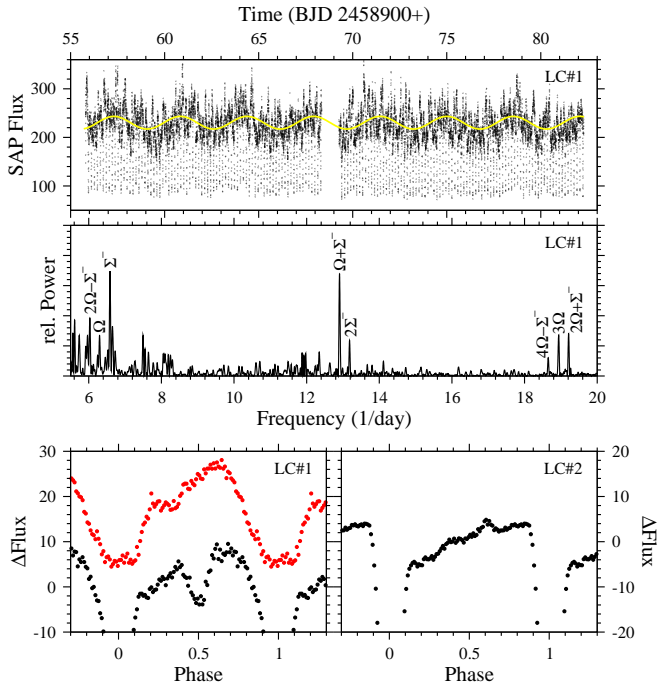}
%      \fbox{\rule[14cm]{14cm}{0cm}}
\caption{{\it Top:} Light curve LC\#1 of LX~Ser. The yellow curve is a least
squares sine fit to the out-of-eclipse flux. {\it Middle:} Power spectrum
of LC\#1. {\it Bottom:} Waveforms of orbital (black) and superhump (red)
variations in LC\#1 (left) and LC\#2 (right).}
\label{lxser}
\end{figure}
%______________________________________________________________

The power spectrum of LC\#2 (not shown), observed 2~yrs later, is very 
different. Apart from low frequency noise it only contains signals at
the orbital frequency and its overtones. Thus, the superhump has subsided.
Moreover, the orbital waveform (lower right frame of Fig.~\ref{lxser})
has also changed a lot. It consists of a single asymmetric hump, the
secondary eclipse having disappeared. The long term ASAS-SN light curve
shows that the average magnitude of LX~Ser during the two TESS observing
epochs was the same. The complete disappearance of the secondary eclipse
is therefore puzzling.

\subsection{EI Ursae Majoris}
\label{EI UMa}

EI~UMa is a well established intermediate polar. The spectroscopic orbital
period of 0.26811(33)~d \citep{Thorstensen86} places it among the longer
period CVs. In optical photometric observations \citet{Kozhevnikov10}
measured a WD spin period of 769.83(10)~sec. This author also gives
a detailed account of the previous optical and X-ray observational history
of the star, to which the reader is referred for further references.  

Light curves from two TESS sectors are available, separated by two years.
They are displayed in the upper frames of Fig.~\ref{eiuma}.
Both of them show episodes of strong outbursts of short duration
($<$0.5~d) in quick succession. Those of LC\#1 have already been
discussed by \citet{Scaringi22a}. Similar event have been observed in several
other intermediate polars and were explained in terms of a magnetic gating
model \citep{Hameury22, Littlefield22} or invoking localized thermonuclear
bursts on the white dwarf \citep{Scaringi22a, Scaringi22b}. More outbursts
of similar nature can be identified in the long term ASAS-SN light curve of
EI~UMa. They will not be discussed in more detail here and have been purged 
from the light curves before further analysis. However, it is noteworthy that 
in both light curves they are superposed upon a temporary brightening of
EI~UMa which lasts for about 10 days. During these epochs the light curves
become noticably more rugged (better seen in Fig.~\ref{eiuma-stacked}).

%--------------------------------------------------------------
\begin{figure*}
	\includegraphics[width=\textwidth]{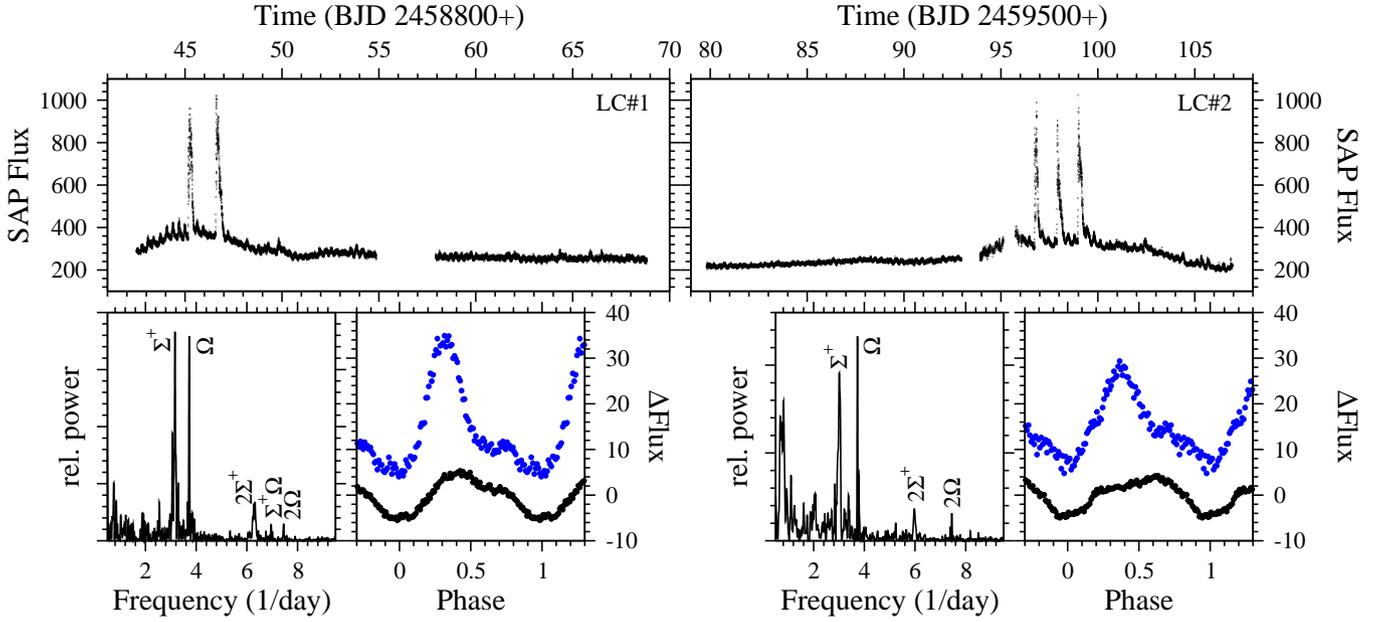}
%      \fbox{\rule[14cm]{14cm}{0cm}}
\caption{{\it Top:} Light curves of EI~UMa drawn on the same flux scale.
{\it Bottom:} Power spectra (left), and superhump (blue)
and orbital (black) waveforms (right) for both light curves.}
\label{eiuma}
\end{figure*}
%______________________________________________________________

Concerning the IP type variations of EI~UMa, and considering that this is
not the main topic of this study, I restrict myself here to state that
the power spectrum only contains weak signals at the WD spin period.
No trace of an orbital sideband signal can be detected. The period
is entirely consistent with the value measured by \citet{Kozhevnikov10}.

While the IP nature of EI~UMa is well known the TESS light curves for the
first time reveal the system also to be a positive superhumper. This was
alread briefly mentioned by \citet{Scaringi22a} but is discussed in more
detail here.
In the lower frames of Fig.~\ref{eiuma} the low frequency parts of the 
power spectra are displayed. Both contain two dominant signals. One is 
due to the orbital motion of EI~UMa and yields a period slightly longer 
but within the error margin of the spectroscopic period of 
\citet{Thorstensen86}. The second one is somewhat structured in both light
curves. This complicates the measurement of an exact period. I interpret it 
as being due to a positive superhump with a slightly varying period.

The time dependent behaviour of the orbital and superhump signals can
be studied in the dynamical power spectra in the second and third
right hand frames of Fig.~\ref{eiuma-stacked}. For comparison, on the
left side the conventional power spectra in a narrow frequency range
around the signals are reproduced. On top and at the bottom the light
curves are drawn, where the rapid outbursts (see above) were removed.
In both light curves the orbital signal vanishes during the brightenings
upon which the outbursts are superposed. In LC\#1 the superhump signal
is strong during the brightening ($P = 0.317$~d), then disappears and 
reappears during what may be a small secondary brightening ($P = 0.323$~d). 
During the second half of the light curve it is missing altogether. 
In LC\#2 the superhump is faintly present from the start, grows in 
strength as the brightness of EI~UMa gradually increases ($P = 0.354$~d), 
and becomes strong, but at a slighly higher frequency ($P = 0.336$)~d, 
during the fully developed brightening, only to vanish when EI~UMa returns 
to its normal brightness. The changing SH periods imply a range of period 
excesses between 0.181 and 0.319. The superhump waveform changes somewhat, 
but its general structure remains the same (lower frames of Fig.~\ref{eiuma}; 
in blue). In contrast, the orbital waveform (black)
evolves significantly between LC\#1 and LC\#2.

%--------------------------------------------------------------
\begin{figure}
	\includegraphics[width=\columnwidth]{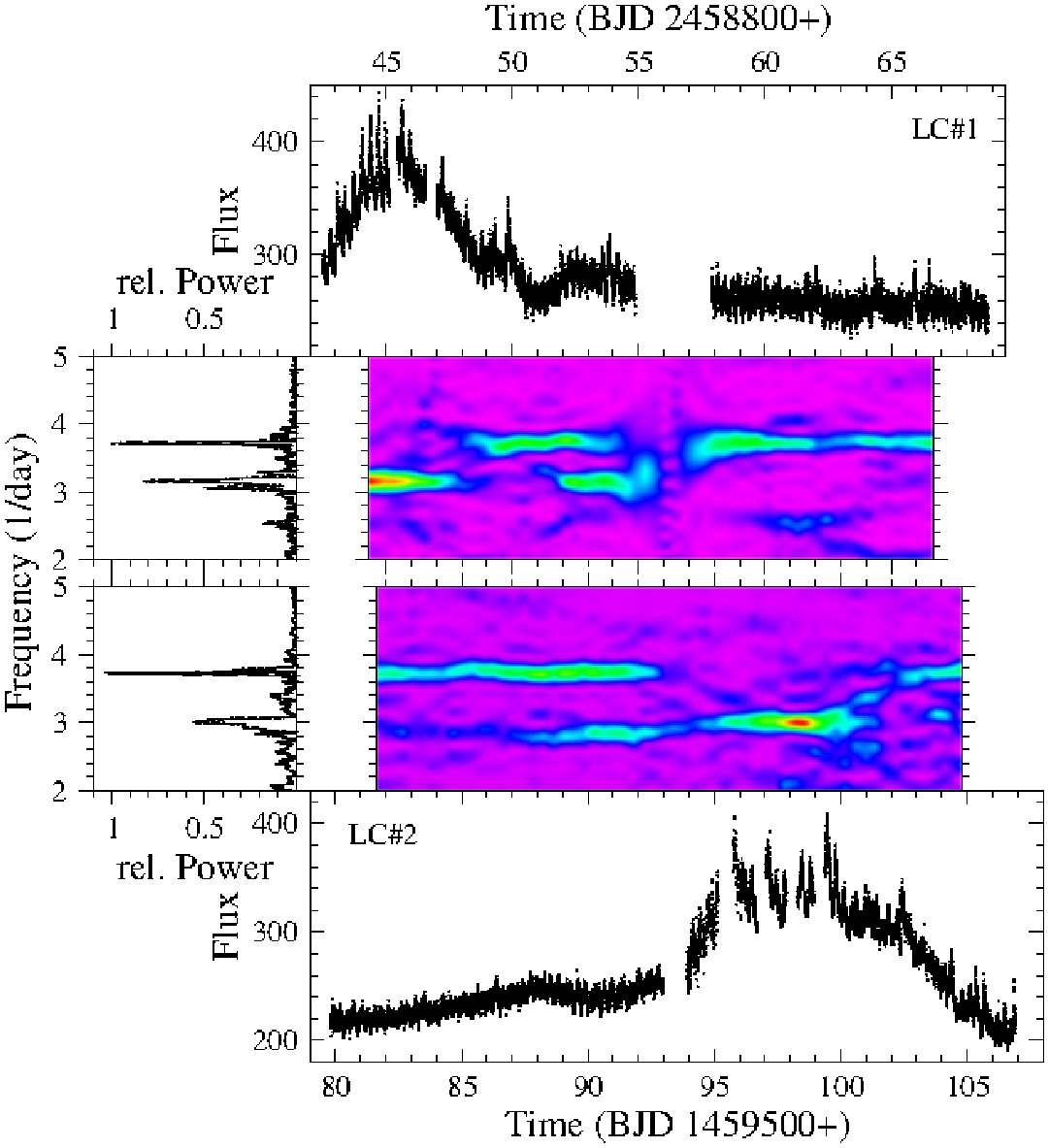}
%      \fbox{\rule[14cm]{14cm}{0cm}}
\caption{Light curves LC\#1 and LC\#2 of EI~UMa and their power 
                 spectra. The right upper and lower frames show the 
                 light curves 
                 from which the rapid ($<$0.5~d) outbursts were removed. The 
                 left frames contain the respective power spectra in a small
                 interval around the orbital and superhump frequencies.
                 The remaining frames show the time resolved power spectra
                 in the same frequency range and aligned in time with the
                 light curves.}
\label{eiuma-stacked}
\end{figure}
%______________________________________________________________
 
\subsection{LN Ursae Majoris}
\label{LN UMa}

While the long-term variations of LN~UMa have been reasonable well
studied and revealed it to be a novalike variable of VY~Scl type
\citep{Hillwig98, Honeycutt04}, the only attempt to find regular variations
in time resolved light curves was performed by \citet{Papadaki09}
and met no success. An orbital period of 0.169~d has been measured 
spectroscopically by \citet{Ringwald93}. This is a 1/day alias of another
period of 0.1444(1)~d, measured by \citet{Hillwig98} and which
the TESS data confirm to be correct (see below).
Just as many other NLs with orbital periods close to the upper rim of the
CV period gap LN~UMa is also a SW~Sex star \citep{Rodriguez-Gil07b}.

TESS observed LN~UMa in six sectors. Some of the data can be combined
which leaves us with four light curves. They are shown in Fig.~\ref{lnuma-lc}
revealing that their character can change a lot over time. I draw
particular attention to LC\#4 which is dominated 
by a brightening with an amplitude of $\approx$0.55~mag, preceded by what 
appears to be the decline from a previous brightening. Similar events
are also observed in the ASAS-SN long term light curve of LN~UMa. Other
systems of the RK sample mentioned in the Introduction show much the same
behaviour which will be investigated in a separate paper.

%--------------------------------------------------------------
\begin{figure}
	\includegraphics[width=\columnwidth]{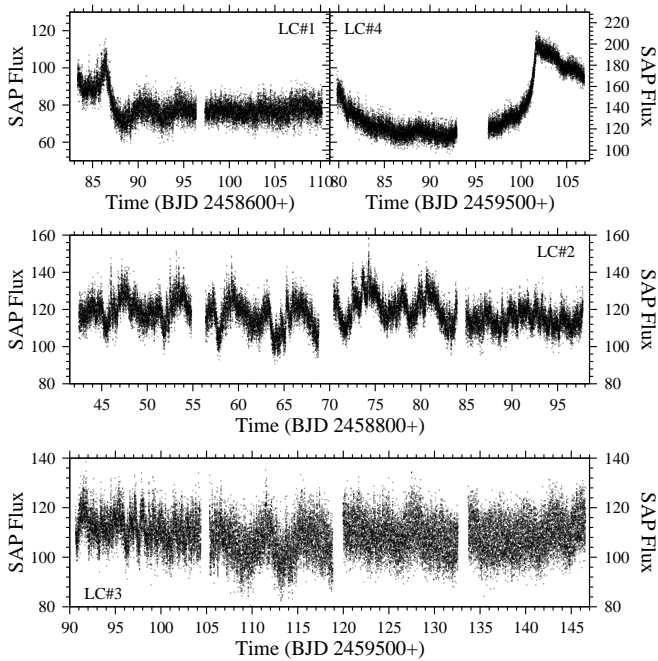}
%      \fbox{\rule[14cm]{14cm}{0cm}}
\caption{The four TESS light curves of LN~UMa. Note that the time scale is
the same but the flux scale has been adapted to the individual light curve.}
\label{lnuma-lc}
\end{figure}
%______________________________________________________________

The power spectra of the four light curves are not all alike. Overall they
indicate the presence of orbital variations and of a negative superhump.
In order to understand the differences
between the individual light curves, Fig.~\ref{lnuma-stacked} shows on 
the left their conventional power spectra in the relevant frequency range,
and on the right the corresponding time resolved spectra.
In LC\#1 only a weak orbital signal is seen apart from random variations.
It is not deteced in LC\#2 and has made room for a much stronger negative
superhump. LC\#3 also does not contain the orbital signal, and the superhumps
are even more developed and increase their strength over time. Finally, in LC\#4
the orbital signal returns and gains in strength towards the end of the light
curve. The superhump is now much weaker and appears to fade away. The
superhump period evolves from 0.13797(2)~d (LC\#2) over 0.1383387(5)~d
(LC\#3) to 0.13754(8)~d (LC\#4). The average period excess is thus
% $\epsilon = -0.046695$ 
$\epsilon = -0.047$. The values listed in Table~3
% Table~\ref{Table: Masterlist}
refer to LC\#3 where the superhump is strongest. 
The orbital variations are most strongly seen in LC\#4 where the period is 
measured to be 0.14471(4)~d, close to the value quoted by \citet{Hillwig98}. 
The orbital (black) and superhump 
(red) waveforms constructed from LC\#4 and LC\#3, respectively, are displayed
in the left frame of Fig.~\ref{lnuma-wf}. The other light curves yield 
waveforms similar in shape, but noisier. Both consist of a single slightly
asymmetric hump.
 
%--------------------------------------------------------------
\begin{figure}
	\includegraphics[width=\columnwidth]{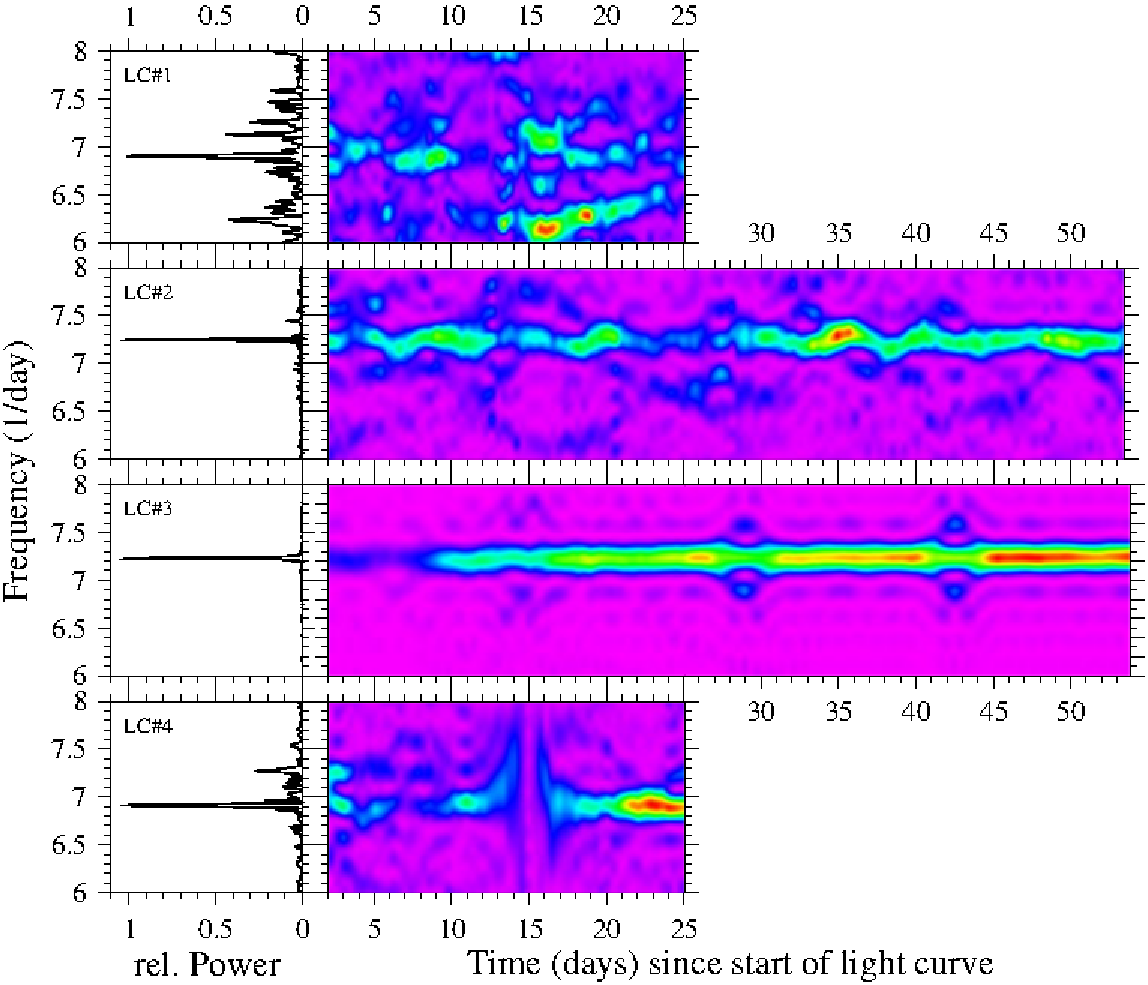}
%      \fbox{\rule[14cm]{14cm}{0cm}}
\caption{Power spectra of the four light curves of LN~UMa in a narrow
range around the orbital and superhump frequencies in the conventional
(left) and the time resolved form (right).}
\label{lnuma-stacked}
\end{figure}
%______________________________________________________________

%--------------------------------------------------------------
\begin{figure}
	\includegraphics[width=\columnwidth]{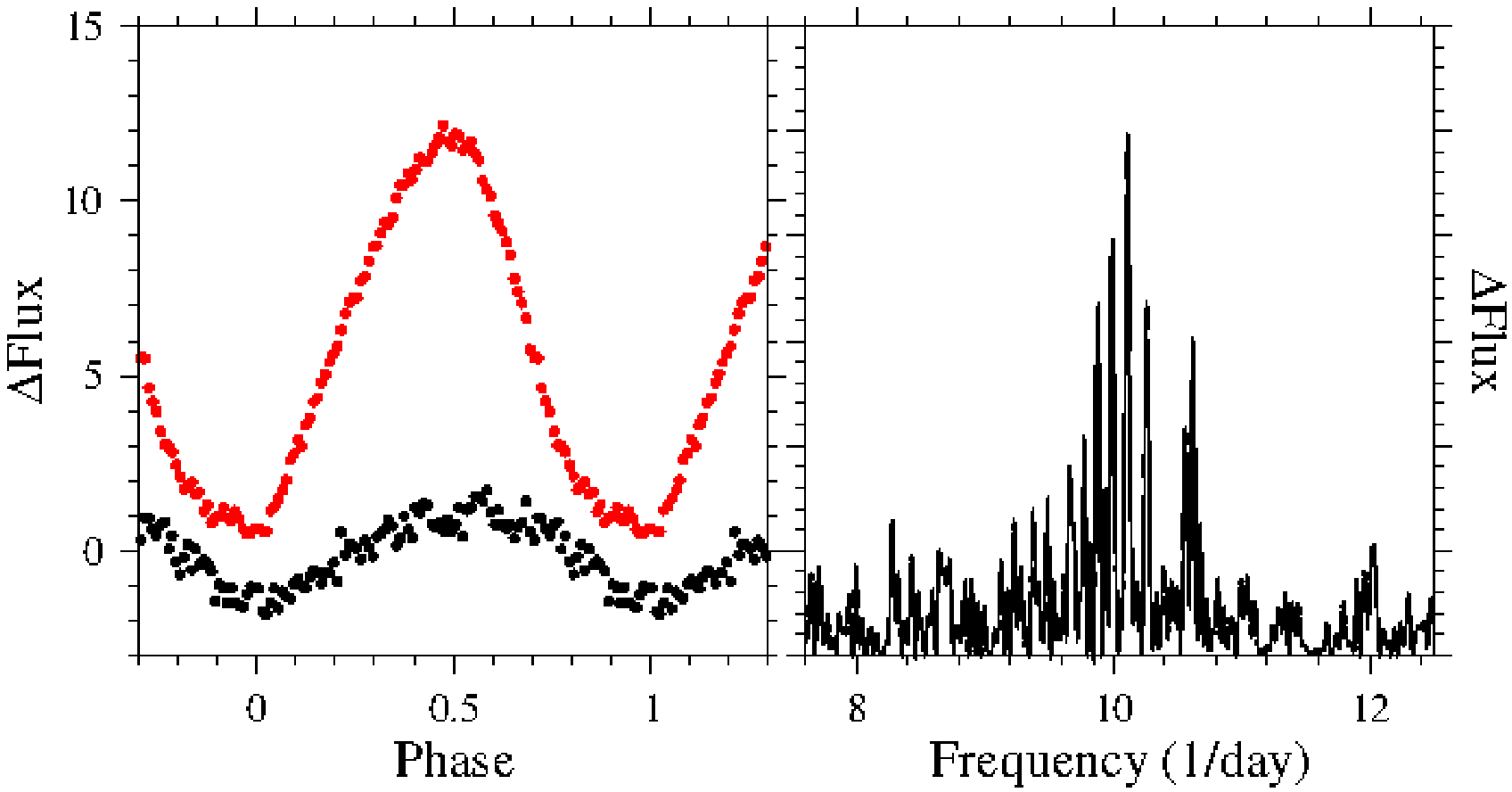}
%      \fbox{\rule[14cm]{14cm}{0cm}}
\caption{{\it Left:} Orbital (black) and superhump (red) waveforms constructed
from LC\#4 and LC\#3, respectively, of LN~UMa. {\it Right:} Part of the
power spectrum of LC\#4 of LN~UMa, containing long period QPOs.}
\label{lnuma-wf}
\end{figure}
%______________________________________________________________

Finally, I draw attention to a concentration of peaks in the power spectrum
of LC\#4 in the frequency range 9--11~d$^{-1}$ (see right frame of 
Fig.~\ref{lnuma-wf}). This phenomenon, also seen in other systems,
will briefly be discussed in Sect.~\ref{Discussion}.

\subsection{CN Velorum}+
\label{CN Vel}

Not many details about the quiescent properties of the 1905 classical nova
CN~Vel have been published. \citet{Tappert13} measured a spectroscopic
orbital period of 0.220(3)~d. 

CN~Vel has a much brighter neighbour, only 22.6~arcsec away, TYC~8619-935-1. 
The magnitude difference is 3.7~mag \citet{GaiaCollaboration20}. Considering 
the coarse spatial resolution of the TESS images it is quite possible that 
the single sector TESS light curve of CN~Vel is heavily contaminated with light 
from this neighbour. However, TYC~8619-935-1 in not known to be variable.
Therefore, the presence of variations similarly found in other CVs but not
typically seen in other variable stars inspires some confidence that at
least these modulations can really be attributed to CN~Vel. 

The light curve is shown in the left frame of Fig.~\ref{cnvel}, while its
conventional power spectrum is reproduced together with its time resolved 
counterpart in 
Fig.~\ref{cnvel-stacked}. In the latter, power is shown on a square root 
scale in order to better visualize fainter structures.
It contains some strong but enigmatic spectral
lines which are identified in the left frame of the figure and are listed in
% Table~\ref{Table: CN Vel frequencies} 
Table~4 where their frequencies and periods
are given. All signals are persistent over the entire duration of the light
curve. Only two lines can be readily interpreted. $1/F_6$ lies only
marginally beyond the formal error limits of the spectroscopic period of
\citet{Tappert13} and is therefore identified as being due to orbital
modulations. The slightly stronger signal $F_5$ must then be considered
as caused by a positive superhump. The noisy waveforms of
both, the SH modulations (blue) and orbital variations (black), are
shown in the right frame of Fig.~\ref{cnvel} and are nearly sinusoidal.

%--------------------------------------------------------------
\begin{figure}
	\includegraphics[width=\columnwidth]{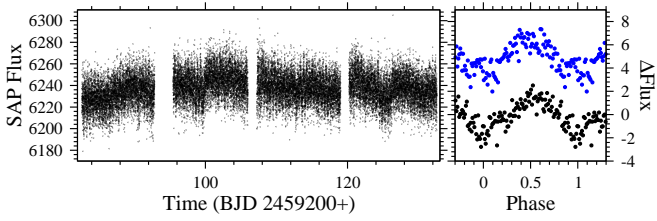}
%      \fbox{\rule[14cm]{14cm}{0cm}}
\caption{{\it Left:} Light curve of CN~Vel, probably heavily contaminated
by the brighter neighbouring star TYC~8619-935-1. {\it Right:} Orbital (black)
and superhump (blue) waveforms.}
\label{cnvel}
\end{figure}
%______________________________________________________________

%--------------------------------------------------------------
\begin{figure}
	\includegraphics[width=\columnwidth]{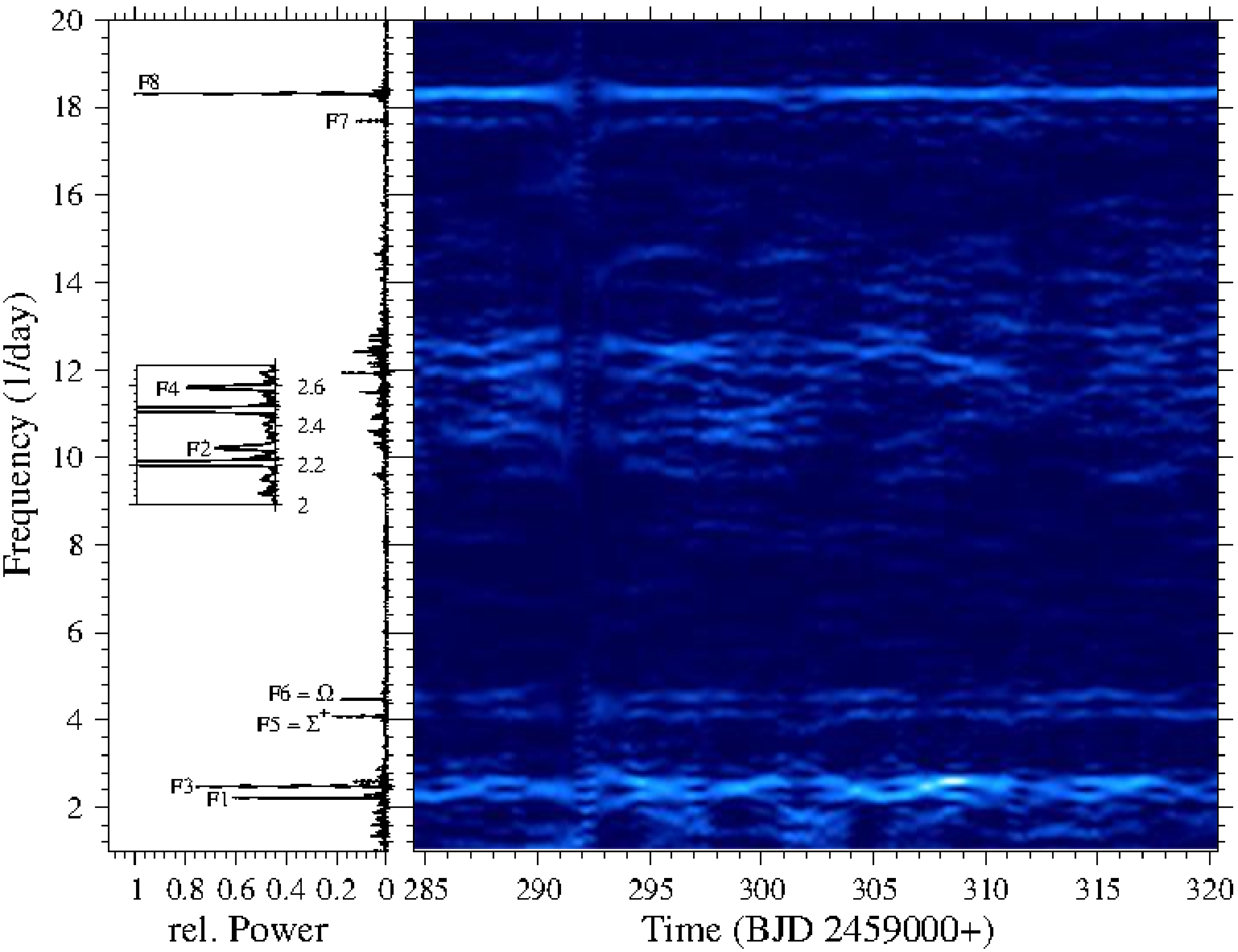}
%      \fbox{\rule[14cm]{14cm}{0cm}}
\caption{{\it Left:} Conventional power spectrum of the light curve of CN~Vel. 
Spectral lines considered significant and persistent over time are identified.
The inset contains a small section of the power spectrum with signals which
are not well resolved in the main frame. {\it Right:} Time resolved power
spectrum of the light curve. Power is scaled by its square root in order to
better visalize fainter structures.}
\label{cnvel-stacked}
\end{figure}
%______________________________________________________________

%--------------------------------------------------------------
\begin{table}
\label{Table: CN Vel frequencies}	
\centering
	\caption{Periodic signals identified in the power spectrum of CN Vel.
Column 1: label as used in Fig.~\ref{cnvel-stacked}; column 2: 
frequency of the signal in 1/day; column 3: Period in days.}
\begin{tabular}{lll}
\hline
Feature & Frequency & Period \\
\hline
$F_1$ & \phantom{0}2.311(1) & 0.4523(3)   \\
$F_2$ & \phantom{0}2.288(4) & 0.4371(8)   \\
$F_3$ & \phantom{0}2.480(1) & 0.40317(2)  \\
$F_4$ & \phantom{0}2.594(3) & 0.3854(4)   \\
$F_5 = F_{\rm pSH}$ & \phantom{0}4.079(1) & 0.24513(6) \\
$F_6 = F_{\rm orb}$ & \phantom{0}4.477(1) & 0.22338(6) \\
$F_7$ & 17.682(1) & 0.0546555(4) \\
$F_8$ & 18.3104(5) & 0.054614(1) \\
\hline
\end{tabular}
\end{table}
%------------------------------------------------------------------

I have no explanation for the other signals. Attempts to find  simple
numerical relations between their frequencies yielded no satisfactory
results. It is, of course, well possible that they are intrinsic to
TYC~8619-935-1 instead of CN~Vel. 
The intriguing time dependent behaviour of the $F_1$
and $F_3$ signals which seem to change their frequencies and even cross
over each other is probably not real but caused by their mutual beating.

Another interesting property of the power spectrum is a concentration of
apparently unstable signals in the frequency range 9.5 -- 13~d$^{-1}$
(periods: 1.8 -- 2.5~h), possibly extending up to 15~d$^{-1}$ (1.6~h).
This seems to put CN~Vel into the same category with several other
NLs exhibiting long period QPOs, as briefly discussed in Sect.~\ref{Discussion}.

\subsection{HS 0229+8016}
\label{HS 0229}

HS~0229+8016 (HS~0229 hereafter) was identified
by \citet{Aungwerojwit05} as a cataclysmic variable among the stars of the
Hamburg Quasar Survey. They measured a spectroscopic 
orbital period of 0.16149(3)~h. Photometric variations compatible with this
period were also detected but did not permit to obtain a more precise value.

For the present study four light curves are available, three of them
encompassing two TESS sectors and the last one spanning only a single 
sector. LC\#2 is reproduced in the upper frame of Fig.~\ref{hs0229}. The
general aspect of the others is similar. The light curves are
dominated by distinct semi-regular brightenings. A similar phenomenology
is also seen in some other systems in the entire sample mentioned in
Sect.~\ref{Introduction} and will be discussed in a separate paper. Here,
I concentrate on the detection of a negative superhump in HS~0229.  

%--------------------------------------------------------------
\begin{figure}
	\includegraphics[width=\columnwidth]{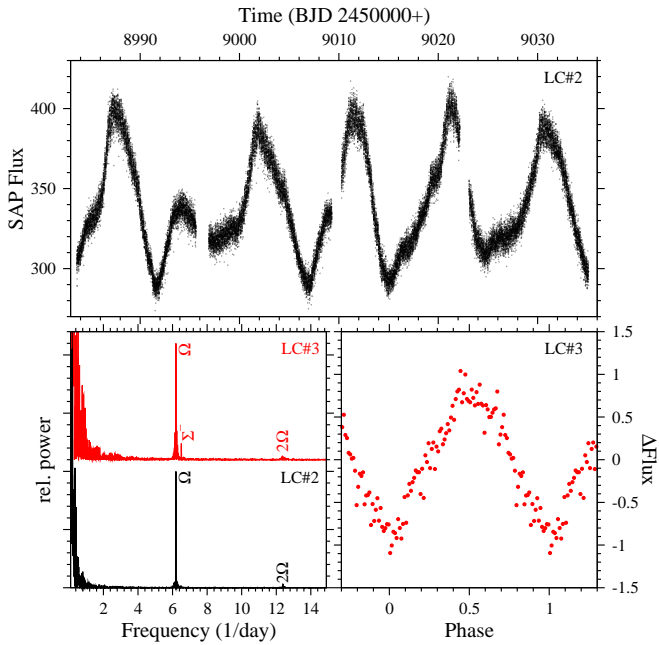}
%      \fbox{\rule[14cm]{14cm}{0cm}}
\caption{Light curve LC\#2 (top), power spectra of LC\#2 (black) and LC\#3
(red) (lower left) and superhump waveform (lower right) of HS~0229.}
\label{hs0229}
\end{figure}
%______________________________________________________________

The power spectra of all light curves contains a strong signal at the
orbital frequency and a much weaker one at its first overtone. The lower
left frame of Fig.~\ref{hs0229} shows as examples the spectra of LC\#2 and
LC\#3. The average value of the peak frequencies permits to improve the
precision of the orbital period to 0.161439(9)~d. The orbital waveform
depends on the phase of the brightenings and will be explored in the 
upcoming paper mentioned above.

Apart from the orbital signal the power spectrum of LC\#3 contains a small
but significant peak at a slightly higher frequency. At a much fainter level
(too faint to be conspicuous if it were not for its stronger appearance in
LC\#3) it is also detected in the other light curves. I take it as an
indication of a negative superhump. It has the same period of
% 0.1536858380 +- 0.0000254534
0.15369(3)~d in LC\#1 -- LC\#3, and a marginally lower one of
0.1534(1) in LC\#4. Its waveform (lower left frame of Fig.~\ref{hs0229}) is
almost sinusoidal.

\subsection{HS 0506+7725}
\label{HS 0506}

HS~0506+7725 (HS~0506 hereafter) is a cataclysmic variable also identified 
in the Hamburg Quasar Survey and studied by \citet{Aungwerojwit05}. 
They detected deep low states (see also the AAVSO long term light curve)
which allow the system be be ranked among the VY~Scl stars. The spectroscopic
orbital period is 0.14770(14)~d. This period was not detected in their
light curves which otherwise are characterized by the strong random 
flickering common to VY~Scl objects \citep{Bruch21}.

TESS observed HS~0605 in eight sectors. The data can be combined into four
2-sector light curves, with an interval of 4 months between LC\#1 and LC\#2, 
and LC\#3 and LC\#4, respectively. LC\#2 and LC\#3 are separated by about
two years. All light curves contain well expressed variations
on times scales shorter than a day, superposed on modulations on longer 
time scales which, while not periodic, appear to have some regularity.
This is best seen in LC\#3 which is displayed together with LC\#2 in the
upper frames of Fig.~\ref{hs0506-stacked}. 

Consequently, the power spectra contain many unstable signals at low
frequencies. As an example, the power spectrum of LC\#2 is reproduced in
the lower left frame of Fig.~\ref{hs0506-stacked}. In none of the spectra
a signal appears at the spectroscopic orbital frequency. Instead, the 
power spectrum of LC\#2 (but not of the other light curves) contains a 
conspicuous signal at a slightly higher frequency corresponding to a period of
% 0.1416855901 +- 0.0001274950
0.1417(1)~d. The time resolved power spectrum (displayed in the third row
of the figure together with the conventional power spectrum in a narrow
frequency range around the signal) reveals that it is not
permanent but appears only in a 16~d interval starting at JD~2449004 and
possible returns right at the end of LC\#2. I interpret it tentatively
as being due
to a temporarily active negative superhump. Within the elevated
noise level the superhump waveform (lower right frame of 
Fig.~\ref{hs0506-stacked}) does not deviate much from a simple sine curve.

%--------------------------------------------------------------
\begin{figure}
	\includegraphics[width=\columnwidth]{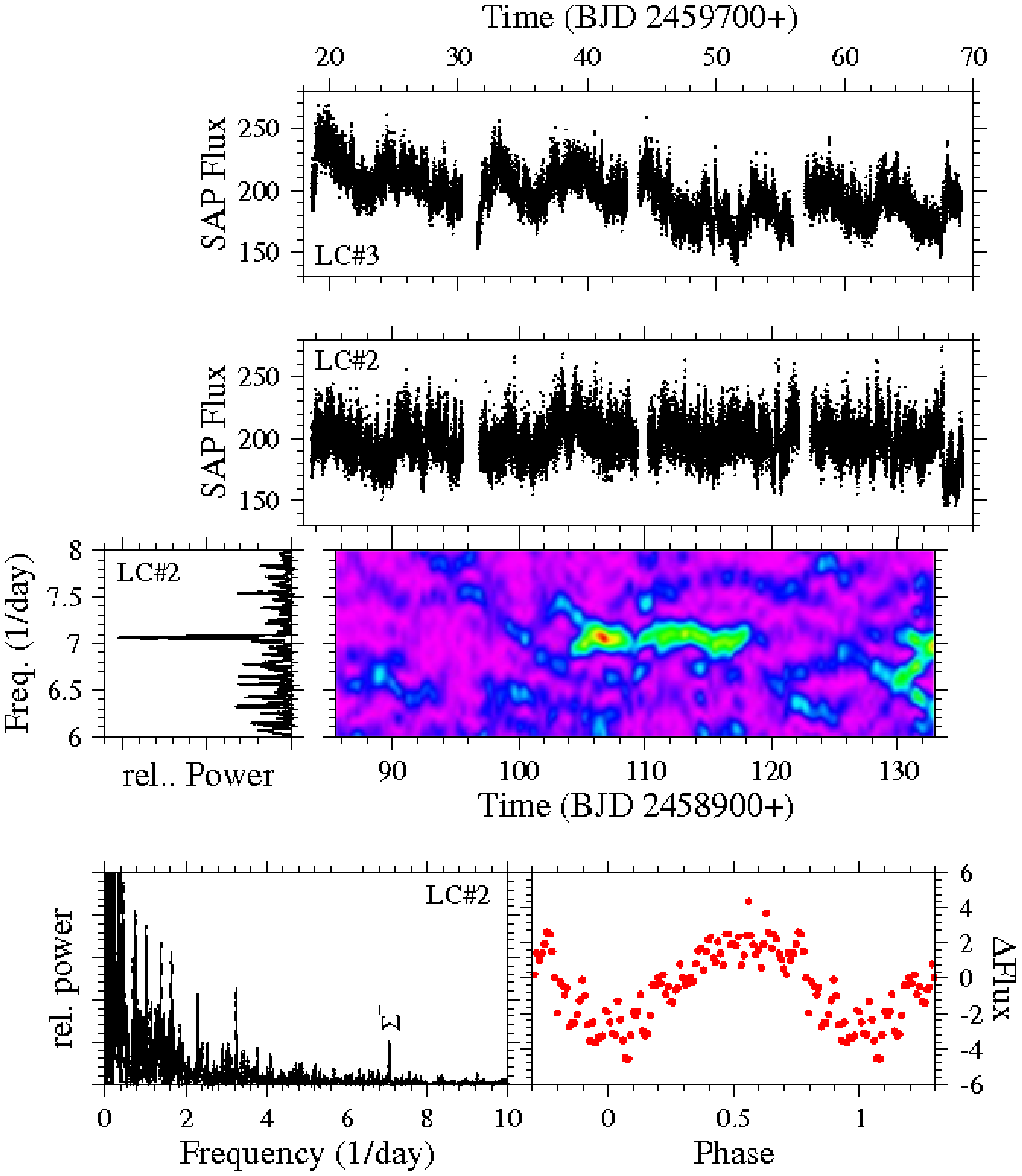}
%      \fbox{\rule[14cm]{14cm}{0cm}}
\caption{{\it Top row:} Light curve LC\#3 of HS~0506.
{\it 2$^{nd}$ and 3$^{rd}$ row:} Light curve of LC\#2 and its power spectrum
in a narrow range around the frequency of a temporarily active negative
superhump in conventional and time resolved forms. {\it Bottom row:}
Power spectrum of LC\#2 in a wider frequency range (left) and the superhump
waveform (right).}
\label{hs0506-stacked}
\end{figure}
%______________________________________________________________

\subsection{HS 0642+5049}
\label{HS 0642}

HS~0652+5049 (HS~0642 hereafter) is still another object listed in 
the Hamburg Quasar Survey (HS) and identified as a cataclysmic variable by
\citet{Aungwerojwit05}. They did not find radial velocity variations in
their spectra, but the power spectrum of the combined light curves of the
nights of October 25, December 8 and December 9, 2004 indicates a period
of 0.156875(16)~d.

Two single sector TESS light curves of HS~0642 are available, separated by 
3 yrs (upper frames of Fig.~\ref{hs0642}). LC\#1 exhibits
clear variations on the time scales of 4 and 1~d, superposed
upon more gradual variations. The yellow curve is
a fit to the data of a fifth order polynomial (to follow the long
term variations) and two sine terms with periods of $P_1 = 4.02$~d and 
$P_2 = 1.09$~d, respectively.
Of course, these long periods also appear as strong peaks in the power 
spectrum (lower left frame of the figure). At
higher frequencies the strongest peak corresponds to a period of
% 0.1579075307 +- 0.0000212531
0.15791(2)~d. Even considering the error limits this is incompatible with 
the orbital
period of \citet{Aungwerojwit05}. But the respective frequency difference
corresponds to a period of 24~d, close to the half the time difference of
45~d of the two observing mission on which their period is based. Thus,
the period difference may well be due to the choice of a wrong alias peak
by \citet{Aungwerojwit05} (see their figure 13). 
Therefore, I consider 0.15791~d as the orbital
period of HS~0642. The star then has a saw-tooth shaped orbital waveform
(black dots in the second lower left frame of Fig.~\ref{hs0642}) with a 
slightly longer rise and a steeper decline.

%--------------------------------------------------------------
\begin{figure*}
	\includegraphics[width=\textwidth]{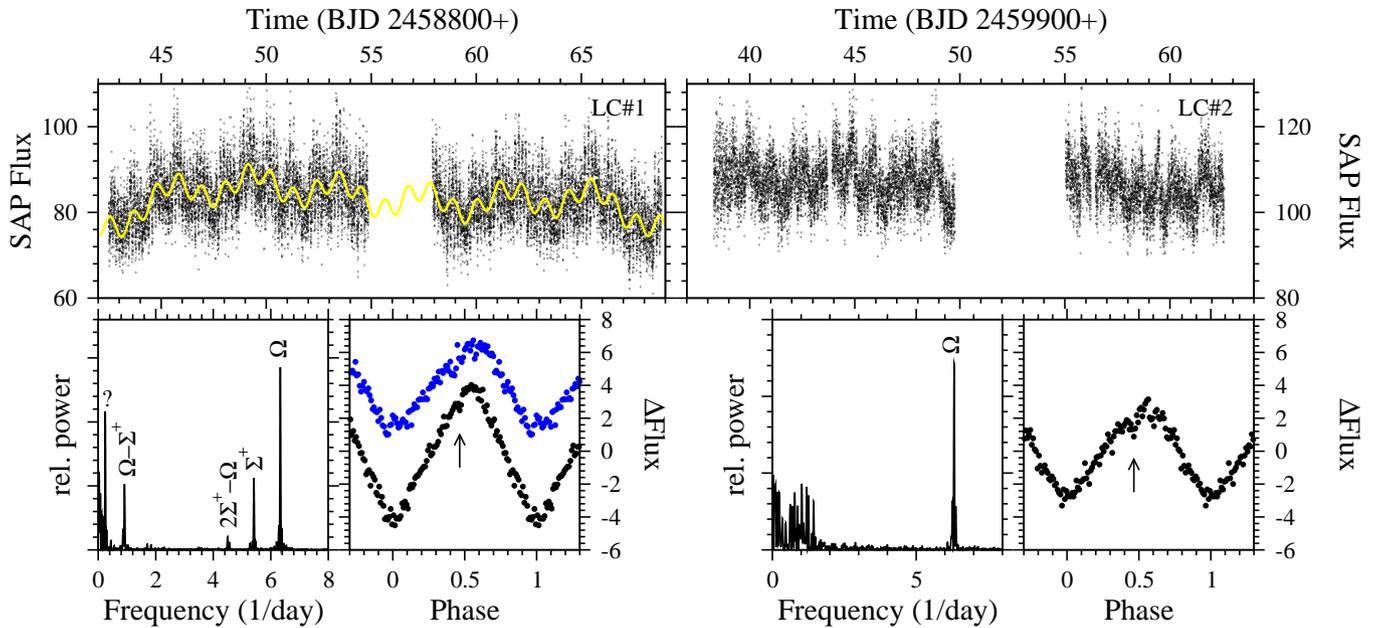}
%      \fbox{\rule[14cm]{14cm}{0cm}}
\caption{{\it Top:} Light curves of HS~0642. The yellow curve in the left
frame is the sum of a fifth order polynomial and two sine waves with
periods of 4.02 and 1.09~h fitted to the data.
{\it Bottom:} Power spectra (left) and waveforms (right) for both light
curves. Black and  blue colours refer to orbital and SH variations,
respectively. The arrows point to a faint but possibly persistent structure 
in the orbital waveform which repeats itself at exactly the same phase in
both light curves.}
\label{hs0642}
\end{figure*}
%______________________________________________________________

Close to the orbital signal in the power spectrum another peak at a lower
frequency can be interpreted as being to due to a positive superhump at
a period of
% 0.1847066581 +- 0.0000573042
0.18471(6)~d (lower left frame of Fig.~\ref{hs0642}). 
A weaker close-by third peak is then identified as 
$2\Sigma^+ - \Omega$, and the 1.09~d period is the beat period between the
orbital and superhump variations. However, at $\epsilon = 0.170$ the
period excess lies much above the Stolz-Schoembs relation. This will
be discussed in Sect.~\ref{Discussion}. The superhump
waveform (blue dots in the figure), being saw-tooth shaped, is quite similar
to the orbital waveform. 

Light curve LC\#2 (upper right frame of Fig.~\ref{hs0642}) also contains
variations on time scale of one to a couple of days. However, in contrast
to LC\#1, they are not periodic. Instead, they generate a region of enhanced
power in the range of 0.5 -- 1.5~d$^{-1}$ (see the power spectrum in the lower
frames of the figure). Otherwise, the power spectrum only contains a strong
signal at the orbital frequency. The superhump has vanished. The orbital
waveform has a lower amplitude than in LC\#1 and a more sinusoidal shape.
The small dip at phase 0.465 (marked by the arrow in the figure)
may not be spurious, because the waveform 
constructed from LC\#1 has a similiar, albeit fainter dip at exactly the
same phase.

At a low power level a concentration of peaks appears in the power spectrum 
of LC\#1 (but not of LC\#2) between frequencies of about 10 and 14~d$^{-1}$ 
(left frame of Fig.~\ref{hs0642-stacked}). Two of these may be identified as 
the first overtones of the superhump and the orbital signal. The others,
however, defy a simple explanation and seem to be unstable as is
illustrated by the time resolved power spectrum in the right hand frame
of the figure. Thus, HS~0642 is another member of the group of systems
which exhibit QPOs on the time scale close to an hour.

%--------------------------------------------------------------
\begin{figure}
	\includegraphics[width=\columnwidth]{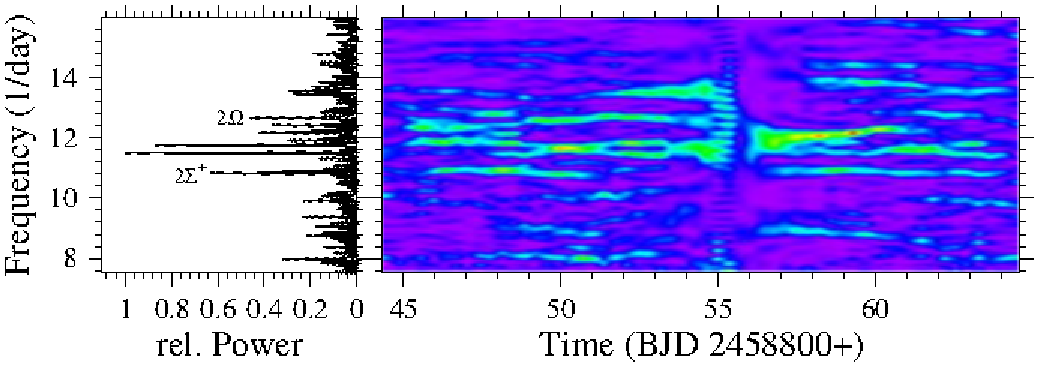}
%      \fbox{\rule[14cm]{14cm}{0cm}}
\caption{Power spectrum of LC\#1 of HS~0642 in a narrow frequency range with
a concentration of unstable signals in conventional (left) and
time resolved form (right).}
\label{hs0642-stacked}
\end{figure}
%______________________________________________________________

This leave just one significant power spectrum signal without explanation,
i.e, the long 4.02~d period in LC\#1. It cannot easily be related to any 
of the other periods in HS~0642 and its nature remains unknown.

\subsection{IGR J08390-4833}
\label{IGR J0839}

Based on its optical spectrum \citet{Kniazev08} identified IGR~J08390-4833
(IGR~J0839 hereafter) as a cataclysmic variable. In Chandra observations
\citet{Sazonov08} detected X-ray pulsations with a period of $1450 \pm 40$~sec
and classified the system as an intermediate polar. Based on XMM-Newton
observations \citet{Bernardini12} refined this period to $1480.8 \pm 0.5$~sec.
A consistent period was also seen in the UV, but the $B$ band data yielded a
period of $1650 \pm 7$~sec which the authors interpreted as the orbital
sideband of the WD spin period, implying a long orbital period of 
IGR~J0839 of $8 \pm 1$~h.

Two 2-sector TESS light curves separated by 2 years are available. They
show that IGR~J0839 is not only an intermediate polar but also exhibits 
negative superhumps, albeit not permanently. LC\#2 (upper frame of
Fig.~\ref{j0839}) contains sinusoidal
variations superposed upon more gradual variations. The yellow curve is 
the sum of a third order polynomial (for the gradual modulation) and a
sine fit to the data. It already suggests being the beat between two
other periods. This is confirmed by the power spectrum (lower left frame
of the figure) which contains two peaks at neighbouring frequencies and 
a signal at their
difference. The lower frequency peak, corresponding to a period of
% 0.2540808469 +- 0.0000248819
0.25408(2)~d, is orbital. This is obvious from the power spectrum of LC\#1
which contains the same signal but not the one at the higher frequency,
as well as from the analysis of the IP type variations (see below). The 
waveform of the orbital variations (black dots in the lower right frame of
the figure) is close to a simple
sine curve. The second peak in the power spectrum corresponds to a period of
% 0.2401317805 +- 0.0000427975
0.24013(4)~h. It can be interpreted as being due to a negative superhump. 
The beat period
between orbit and superhump of 4.38(2) is reflected in the sine wave in
the light curve. The superhump waveform (red dots in Fig.~\ref{j0839}) consists
of a single hump with a structured maximum. The superhump is not permanent 
because no trace of it can be detected in LC\#1.

Although not the topic of this study, a brief discussion of the IP type
variations in J0839 is in order. They are better expressed in LC\#1 than 
in LC\#2. The power spectrum (insert in the lower left frame of 
Fig.~\ref{j0839}) contains only a faint signal
at the frequency of the X-ray pulsations seen by \citet{Sazonov08} and
\citet{Bernardini12}. Their period can be determined with much higher
precision than based on the previous observations. I measured 
% 1482.33917 +- 0.04933
1482.34(5)~sec. Note that this is significantly longer than expected 
considering the error margin quoted by \citet{Bernardini12}. At optical
wavelengths ($B$ band) these authors detected a longer period of 1560(7)~sec
which they take to be the orbital sideband of the WD spin period. No
peak is seen at the corresponding frequency in the TESS power spectra. 
However, a signal much stronger than the WD spin signal corresponds to a
period of
% 1589.603448 +- 0.009097
1589.603(9)~sec. Adopting the orbital period derived above it is 
undoubtedly the orbital sideband $\omega - \Omega$ of the spin. 
Another signal can be identified as
$\omega - 2\Omega$. The same is true for the first overtone at 
$2\omega - 2\Omega$ (beyond the limits of the figue) 
but not for the spin signal at $2\omega$. I cannot offer an explanation for
the remarkable difference of almost 30~sec, much beyond the error
margin, between the sideband period of 1560(7)~sec quoted by 
\citet{Bernardini12} and the 1589.603(9)~sec period measured in the TESS data. 

%--------------------------------------------------------------
\begin{figure}
	\includegraphics[width=\columnwidth]{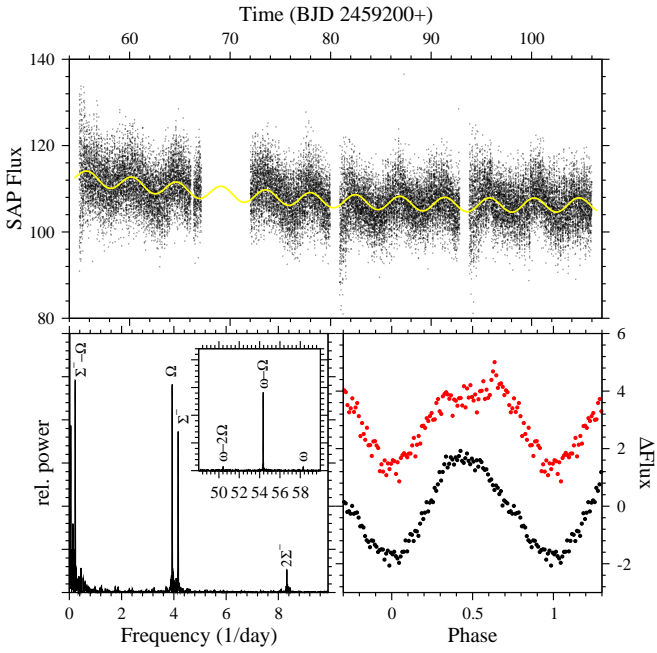}
%      \fbox{\rule[14cm]{14cm}{0cm}}
\caption{Light curve (top), power spectrum (lower left) and waveforms
(lower right) of the orbital (black) and superhump variations (red) of
LC\#2 of IGR~J0839. The insert in the lower left frame contains the power
spectrum of LC\#1 at higher frequencies showing the (weak) WD spin signal
and its orbital sidebands. The continuous yellow curve in the upper frame is
the sum of a low order polynomial (to follow the gradual variations) and a
least squares sine fit to the data.}
\label{j0839}
\end{figure}
%______________________________________________________________

\subsection{H$\alpha$ 103959.96-470126.1}
\label{Ha 1039}

H$\alpha$~103959.96-470126.1 (H$\alpha$~1039 hereafter) was
identified as a cataclysmic variable by \citet{Pretorius08} in a 
sample of H$\alpha$ emission line star from the AAO/UKST SuperCOSMOS 
H$\alpha$ Survey. They measured a spectroscopic orbital period of
0.1577(2)~d.

TESS observed H$\alpha$~1039 in two subsequent sectors. Just as in the
case of KQ~Mon (Sect.~\ref{KQ Mon}) \citet{Stefanov23} published their
analysis of the same data when this study was in preparation. Therefore,
the reader is referred to that paper for further details. I only add 
that the superhump waveform consists of a somewhat skewed single hump.

\subsection{H$\alpha$ 112921.67-535543.6}
\label{Ha 1129}

H$\alpha$~112921.67-535543.6 (H$\alpha$~1129 hereafter) is another
H$\alpha$ emission line star identified as a cataclysmic variable by 
\citet{Pretorius08}. The spectroscopic orbital period is 0.153546(2)~d.

The single sector TESS light curve (upper frame of Fig.~\ref{ha1129})
indicates that this system
exhibits negative and possibly simultaneously positive superhumps.
It contains only slight variations on time scales of days above the noise
level. The power spectrum (lower left frame of the figure) is dominated 
by a signal at the orbital frequency and its weaker first overtone. 
The waveform (lower right frame; back dots)
consists of a single flat-topped hump. An amplified view of the orbital
signal reveals some satellite lines (inset in the lower left frame of 
Fig.~\ref{ha1129}).
Apart from the usual side-lobes close to the main peak, unavoidable in
the power spectra of finite data sets, another maximum
at a higher frequency, corresponding to a period of
% 0.1479624808 +- 0.0000753199
0.14796(8)~d, can be identified as being due to a negative superhump.
A weaker peak close to 7.01~d$^{-1}$ is then the 
$2\Sigma^- - \Omega$ signal. The beat between orbital and superhump
period is 4.1~d. Considering the uncertainties caused by the noise in the
light curve a low frequency signal at 3.9~d may be a manifestation of
the disk precession period. The superhump waveform (lower right frame of
the figure; red dots) is noisy and undistinguishable from a simple sine 
curve. 

%--------------------------------------------------------------
\begin{figure}
	\includegraphics[width=\columnwidth]{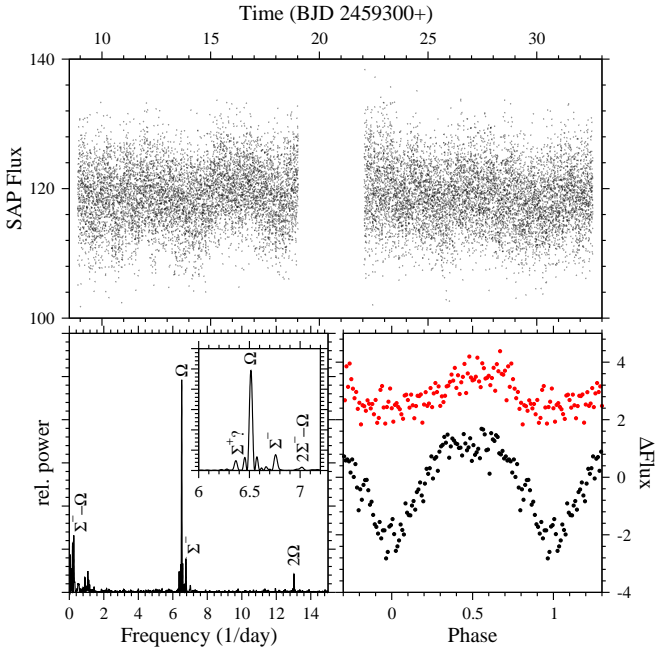}
%      \fbox{\rule[14cm]{14cm}{0cm}}
\caption{Light curve (top), power spectrum (lower left) and waveforms
(lower right) of the orbital (back) and superhump variations (red) of
H$\alpha$~1129. The insert in the lower left frame contains a blow-up
of the range around the orbital and superhump frequencies.}
\label{ha1129}
\end{figure}
%______________________________________________________________

A power spectrum peak below the frequency of the orbital signal
(inset in Fig.~\ref{ha1129}) corresponds to a period of 
% 0.1570744067+- 0.0001003117
0.1571(1)~d. Could this be caused by a positive superhump? The implied
period excess of 0.023 is small but not drastically so considering the 
scatter in the superhump period -- period excess relation 
(see Fig.~\ref{sh-epsilon-rel}). But then, H$\alpha$~1129 would be the 
only system with positive and negative superhumps (simultaneously or not) 
where the period excess is smaller for the pSH than for the nSH. Therefore, 
I leave open the question whether the 0.1571~d period is caused by a 
superhump or some other mechanism. 

\subsection{ASASSN-14ix}
\label{ASASSN-14ix}

The transient ASASSN-14ix remains an almost unstudied object. The only
available information consists of informal communications
\citep{Hambsch14a, Hambsch14b} reporting deep eclipses and an orbital
period of 0.1444610954(1)~d\footnote{The formal accuracy of this value
must be grossly overestimated. It would lead to an error of eclipse epochs
of merely 2.2~sec over a time base of 100~years!}. The single sector TESS 
light curve (upper frame of Fig.~\ref{asassn-stacked})
contains a $\approx$1~mag outburst and what appears
the be the decline from a second outburst. Thus, ASASSN-14ix may be
be a NL with brightenings such as, e.g., FY~Per (see Sect.~\ref{FY Per})
or HS~0229 (Sect.~\ref{HS 0229}) or even an outright dwarf nova.

The strongest signal in the power spectrum (lower left frame of 
Fig.~\ref{asassn-stacked}) corresponds to a period of
% 0.1381602734 +- 0.0001518212
0.1382(2)~d, shorter than the orbital period. I take this as the manifestation
of a negative superhump.
The dynamical power spectrum shown in the second
row of the figure together with the conventional spectrum in a small range
around the orbital and superhump frequencies shows that the superhump signal
subsides during the outburst. It persists during quiescence,
vanishes abruptly right at the onset of the outburst, but does not wait until
the end of the outburst to reappear. Instead, it restarts already just after 
outburst maximum.

%--------------------------------------------------------------
\begin{figure}
	\includegraphics[width=\columnwidth]{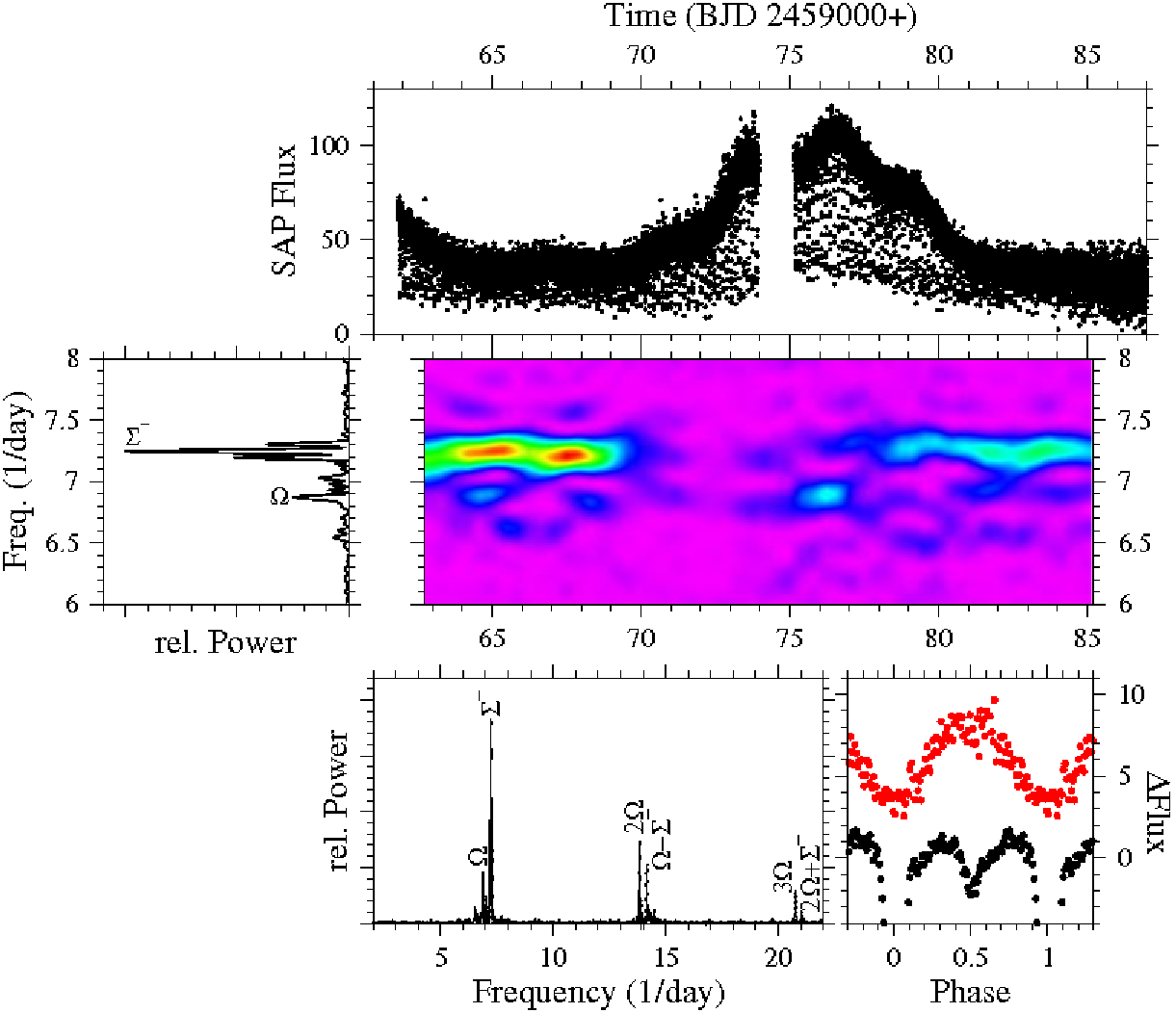}
%      \fbox{\rule[14cm]{14cm}{0cm}}
\caption{{\it Top:} Light curve of ASASSN-14ix. {\it Middle:} Power spectrum
of the light curve (eclipses masked) in a narrow range around the
orbital and superhump frequencies in the conventional (left) and time
resolved forms (right). {\it Bottom:} Power spectrum in a wider frequency
range (left) and the orbital (black) and superhump (red) waveforms (right).}
\label{asassn-stacked}
\end{figure}
%_____________________________________________________________

The superhump waveform (lower right frame of Fig.~\ref{asassn-stacked}; in red)
cannot be distinguished from a simple sinusoid. The orbital waveform (in black)
contains a well expressed secondary eclipse. Except for the depth of the 
primary eclipse it does not change significantly between quiescence and 
outburst. 

\section{Discussion}
\label{Discussion}

In Paper~II, I presented a census of superhumps in NLs. 
The current study already supersedes it, enlarging the number of known
SH systems among NLs by almost 50\%. Moreover, SHs in other NLs have 
recently been reported in the literature. These are listed in Table~5.
% Table~\ref{Table: Additional superhumpers}. 
Some basic statistics of the entire sample of the superhump systems is 
given in Table~6,  
% Table~\ref{Table: Statistics}, 
noting that in some of the newly added
systems the presence of SHs still needs confirmation [LZ~Mus (nSH);
H$\alpha$~1129 (pSH)], or their nature as either pSH or nSH requires 
clarification (DK~Lac). While the previous census showed that the number
of positive and negative superhump systems is approximately balanced, the
new detections revealed 2.4 times more negative than positive SH systems.
I cannot think of an observational bias to explain the preponderance of
negative SH systems among the new detections. On the other hand, the lower
than proportional growth rate of NLs showing both, pSHs and nSHs, or even
both simultaneously, may be due to insufficient temporal coverage of the
additional systems.  

%--------------------------------------------------------------
\begin{table*}
\label{Table: Additional superhumpers}	
\centering
	\caption{Additional NLs identified as superhump systems.}

\begin{tabular}{lllllll}
\hline
Name &
orb. period &
nSH &
    &
pSH &
    &
Reference \\
  &
(d) &
period (d) &
$\epsilon$ &
period (d) &
$\epsilon$ &
 \\
\hline
V598 Pup                   &     % Name of star
0.162874                   &     % Orbital period (d)
0.155                      &     % Negative superhump period
$-0.048$                   &     % period excess (nSH)
--                         &     % Positive superhump period
--                         &     % period excess (pSH)
1                          \\    % Reference
YZ Ret                     &     % Name of star
0.1324539                  &     % Orbital period (d)
                           &     % Negative superhump period
                           &     % period excess (nSH)
0.1384                     &     % Positive superhump period
0.045                      &     % period excess (pSH)
1                          \\    % Reference
BG Tri                     &     % Name of star
0.15845(10)                &     % Orbital period (d)
                           &     % Negative superhump period
                           &     % period excess (nSH)
0.1727(14)                 &     % Positive superhump period
0.090                      &     % period excess (pSH)
2                          \\    % Reference
Gaia DR3 4684361817175293440 &     % Name of star
0.15401(53)                &     % Orbital period (d)
0.14750(52)                &     % Negative superhump period
$-0.042$                   &     % period excess (nSH)
--                         &     % Positive superhump period
--                         &     % period excess (pSH)
3                          \\    % Reference
Gaia DR3 5931071148325476992 &     % Name of star
0.14827(46)                &     % Orbital period (d)
0.14248(43)                &     % Negative superhump period
$-0.039$                   &     % period excess (nSH)
--                         &     % Positive superhump period
--                         &     % period excess (pSH)
3                          \\    % Reference
HBHA 4204-09               &     % Name of star
0.14128(22)                &     % Orbital period (d)
0.13657(22)                &     % Negative superhump period
$-0.033$                   &     % period excess (nSH)
--                         &     % Positive superhump period
--                         &     % period excess (pSH)
3                          \\    % Reference
SDSS J090113.51+144707.6   &     % Name of star
0.14631(17)                &     % Orbital period (d)
0.13991(17)                &     % Negative superhump period
$-0.044$                   &     % period excess (nSH)
--                         &     % Positive superhump period
--                         &     % period excess (pSH)
3                          \\    % Reference
\hline
\multicolumn{7}{l}{
References: 
1 = \citet{Schaefer22};
2 = \citet{Stefanov22};
3 = \citet{Stefanov23}} \\
\end{tabular}
\end{table*}
%------------------------------------------------------------------

%--------------------------------------------------------------
\begin{table}
\label{Table: Statistics}	
\centering
	\caption{Basic statistics of superhump systems among NLs.}

\begin{tabular}{lc}
\hline
   & Number \\
\hline
Total number of superhump systems & 74 \\
Positive superhump systems   & 37 \\
Negative superhump systems   & 55 \\
Systems with both, pSHs and nSHs & 20 \\
Systems with simultaneous pSHs and nSPs & 11 \\
\hline
\end{tabular}
\end{table}

The new data points
warrant an update on the relationship between the superhump period excess
$\epsilon = \left( P_{\rm SH} - P_{\rm orb} \right) / P_{\rm orb}$ and
the superhump period $P_{\rm SH}$ \citep[i.e., the Stolz-Schoembs relation,][]
{Stolz84} for NLs.
This is shown in the left frame of Fig.~\ref{sh-epsilon-rel} for positive,
and in the right frame for negative superhumps. The black dots represent
data listed in table~4 of Paper~II, while the red ones are the additional
data points from Tables~3 and 5.
% Tables~\ref{Table: Masterlist} and \ref{Table: Additional superhumpers}. 
The small orange dots in the left
frame are taken from table~9 of \citet{Patterson05} to serve as reference. 
Most of them represent SU~UMa type dwarf novae in outburst which follow the
Stolz-Schoembs relation much more closely than NLs. Moreover, the gradient of 
the relation is much steeper for dwarf novae than for NLs. The range of 
pSH periods measured in EI~UMa (Sect.~\ref{EI UMa}) and the two distinct 
nSH values of NS~Cnc (Sect.~\ref{NS Cnc}), respectively, lead to 
points in the diagrams which are connected by a straight line. 

%--------------------------------------------------------------
\begin{figure*}
	\includegraphics[width=\textwidth]{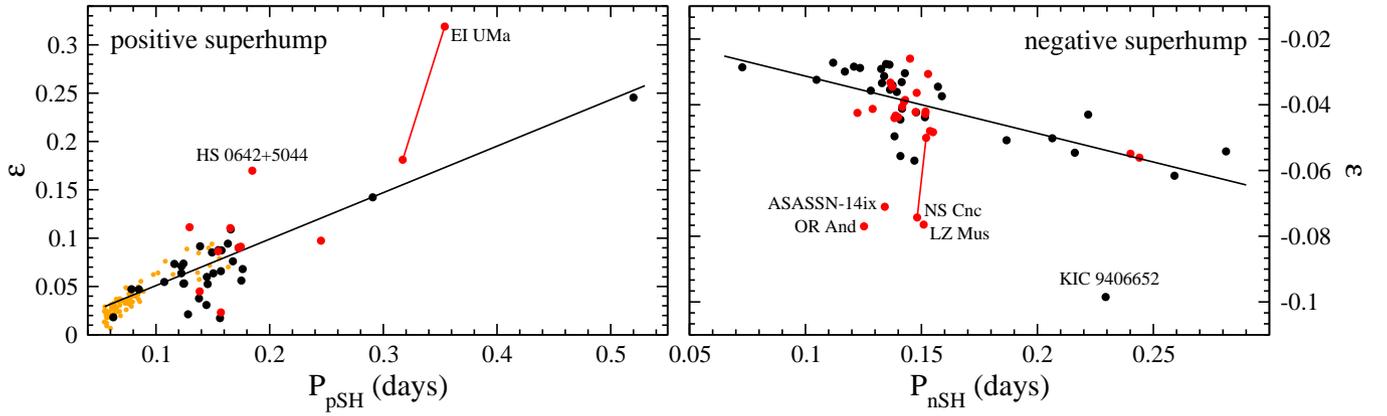}
%      \fbox{\rule[14cm]{14cm}{0cm}}
\caption[]{Period excess $\epsilon$ of superhumps as a function of the 
          superhump period $P_{\rm SH}$ (Stolz-Schoembs relation). 
          {\it Left:} Stolz-Schoembs relation for positive superhumps.
          Data listed in table~4 of Paper~II are shown as black dots,
          while the red dots represent data from Tables~3 and 5 of this
          study. The extremes of the range of $\epsilon$ values measured 
          in EI~UMa are connected by a red line. The black line is a linear
          least squares fit to all data, excluding six low credential
          superhump systems identified in Paper~II plus H$\alpha$~1129, 
          the strongly deviating HS~0642+5044, and the
          high $\epsilon$ point of EI~UMa. For comparison,
          data taken from table~9 of \citet{Patterson05} are
          plotted as smaller orange dots.
          {\it Right:} The same for negative superhumps. The red line connects
          the two SH period excess values 
          measured in NS~Cnc. The least squares
          fit to the data does not include the individually labelled stars
          and the low credential system RW~Tri (Paper~II).}
\label{sh-epsilon-rel}
\end{figure*}
%______________________________________________________________

Updated numerical expressions for the Stolz-Schombs relation were calculated
as a least squares linear fit to the data, excluding the systems identified
as low credential superhumpers in Paper~II\footnote{But note the almost perfect 
alignment to the Stolz-Schoembs relation of the lone point corresponding 
to RZ~Gru ($P_{\rm pSH} = 0.5022$; $\epsilon = 0.2455$), considered to have 
low credentials in Paper~II and which was thus not included in the fit.},
as well as the pSH data point for
H$\alpha$~1129 which requires confirmation, the high $\epsilon$ point of 
EI~UMa, the low $\epsilon$ point of NS~Cnc, and the other individually 
identified stars in the figure. Expressing
the relation in the original form as $\epsilon$ versus $P_{\rm SH}$, or 
alternatively as a function of $P_{\rm orb}$, we have for
positive superhumps:
%\begin{eqnarray}
%\epsilon = 0.002746409(003349195) + 0.4810257(02047368) \times P_{\rm pSH} \\
%\epsilon = -0.0004903899(0.003863089) + 0.5403358(02582897) \times P_{\rm orb}
%\end{eqnarray}
%
%{\parindent0ex and for negative superhumps}
%\begin{eqnarray}
%\epsilon & = & -0.0138330(9311) - 0.174345(6001) \times P_{\rm nSH} \\
%\epsilon & = & -0.0140166(9248) - 0.165186(5696) \times P_{\rm orb}
%\end{eqnarray}
%
\begin{eqnarray}
\epsilon & = & 0.003(3) + 0.481(20) \times P_{\rm pSH} [{\rm h}]\\
\epsilon & = & 0.000(4) + 0.540(26) \times P_{\rm orb} [{\rm h}]
\end{eqnarray}

{\parindent0ex and for negative superhumps}
\begin{eqnarray}
\epsilon & = & -0.014(9) - 0.174(6) \times P_{\rm nSH} [{\rm h}]\\
\epsilon & = & -0.014(9) - 0.165(6) \times P_{\rm orb} [{\rm h}]
\end{eqnarray}

It is striking that the negative superhumps in five stars (labelled in 
Fig.~\ref{sh-epsilon-rel}; right) have a period excess much below the sequence
defined by the other systems. They almost appear to form a separate parallel 
lower sequence. The ratio between the observed and calculated (Eq.~3) 
period excess is 2.17 for OR~And, 1.92 (ASASSN-14ix), 1.87 (NS~Cnc),
1.91 (LZ~Mus) and 1.78 (KIC~940652). Similarly, the high $\epsilon$ point of 
the positive
superhump system EI~UMa lies 1.870, times above the expected values (while the
low point is not far from the overall relation). The corresponding value
for HS~0642 is 1.88. Is it a coincidence that all these deviating period
excesses are quite close to two times the values predicted by the 
Stolz-Schoembs relation? Doubling the period excess would occur if the
disk precession period increases by a similar factor (the exact factor
depends somewhat on the orbital period). This is dificult to envision.
Alternatively, the observational manifestation of the superhumps may change
such that their period changes by a factor of two. This may happen if the 
SH light source switches on and off, or becomes visible to the observer, 
either one or two times during each synodic orbit of the secondary
around the eccentric or tilted accretion disk. However, I am not aware of 
any theoretical work or of simulations which support such a scenario.
In this context it may be worthwhile to remind the reader of other
unexpected relations between orbital, superhump and beat periods involving
small integral numbers within their formal error margins, namely the 
supraorbital variations of V603~Aql \citep{Bruch18} and V1974~Cyg 
\citep[][and Paper~II]{Semeniuk94} on twice, and possibly of RZ~Gru
(Paper~I) on four times the calculated beat period between orbit and 
superhump. In the other extreme, supraorbital periods of half 
(OR~And; Sect.~\ref{OR And}) or a quarter (DK~Lac; Sect.~\ref{DK Lac}) 
of the calculated beat period are seen. Is this just numerology, or do 
these facts really tell us something about the physics of superhumps?

% Doubling the period excess means reducing the
% disk precession period by a factor of two (which may be dificult to
% envision) or changing the observational manifestation of the superhump
% such that its period doubles. However, in view of the few system which 
% exhibit this appearent doubling of the period excess it is probably still
% idle to speculate about the reality and possible explanations of this effect.
% But it is intriguing that in some systems, most clearly in V603~Aql
% \citep{Bruch18}, a so far unexplained supraorbital period of exactly twice 
% the disk precession period predicted from the observed orbital and superhump 
% periods has been detected. 

As is well known, theory predicts an upper limit to the mass ratio of CVs
which can develop positive superhumps. The exact value is contested. Numbers
between $q = M_{\rm sec} / M_{\rm prim} = 0.22$ and 0.39 are quoted in the
literature \citep{Whitehurst91, Pearson06, Smak20}. Since on average the 
secondary star mass of CVs but not the WD mass increases with the orbital 
period, the average mass ratio also 
increases with period. Consequently, long period CVs should not exhibit 
positive superhumps. Yet, Fig.~\ref{sh-epsilon-rel} contains four stars with 
an orbital period above 0.2~d. Two of these, KIC~9406653 and GZ~Gru, have
already been discussed in Paper~II. The others are CN~Vel and EI~UMa.
KIC~9406653 is a double lined 
spectroscopic binary with a securely measured mass ratio of $q=0.83$
\citep{Gies13}. In the other cases no direct measurements of $q$ are
available. But the semi-periodic mass-period relation of \citet{Knigge11}
permits to estimate the secondary star mass or to provide a lower limit.
For CN~Vel, $M_{\rm sec}$ is estimated to be 0.58~$M_\odot$. The orbital period 
of the remaining two systems lies above the upper limit of the list provided 
by \citet{Knigge11} which ends at a period of 0.2468~d, corresponding to 
$M_{\sec}=0.7\, M_\odot$. I take this as a lower limit to the secondary
star mass of EI~UMa and RZ~Gru. Assuming the primary star mass to be equal
to the Chandrasekhar mass, lower limits of $q>0.39$ for CN~Vel
and $q>0.48$ for EI~UMa and RZ~Gru\footnote{In Paper~II, too low a secondary 
star mass was erroneously assumed for RZ~Gru, leading to a more favourable
but wrong conclusion about the compatibility of its superhumps with theory.}
are derived. Thus, while superhumps in CN~Vel are just barely compatible 
with theory, this is not true for the other three long period systems 
exhibiting positive superhumps, aggravating the problem outlined in Paper~II.  

In addition to superhumps, the power spectra of several of the stars of this 
study as well as in Papers~I and II contain enhanced power within a more or
less broad frequency range indicative of QPOs with periods on time scales of 
the order of an hour. 
% Table~\ref{Table: Long period QPOs} 
Table~7 lists those systems together with the approximate range of periods. 
In addition to hourly time scale QPOs, HS~0642 also shows QPOs with 
quasi-periods of the order of a day, as does MV~Lyr. QPOs in this period range
would be difficult to detect in terrestrial observations but they show up in 
the long TESS (or Kepler) data trains. I am not aware that this phenomenon has
been discussed in the literature. Since it is also not the topic of this 
study, I will refrain here from looking deeper into this issue. But it may
well deserve some specific consideration in a dedicated publication.

%--------------------------------------------------------------
\begin{table}
\label{Table: Long period QPOs}	
\centering
	\caption{NLs exhibiting long period QPOs.}

\begin{tabular}{lll}
\hline
Name   & Period range (h) & Reference \\
\hline
UU Aqr & \phantom{1}2.15 \ldots \phantom{3}2.40 & Paper~II \\
TT Ari & \phantom{1}0.25 \ldots \phantom{3}0.30 & Paper~I \\
BZ Cam & \phantom{1}1.70 \ldots \phantom{3}3.00 & Paper~II  \\
LS Cam & \phantom{1}0.80 \ldots \phantom{3}0.85 & This work \\
V592 Cas & \phantom{1}0.15 \ldots \phantom{3}0.35 & Paper~II \\
V751 Cyg & \phantom{1}0.35 \ldots \phantom{3}0.40 & Paper~II \\
BB Dor & \phantom{1}0.25 \ldots \phantom{3}0.80 & Paper~II \\
RZ Gru & \phantom{1}0.35 \ldots \phantom{3}1.20 & Paper~I \\
V533 Her & \phantom{1}0.30 \ldots \phantom{3}0.70 & Paper~I \\
BH Lyn & \phantom{1}3.40 \ldots \phantom{3}3.80 & Paper~II \\
MV Lyr & 15\phantom{.00} \ldots 23 & Paper~I \\
CP Pup & \phantom{1}1.45 \ldots \phantom{3}1.65 & Paper~I \\
LN UMa & \phantom{1}2.20 \ldots \phantom{3}2.70 & This work \\
CN Vel & \phantom{1}1.85 \ldots \phantom{3}2.50 & This work \\
HS 0642 & \phantom{1}1.70 \ldots \phantom{3}2.40 & This work  \\
HS 0642 & 12\phantom{.00} \ldots 36 & This work \\
\hline
\end{tabular}
\end{table}

\section{Summary}
\label{Summary}

The analysis of the TESS light curves of all stars classified as novalike
variables or old nova in the Ritter \& Kolb catalogue (excluding AM~Her type
stars), and the TESS data of which have not been the subject of previous
studies, revealed 19 systems to exhibit previously undetected superhumps. 
The scope of this study is exploratory, limited to the detection and the
characterization of the basic properties of the superhumps of the new SH 
systems (i.e., their periods, their persistance or coming and going, their 
waveforms), including also the SH properties of four previously known or 
recently detected superhump systems.
In many cases, a deeper investigation of the superhumps, their behaviour and 
their consequences for the structure of the individual binaries is warranted,
but must await more specific studies than the present compilation can provide.  

As a by-product the TESS data permitted to correct or to improve the
orbital periods of more than half of the investigated systems.
Additionally, further interesting
features observed in the light curves are briefly mentioned. Together with
similar properties seen in other stars of the entire Ritter \& Kolb sample
they will be the topic of future studies. 

The new detections listed here, together with other recent discoveries of
superhumps in NLs reported in the literature, all of them also based on TESS
data, elevate the known number of SH systems among non dwarf novae CVs by
more than 50\%, emphasizing the enormous potential of long, (almost) 
uninterrupted data trains to reveal even low amplitude periodic variations 
on time scales comparable to typical nightly terrestrial light curves
in the presence of often much stronger irregular variability auch as 
flickering, or in noisy light curves of faint stars. They also show that
superhumps in NLs are more common than previously known.

\section*{Acknowledgements}

This paper is based on data collected by the TESS and Kepler missions and 
obtained from the MAST data archive at the Space Telescope Science Institute 
(STScI). Funding for the missions is provided by the NASA Explorer Program
and the NASA Science Mission Directorate for TESS and Kepler, respectively. 
STScI is operated by the Association of Universities for Research in 
Astronomy, Inc., under NASA contract NAS 5-26555. Supportive data were
obtained from the data archives operated by the American Association of
Variable Star Observers (AAVSO) and the All-Sky Automated Search for 
Supernovae (ASAS-SN) project.

\section*{Data availability}

All data used in the present study are publically available at the
Barbara A.\ Mikulski Archive for Space Telescopes 
(MAST:\\ https://mast.stsci.edu/portal/Mashub/clients/MAST/\\Portal.html),
the AAVSO web site (https://www.aavso.org) and the ASAS-SN web site
(https://asas-sn.osu.edu).

%%%%%%%%%%%%%%%%%%%%%%%%%%%%%%%%%%%%%%%%%%%%%%%%%%%%%%%%%%%%%%%%%%%%%%%%

% Don't change these lines
\bsp	% typesetting comment
\label{lastpage}

\begin{thebibliography}{99}
\bibitem[\protect\citeauthoryear{Aungwerojwit et al.}{2005}]{Aungwerojwit05}
        Aungwerojwit A., G\"ansicke B.T., Rodr\'{\i}guez-Gil P., et al.,
        2005, A\&A, 443, 995
\bibitem[\protect\citeauthoryear{Barlow et al.}{2022}]{Barlow22}
        Barlow B.N., Corcoran K.A., Parker I.M., et al., 2022, ApJ, 928, 20
\bibitem[\protect\citeauthoryear{Bernardini et al.}{2012}]{Bernardini12}
        Bernardini F., de Martino D., Falanga M., et al., 
        2012, A\&A, 542, A22
\bibitem[\protect\citeauthoryear{Bond}{1979}]{Bond79}
        Bond H.E., 1979, {\it White Dwarfs and Variable Degenerate Stars}, 
        ed. H.M.\ van Horn, V.\ Weidemann, \& M.P.\ Savedoff,
        IAU Coll.\ 53, p.\ 495
\bibitem[\protect\citeauthoryear{Borucki et al.}{2010}]{Borucki10}
        Borucki W.H., Koch D., Basrik G., et al., 2010, Science, 327, 977
\bibitem[\protect\citeauthoryear{Bruch}{2021}]{Bruch21}
        Bruch A., 2021, MNRAS, 503, 953
\bibitem[\protect\citeauthoryear{Bruch}{2022}]{Bruch22}
        Bruch A., 2022, MNRAS, 514, 4718
\bibitem[\protect\citeauthoryear{Bruch}{2023}]{Bruch23}
        Bruch A., 2023, MNRAS, 519, 352
\bibitem[\protect\citeauthoryear{Bruch \& Cook}{2018}]{Bruch18}
        Bruch A., \& Cook L.M., 2018, New Astr., 63, 1
\bibitem[\protect\citeauthoryear{Deeming}{1975}]{Deeming75} 
        Deeming T.J., 1975, Ap\&SS, 39, 137
\bibitem[\protect\citeauthoryear{Dobrzycka et al.}{1998}]{Dobrzycka98}
        Dobrzycka D, Dobrzycki A., Engels D., \& Hagen H,-J.,
        1998, AJ, 115, 1634
\bibitem[\protect\citeauthoryear{G\"ansicke et al.}{2002}]{Gaensicke02}
        G\"ansicke B.T., Hagen H.-J., Kube J., et al., 2002, 
        {\it The Physics of Cataclysmic Variables and
        Related Objects}, eds. B.T.\ G\"ansicke, K.\ Beuermann, \& K.\ Reinsch,
        ASP Comf.\ Ser., 161, p.\ 623
\bibitem[\protect\citeauthoryear{Gaia Collaboration}{2020}]{GaiaCollaboration20}
        Gaia Collaboration, 2022, A\&A, 649, A1
\bibitem[\protect\citeauthoryear{Gies et al.}{2013}]{Gies13}  
        Gies D.R., Guo Z., Howell, S.B., 2013, ApJ, 775, 64
\bibitem[\protect\citeauthoryear{G\"ulse\c{c}en \& Esenoglu}{2014}]{Gulsecen14}
        G\"ulse\c{c}en H., \& Esenoglu H., 2014, New Astr., 144, 81
\bibitem[\protect\citeauthoryear{Hambsch}{2014a}]{Hambsch14a}
        Hambsch F.-J., 2014a, 
        http://ooruri.kusastro.kyoto-u.ac.jp/mailarchive/vsnet-alert/18036
\bibitem[\protect\citeauthoryear{Hambsch}{2014b}]{Hambsch14b}
        Hambsch F.-J., 2014b, 
        http://ooruri.kusastro.kyoto-u.ac.jp/mailarchive/vsnet-alert/18077
\bibitem[\protect\citeauthoryear{Hameury}{2020}]{Hameury20}
        Hameury J.-M., 2020, AdSpR, 66, 1004
\bibitem[\protect\citeauthoryear{Hameury et al.}{2022}]{Hameury22}
        Hameury J.-M., Lasota, J.-P., \& Shaw A.W., 2022, A\&A, 665, A7
\bibitem[\protect\citeauthoryear{Hillwig et al.}{1998}]{Hillwig98}
        Hillwig T.C., Robertson J.W., \& Honeycutt R.K., 1998, AJ, 115, 2044
\bibitem[\protect\citeauthoryear{Honeycutt}{2001}]{Honeycutt01}
        Honeycutt R.K., 2001, PASP, 113, 473
\bibitem[\protect\citeauthoryear{Honeycutt \& Kafka}{2004}]{Honeycutt04}
        Honeycutt R.K., \& Kafka S., 2004, AJ, 128, 1279
\bibitem[\protect\citeauthoryear{Kato et al.}{2001}]{Kato01}
        Kato T., Uemura M., Ishioka R., \& Kinnunen T., 2001, PASJ, 53, 1185
\bibitem[\protect\citeauthoryear{Katysheva \& Shugarov}{2008}]{Katysheva08}
        Katysheva N.A., \& Shugarov S.Yu. 2008, ASP Conf.\ Ser., 372,
        15th European Workshop on White Dwarfs, eds. R.\ Napivotzki \& 
        M.R.\ Burleigh, 523
\bibitem[\protect\citeauthoryear{Kniazev et al.}{2008}]{Kniazev08}
        Kniazev A., Revnivtsev M., Sazonov S., Burenin A., \& Tekola A.,
        2008, ATel, 1488
\bibitem[\protect\citeauthoryear{Knigge et al.}{2011}]{Knigge11}
        Knigge C., Baraffe I., \& Patterson J., 2011, ApJS, 194, 28 
\bibitem[\protect\citeauthoryear{Kozhevnikov}{2010}]{Kozhevnikov10}
        Kozhevnikov V.P., 2010, Astr.\ Lett., 36, 554
\bibitem[\protect\citeauthoryear{Li et al.}{2017}]{Li17}
        Li K, Hu S.-M., Zhou J.-L., et al., 2017, PASJ, 69, 28
\bibitem[\protect\citeauthoryear{Liller}{1980}]{Liller80}
        Liller M.H., 1980, IBVS, 1743
\bibitem[\protect\citeauthoryear{Littlefield et al.}{2022}]{Littlefield22}
        Littlefield C., Lasota J.-P., Hameury J.-M., et al.,
        2022, ApJL, 924, 8
\bibitem[\protect\citeauthoryear{Lomb}{1977}]{Lomb76}
        Lomb N.T., 1976, Ap\&SS, 39, 447
\bibitem[\protect\citeauthoryear{Mickaelian et al.}{2002}]{Mickaelian02}
        Mickaelian A.M., Balayan S.K., Ilovaisky S.A., et al.,
        2002, A\&A, 381, 894
\bibitem[\protect\citeauthoryear{Papadaki et al.}{2009}]{Papadaki09}
        Papadaki C., Boffin H.M.J., Stanishev V., et al., 
        2009, J.\ Astr.\ Data, 15,1
\bibitem[\protect\citeauthoryear{Patterson et al.}{2002}]{Patterson02} 
        Patterson J., Fenton W.H., Thorstensen J.R., et al., 2002, 
        PASP, 114, 1364 
\bibitem[\protect\citeauthoryear{Patterson et al.}{2005}]{Patterson05} 
        Patterson J., Kemp J., Harvey D.A., et al., 2005, PASP, 117, 1204
\bibitem[\protect\citeauthoryear{Pearson}{2006}]{Pearson06} 
        Pearson K.J., 2006, MNRAS, 371, 235
\bibitem[\protect\citeauthoryear{Peters \& Thorstensen}{2006}]{Peters06}
        Peters C.S., \& Thorstensen J.R., 2006, PASP, 118, 687
\bibitem[\protect\citeauthoryear{Piirola et al.}{2008}]{Piirola08}
        Piirola V., Vornanen T., Berdyugin A., \& Coyne G.V.,
        2008, ApJ, 684, 553
\bibitem[\protect\citeauthoryear{Pretorius \& Knigge}{2008}]{Pretorius08}
        Pretorius M.L., \& Knigge C., 2008, MNRAS, 385, 1471
\bibitem[\protect\citeauthoryear{Rawat et al.}{2022}]{Rawat22}
        Rawat N., Pandey J.C., Joshi A., \& Yadawa U., 
        2022, MNRAS, 512, 6054
\bibitem[\protect\citeauthoryear{Retter et al.}{1996}]{Retter99} 
        Retter A., Leibowitz E.M., \& Naylor T., 1999, MNRAS, 308, 140
\bibitem[\protect\citeauthoryear{Retter et al.}{1998}]{Retter98}
        Retter A., Liller W., \& Garradd G., 1998, IAU Circ., 7124
\bibitem[\protect\citeauthoryear{Ricker et al.}{2014}]{Ricker14}
        Ricker G.R., Winn J.N., Vanderspek R., et al., 2014, 
        J.\, Astr.\, Tel.\, Instr.\, \& Systems, 1, 014003
\bibitem[\protect\citeauthoryear{Ringwald}{1993}]{Ringwald93}
        Ringwald F.A., 1993, {\it The Cataclysmic Variables from the 
        Palomar-Green Survey}, PhD thesis, Dartmouth College
\bibitem[\protect\citeauthoryear{Ritter \& Kolb}{2003}]{Ritter03}
        Ritter H., \& Kolb U., 2003, A\&A, 404, 301
\bibitem[\protect\citeauthoryear{Rodr\'{\i}guez-Gil et al.}{2007a}]
        {Rodriguez-Gil07a}
        Rodr\'{\i}guez-Gil P., G\"ansicke B.T., Hagen H.-J., et al.,
        2007a, MNRAS, 377, 1747
\bibitem[\protect\citeauthoryear{Rodr\'{\i}guez-Gil et al.}{2009}]
        {Rodriguez-Gil09}
        Rodr\'{\i}guez-Gil P., Mart\'{\i}nez-Pais I.G., \&
        de la Cruz Rodr\'{\i}guez I.G., 2009, MNRAS, 395, 973
\bibitem[\protect\citeauthoryear{Rodr\'{\i}guez-Gil et al.}{2007b}]
        {Rodriguez-Gil07b}
        Rodr\'{\i}guez-Gil P., Schmidtobreick L., \& G\"ansicke B.T.,
        2007b, MNRAS, 374, 1359
\bibitem[\protect\citeauthoryear{Rodr\'{\i}guez-Gil \& Torres}{2005}]
        {Rodriguez-Gil05}
        Rodr\'{\i}guez-Gil P., \& Torres M.A.P., 2005, A\&A, 431, 289
\bibitem[\protect\citeauthoryear{Savitzky \& Golay}{1964}]{Savitzky64}
        Savitzky A., \& Golay M.J.E., 1964, Analytical Chemistry, 36, 1627
\bibitem[\protect\citeauthoryear{Sazonov}{2008}]{Sazonov08}
        Sazonov S., Revnivtsev M., Burenin R., et al., 2008, A\&A, 487, 509
\bibitem[\protect\citeauthoryear{Scargle}{1982}]{Scargle82}
        Scargle J.D., 1982, ApJ, 263, 853
\bibitem[\protect\citeauthoryear{Scaringi et al.}{2022a}]{Scaringi22a}
        Scaringi S., Groot P.J., Knigge C., et al., 2022a, Nature, 604, 447
\bibitem[\protect\citeauthoryear{Scaringi et al.}{2022b}]{Scaringi22b}
        Scaringi S., Groot P.J., Knigge C., et al., 2022b, MNRAS, 514, L11
\bibitem[\protect\citeauthoryear{Schaefer}{2022}]{Schaefer22}
        Schaefer B.E., 2022, MNRAS, 517, 3640
\bibitem[\protect\citeauthoryear{Schmidtobreick et al.}{2005}]{Schmidtobreick05}
        Schmidtobreick L., Tappert C., Galli L., \& Whiting A., 
        2005, IBVS, 5627
\bibitem[\protect\citeauthoryear{Schwarzenberg-Czerny}{1991}]
        {Schwarzenberg-Czerny91} 
        Schwarzenberg-Czerny A., 1991, MNRAS, 253, 198
\bibitem[\protect\citeauthoryear{Semeniuk et al.}{1994}]{Semeniuk94} 
        Semeniuk I., Pych W., Olech A., \& Ruszkowski M., 1994 Acta Astr.,
        44, 277
\bibitem[\protect\citeauthoryear{Shafter \& Ulrich}{1982}]{Shafter82}
        Shafter A.W., \& Ulrich R.K., 1982, Bull.\ AAS, 14, 880
\bibitem[\protect\citeauthoryear{Shears et al.}{2016}]{Shears16}
        Shears J., G\"ansicke B.T., Rodr\'{\i}guez-Gil P., et al.,
        2016, J.\ Brit.\ Astron.\ Assoc., 126, 42
\bibitem[\protect\citeauthoryear{Sing et al.}{2004}]{Sing04}
        Sing D.K., Howell S.B., Szkody P., \& Cordova F.A., 
        2004, PASP, 116, 1056
\bibitem[\protect\citeauthoryear{Sion \& Guinan}{1982}]{Sion82}
        Sion E.M., \& Guinan E.G., 1982, {\it Advances in Ultraviolet 
        Astronomy. Four Years of IUE Research}, ed. Y.\ Kondo, J.M.\ Mead, 
        \& R.D. Chapman, NASA CP-2238, p.\ 460 
\bibitem[\protect\citeauthoryear{Smak}{2020}]{Smak20} 
        Smak, J. 2020, Acta Astron., 70, 313
\bibitem[\protect\citeauthoryear{Stefanov \& Stevanov}{2023}]{Stefanov23}
        Stefanov S.Y., Latev G., Boeva S., \& Moyseev M., 
        2022, MNRAS, 516, 2775
\bibitem[\protect\citeauthoryear{Stefanov et al.}{2022}]{Stefanov22}
        Stefanov S.Y., Stefanov A.K., 2023, MNRAS, 520, 3187
\bibitem[\protect\citeauthoryear{Stolz \& Schoembs}{1984}]{Stolz84} 
        Stolz V., Schoembs R., 1984, A\&A, 132, 187
\bibitem[\protect\citeauthoryear{Szkody et al.}{2004}]{Szkody06}
        Szkody P., Henden A., Ag\"ueros M., et al., 2006, AJ, 131, 973
\bibitem[\protect\citeauthoryear{Tappert et al.}{2013}]{Tappert13}
        Tappert C., Schmidtobreick L., Vogt N., \& Ederoclite A.,
        2013, MNRAS, 436, 2412
\bibitem[\protect\citeauthoryear{Thorstensen}{1986}]{Thorstensen86}
        Thorstensen J.R., 1986, AJ, 91, 940
\bibitem[\protect\citeauthoryear{Thorstensen et al.}{2017}]{Thorstensen17}
        Thorstensen J.R., Ringwald A.F., Taylor C.J., et al.,
        2017, RNAAS, 1, 29 
\bibitem[\protect\citeauthoryear{Thorstensen et al.}{2015}]{Thorstensen15}
        Thorstensen J.R., Taylor C.J., \& Peters C.S., 2015, AJ, 149, 128
\bibitem[\protect\citeauthoryear{Warner}{1995}]{Warner95}
        Warner B., 1995, {\it Cataclysmic Variable Stars}, Cambridge 
        University Press, Cambridge
\bibitem[\protect\citeauthoryear{Warner}{2002}]{Warner02}
        Warner B., \& Woudt P.A., 2002, PASP, 114, 1222
\bibitem[\protect\citeauthoryear{Wenzel}{1987}]{Wenzel87}
        Wenzel W., 1987, IBVS, 3086
\bibitem[\protect\citeauthoryear{Whitehurst \& King}{1991}]{Whitehurst91} 
        Whitehurst R., King A., 1991, MNRAS, 249, 25
\bibitem[\protect\citeauthoryear{Wolfe et al.}{2013}]{Wolfe13}
        Wolfe A., Sion E.M., \& Bond H.E., 2013, AJ, 145, 168
\bibitem[\protect\citeauthoryear{Worpel et al.}{2020}]{Worpel20}
        Worpel H., Schwope A.D., Traulsen I., Mukai K., \& Ok S.,
        2020, A\&A, 639, A17
\end{thebibliography}
\end{document}